\documentclass[aps, pra,twocolumn,notitlepage, showpacs,superscriptaddress,10pt,floatfix]{revtex4-2}

\usepackage{graphicx}
\usepackage{grffile}
\usepackage{amsthm}
\usepackage{epsfig}
\usepackage{amsmath,amssymb,amsfonts,mathtools,bbm,MnSymbol,mathrsfs,xfrac}
\usepackage{longtable}
\usepackage{tabularx}
\usepackage{multirow}
\usepackage{makecell}
\usepackage{colortbl}
\usepackage[table,svgnames,xcdraw]{xcolor}
\usepackage{array, graphicx}
\usepackage[breaklinks=true,colorlinks=true,linkcolor=blue,urlcolor=blue,citecolor=blue]{hyperref}
\usepackage{comment}
\usepackage{color}
\usepackage[makeroom]{cancel}

\usepackage{tikz}
\usetikzlibrary{decorations.pathreplacing,calligraphy}
\usepackage{pgfplots}
\usepackage{tikz-3dplot}
\tdplotsetmaincoords{60}{115}
\pgfplotsset{compat=newest}

\usepackage{soul}
\pdfoutput=1

\renewcommand{\[}{\begin{equation}}
\renewcommand{\]}{\end{equation}}

\newcommand{\ket}[1]{|#1\rangle}
\newcommand{\bra}[1]{\langle#1|}
\newcommand{\braket}[2]{\langle#1|#2\rangle}
\newcommand{\pro}[2]{|#1\rangle\!\langle#2|}
\newcommand{\mean}[1]{\left\langle#1\right\rangle}
\newcommand{\abs}[1]{\left|#1\right|}

\newcommand{\tr}{\mathrm{tr}}

\newcommand{\norm}[1]{\left\lvert\left\lvert#1\right\rvert\right\rvert}

\newcommand{\R}{{\hat{\rho}}}
\renewcommand{\S}{{\hat{\sigma}}}
\newcommand{\I}{{\hat{I}}}

\newcommand{\HS}{\mathcal{H}}

\newcommand{\U}{{\hat{U}}}  
\newcommand{\h}{{\hat{H}}}
\newcommand{\V}{{\hat{V}}}

\newcommand{\ham}{{\hat{H}}}

\definecolor{mygray}{gray}{0.6}

\newcommand{\w}{w}
\theoremstyle{definition}

\newtheorem{theorem}{Theorem}

\newtheorem{corollary}{Corollary}

\definecolor{dfcol}{cmyk}{1, 0.2108, 0.13, 0.3}
\newcommand{\df}[1]{\ifthenelse{\boolean{}}{\textcolor{dfcol}{[{\bf DF}: #1]}}{}}

\usepackage{physics}
\usepackage[page]{appendix}

\pdfoutput=1

\begin{document}


\title{
Minimal time required to charge a quantum system
}

\author{Ju-Yeon Gyhm}
\email{kjy3665@snu.ac.kr}
\affiliation{Department of Physics and Astronomy, Seoul National University, 1 Gwanak-ro, Seoul 08826, Korea}

\author{Dario Rosa}
\email{dario\_rosa@ibs.re.kr}
\affiliation{Center for Theoretical Physics of Complex Systems, Institute for Basic Science (IBS), Daejeon - 34126, Korea}
\affiliation{Basic Science Program, Korea University of Science and Technology (UST), Daejeon - 34113, Korea}

\author{Dominik \v{S}afr\'{a}nek}
\email{dsafranekibs@gmail.com}
\affiliation{Center for Theoretical Physics of Complex Systems, Institute for Basic Science (IBS), Daejeon - 34126, Korea}

\date{\today}

\begin{abstract}
We introduce a quantum charging distance as the minimal time that it takes to reach one state (charged state) from another state (depleted state) via a unitary evolution, assuming limits on the resources invested into the driving Hamiltonian. 
For pure states it is equal to the Bures angle, while for mixed states, its computation leads to an optimization problem. Thus, we also derive easily computable bounds on this quantity. The charging distance tightens the known bound on the mean charging power of a quantum battery, it quantifies the quantum charging advantage, and it leads to an always achievable quantum speed limit. In contrast with other similar quantities, the charging distance does not depend on the eigenvalues of the density matrix, it depends only on the corresponding eigenspaces. This research formalizes and interprets quantum charging in a geometric way, and provides a measurable quantity that one can optimize for to maximize the speed of charging of future quantum batteries.
\end{abstract}

\maketitle

\section{Introduction}

The study of isolated quantum systems has been always viewed as an interesting subject in theoretical physics: to name just a single, prominent, example one can think of the well-known problem of defining thermalization in an isolated quantum setup.
In this respect, the celebrated Eigenstate Thermalization Hypothesis has provided the first mechanism to describe thermalization in isolated, and so governed by \emph{unitary} evolution, settings~\cite{Deutsch_2018, PhysRevE.90.052105, PhysRevE.97.012140, PhysRevE.99.042139, PhysRevLett.123.230606, PhysRevLett.130.140402}.

Until very recently, isolated quantum systems have been regarded as useful playgrounds to study ideal and academically relevant scenarios~\cite{Mori_2018, PhysRevB.96.020406, PhysRevA.69.052315}.
On the other hand, it is often assumed that \emph{realistic} quantum systems cannot be practically isolated, while  they are forced to interact with their surroundings~\cite{1989101}.
For this reason, realistic scenarios are usually investigated within the \emph{open quantum systems} paradigm~\cite{JOHNSON1985405, Rotter_2015, REDFIELD19651}.

However, recent tremendous experimental advances in, for instance, quantum simulators are forcing us to reconsider this assumption~\cite{PhysRevA.20.1521}. 
It is by now possible to build and manipulate small quantum systems in almost perfect isolation from their surroundings (at least up to a certain time), thus realizing experimental examples of quantum systems which can be viewed, to a large extent, as isolated rather than open~\cite{PhysRevResearch.3.033043, Bernon2013, Léonard2023}.
As a consequence, we are dealing with a substantial growth of interest in studying isolated quantum systems with a view to practical applications, and in exporting theoretical tools from the open systems paradigm to its isolated counterparts.

These experimental developments, in turn, have been at the core of the recent surge of interest towards the so-called \emph{quantum technologies}, \textit{i.e.}, small and (at least to a large extent) isolated quantum systems which can be used to perform a given task~\cite{dowling2003quantum}.
The interest in developing such technologies relies on the possibility of using their quantum features -- like coherence and entanglement -- to reach performances unobtainable by means of classical systems only, thus realizing a \textit{quantum advantage}~\cite{Daley2022, doi:10.1137/S0097539795293172}. 

Following this idea, several concrete examples of quantum technologies have been introduced, with the most notable example provided by quantum computers~\cite{aaronson2008limits, cao2018quantum, menges2016temperature}.
Other examples include quantum teleportation~\cite{PhysRevA.93.012311}, quantum simulation~\cite{RevModPhys.86.153}, quantum cryptography~\cite{pirandola2020advances}, quantum sensors~\cite{LIGO2013,degen2017quantum,cheiney2018navigation}, and quantum batteries~\cite{campaioli2023colloquium}, the latest being the main focus of this paper.

Quantum batteries, as the name suggests, are quantum mechanical systems that can be conveniently used to accumulate energy in their excited states and release it when necessary~\cite{Binder_2015, PhysRevE.87.042123}.
As customary when dealing with quantum technology, the effects that quantum mechanical ingredients, like entanglement, play on energy accumulation and delivery have been studied in certain detail in the last few years.
To name some of them, quantum effects have been proven to be beneficial in work extraction~\cite{PhysRevE.87.042123, PhysRevLett.111.240401, tirone2023work}, energy storage~\cite{PhysRevE.99.052106, PhysRevResearch.2.023095, PhysRevApplied.14.024092}, available energy~\cite{allahverdyan2004maximal, touil2022ergotropy, safranek2023work, yang2023battery, mula2023ergotropy, erdman2022reinforcement, morrone2023charging, morrone2023daemonic, PhysRevLett.131.030402}, charging stability~\cite{Friis2018precisionwork,Rosa2020, PhysRevB.100.115142, PhysRevE.100.032107, shaghaghi2022micromasers, shaghaghi2023lossy, rodríguez2023aidiscovery}, and charging power~\cite{ PhysRevLett.125.236402,  PhysRevResearch.5.013155, PhysRevLett.120.117702, PhysRevResearch.4.013172, PhysRevB.98.205423, PhysRevLett.122.047702, PhysRevLett.118.150601, seah2021quantum, PhysRevLett.128.140501, PhysRevA.97.022106, gyhm2023beneficial, Mohan2021, PhysRevLett.125.040601}.
Focusing on this last figure of merit, the goal is to charge a battery in the shortest possible time --  which is one of the bottlenecks in preventing the widespread use of battery-based renewable technologies. Similarly, one might be interested in discharging a charged battery in the fastest possible time, for instance, to create high energy currents. This is the situation required, for example, in nuclear fusion or high-energy pulse lasers. 

To achieve these goals, one has to find the shortest time required to reach one state from another, i.e. from the quantum state representing  the discharged battery (the \emph{discharged state}) to the state representing the charged battery (the \emph{charged state}) and vice versa. 

Given the experimental successes in dealing with isolated systems delineated above, quantum batteries are ideally isolated systems~\cite{PhysRevA.106.042601, doi:10.1126/sciadv.abk3160}. Thus, we are interested in the smallest amount of time to reach one state from another using \emph{unitary evolution} only.


In this paper, we define a notion of the minimal time required to reach by unitary evolution a target state from another, motivated by finding the minimal time required to charge a quantum battery. These results will be applied to study the notion of \textit{quantum advantage in charging}, by which it was suggested that quantum charging leads to a shortcut in reaching the charged state, by evolving the state through entangled states, even though both initial and final states are product states. Classical charging does not create entanglement and does not lead to this advantage. 
Our results will provide a quantitative description of this shortcut phenomenon.

We will also sketch some applications of these results in other quantum tasks, such as quantum computing and quantum speed limits (QSL).
The latter follows since the minimal charging time can be used to define a quantum speed limit optimized for the unitary evolution that is achievable by definition. 

The paper is organized as follows: Section \ref{sec:quantum_distance_def} defines the quantum charging distance, which satisfies the axioms for a well-defined notion of a distance. In Section \ref{sec:alternative_expressions}, we show that the original definition can be rewritten in a form more suitable for computations. In Section \ref{sec:bounds}, we provide upper and lower bounds, and in Section~\ref{sec:examples}, we provide two examples where this formalism is applied.
In Section~\ref{sec:quantum_batteries}, we show how our results apply to the charging power of quantum batteries, tightening the previously known bound, and how the quantum charging distance allows for a precise definition of ``shortcuts'' in the space of quantum states as responsible for the presence of quantum charging advantage. 
In Section~\ref{sec:QSL}, we apply the quantum charging distance to derive a quantum speed limit that is optimized for unitary evolution and achievable by construction.
Finally, in Section \ref{sec:conclusions} we conclude the paper and present further directions to explore.

\section{Definition of quantum charging distance}
\label{sec:quantum_distance_def}

Consider two density matrices $\R$ and $\S$ with the same eigenspectra.  By definition, these can be connected via a unitary evolution that does not change the spectra and is generated by some (possibly time-dependent) Hamiltonian. We define the quantum charging distance as the minimal time it takes for the initial state $\R$ to evolve into the final state $\S$ by the potentially time-dependent Hamiltonian $\V_t$ normalized to one, i.e.,
\[\label{Def:D_Her}
D(\R,\S)=\min_{\V_t:\norm{\V_t}=1}T,
\]
where $\S=U \R U^\dag$ and
$
U=\mathrm{Texp}\left(-i\int_0^{T
} \V_t\, dt\right).
$ $\V_t$ is called the driving Hamiltonian and $\norm{~}$ denotes the operator norm, which is defined as the largest absolute eigenvalue.

The motivation for this definition is to find the minimal time it takes to charge a quantum system, considering the standard condition of the driving Hamiltonian used extensively in the quantum batteries literature~\cite{PhysRevLett.128.140501,PhysRevE.87.042123, Campaioli2018, Crescente_2020, PhysRevE.100.032107}. This condition on the norm also motivates the name ``quantum charging distance''. However, we would like to emphasise that at this point, until Section~\ref{sec:quantum_batteries}, there is no inherent notion of a battery or a charging process (i.e., changing the mean energy of the system). 
Operator norm $\|\V_t\|$ is used to set all potential Hamiltonians on an equal footing since the evolution (i.e., charging) speed can be trivially increased simply by rescaling eigenvalues of this Hamiltonian to increase this norm. In other words, $N$ times larger norm leads to $N$ times faster charging speed. We can view this norm as analogous to the electric potential, which mathematically acts similarly---doubling the potential doubles the electric current, given the same resistance. 
Physically, it is related to the energy invested into the creation of the Hamiltonian.
Thus, setting $\|\V_t\|=1$ intuitively accounts for investing the same resources into the creation of different driving Hamiltonians as measured by this norm.

First, we show that the definition above is indeed a proper distance measure.

\begin{theorem}\label{thm1}
$D$ defined above is a distance on the set of density matrices with equal spectra. In other words, it is well-defined, positive, symmetric, and it satisfies the triangle inequality.
\end{theorem}
\begin{proof}
See Appendix~\ref{app:1}.
\end{proof}

\section{Alternative expressions}
\label{sec:alternative_expressions}

Although $D$ is a well-defined distance, the original definition yields little idea of how to compute it. Therefore, we provide several alternative formulas that are more tractable.


We found it challenging to find an optimal time-dependent Hamiltonian, denoted as $\V_t$, that achieves the minimum required time as per the definition. However, we have discovered that we do not need to grapple with a time-dependent Hamiltonian. The first formula we are about to present demonstrates that we can consistently select a time-independent Hamiltonian that accomplishes the minimal evolution time. 


\begin{theorem}\label{theorem:D_equal_lnU}
The charging distance is equal to
    \[\label{eq_Unitary}
    D(\R,\S)= \min_{U:U\R U^{\dag}=\S} \|i \ln(U)\|.
    \]
In the above, the minimum is taken over all unitary operators $U$ that transform the density matrix $\R$ into $\S$. The unitary operator $U_{\mathrm{opt}}$ that attains this minimum is referred to as the optimal unitary, as it achieves the transformation $\R\rightarrow\S$ in the shortest possible time. Furthermore, the optimal driving Hamiltonian $\V_{\mathrm{opt}}$ such that $U_t=\exp(-i \V_{\mathrm{opt}} t)$ generates the shortest distance, is time-independent and given by $\V_{\mathrm{opt}}=i \ln(U_{\mathrm{opt}})/D(\R,\S)$.
\end{theorem}

\begin{proof}See Appendix~\ref{app:2}.
\end{proof}


Note that the driving Hamiltonian $\V_{\mathrm{opt}}$ above is optimal among all operators with norm one, see discussion of the effect of this norm in Section~\ref{sec:quantum_distance_def}, and the minimal time to achieve the target state for any unnormalized Hamiltonian in Section~\ref{sec:QSL}.

Having established that we need to consider only time-independent Hamiltonians, we encounter a remaining complexity in the form of the minimization process. This challenge arises from the requirement to optimize over all unitary operators that satisfy $\S=U\R U^\dag$. 

For pure states, however, we can perform this minimization and obtain the following analytic expression.


\begin{theorem}\label{thm:pure_states}
    For the pure states $\R=\pro{\psi}{\psi}$ and $\S=\pro{\varphi}{\varphi}$, the charging distance is given by
    \[\label{C_U_pure}
    D(\ket{\psi},\ket{\varphi})=\arccos{\abs{\braket{\psi}{\varphi}}}.
    \]
An optimal charging unitary is
\[
\label{eq:optimal_U}
U_{\mathrm{opt}}=\pro{\varphi}{\psi}e^{i\phi_1}+\pro{\varphi_\perp}{\psi_\perp}e^{i\phi_2}+\hat{I}_{d-2},
\]
generated by the corresponding optimal Hamiltonian
\[
\V_{\mathrm{opt}}=ie^{i(\phi_1-\phi_2)}\pro{\psi_\perp}{\psi}-ie^{i(\phi_2-\phi_1)}\pro{\psi}{\psi_\perp}.
\]
There, $\ket{\psi_\perp}$ and $\ket{\varphi_\perp}$ are states orthogonal to $\ket{\psi}$ and $\ket{\varphi}$, respectively, in the two-dimensional plane spanned by the latter two vectors. $\hat{I}_{d-2}$ is the identity matrix in the other orthogonal dimensions.  For non-orthogonal states, $\abs{\braket{\varphi}{\psi}}\neq 0$, we have $e^{i\phi_1}=\frac{\braket{\varphi}{\psi}}{\abs{\braket{\varphi}{\psi}}}$ and $e^{i\phi_2}=\frac{\braket{\varphi_\perp}{\psi_\perp}}{\abs{\braket{\varphi_\perp}{\psi_\perp}}}$. On the other hand, for orthogonal states, $\abs{\braket{\varphi}{\psi}}= 0$, we have $e^{i\phi_1}=-\braket{\varphi_\perp}{\psi}$ and $e^{i\phi_2}=\braket{\varphi}{\psi_\perp}$.

\end{theorem}

\begin{proof}See Appendix~\ref{app:3}.
\end{proof}

Interestingly, the result for pure states, eq.~\eqref{C_U_pure}, can also be derived from the well-known Mandelstam-Tamm bound~\cite{bengtsson2017geometry,PhysRevLett.103.160502}, especially then~\cite{uffink1993}, which contains standard deviation in energy as one of its parameters. In contrast, the charging distance is formulated using the operator norm, which is always larger than the standard deviation. After the optimization for the minimal time, the two bounds coincide, as we show in Appendix~\ref{app:PurefromMT}.


Unlike for pure states, there is no general analytic solution for mixed states. However, even in the most general case, we can still significantly reduce the complexity of the minimization to optimization over only local subspaces.

\begin{theorem}\label{theorem:deg_mixed}
    Consider two density matrices with the same spectra and spectral decompositions
    \[\label{eq:spectral_dec}\R=\sum_{i=1}^{m}r_i\sum_{k=1}^{n_i}\pro{r_i^k}{r_i^k},\quad\S=\sum_{i=1}^m r_i\sum_{k=1}^{n_i}\pro{s_i^k}{s_i^k}.
    \]
    $d$ denotes the dimension of the Hilbert space, $n_i$ dimensions of the eigenspaces, $\sum_{i=1}^m n_i=d$, $r_i$ are eigenvalues that differ for different $i$, and both $\{\ket{r_i^k}\}$ and $\{\ket{s_i^k}\}$ are orthonormal bases. Then
    \[\label{eq:formula_mixed_states}
    D(\R,\S)=\min_{U_1,U_2,\cdots, U_m}\left\|i\ln( \bigoplus_{i=1}^m U_i \sum_{i=1}^m\sum_{k=1}^{n_i}\pro{s_i^k}{r_i^k})\right\|,
    \]
    where $U_i$ is a unitary operator acting on the subspace-eigenspace $\mathcal{H}_i = \mathrm{span} \{\ket{s_i^k}\}_{i=1}^{n_i}$.
\end{theorem}
\begin{proof}
    See Appendix~\ref{app:4}.
\end{proof}

For non-degenerate mixed states, this theorem can be further simplified to a minimization over $d$ angles.


\begin{corollary}\label{Coro:nond_mixed}
    When $\R=\sum_{i=1}^d r_i\pro{r_i}{r_i}$ and $\S=\sum_{i=1}^d r_i\pro{s_i}{s_i}$ are non-degenerate $d$-dimensional density matrices of the same spectra, the distance between them is equal to
    \[\label{eq:formulanondegenerate}
    \begin{split}
D(\R,\S)=\min_{\phi_1,\phi_2\cdots\phi_d}\left\|i\ln(\sum_{i=1}^d e^{i\phi_i}\pro{s_i}{r_i})\right\|.
    \end{split}
    \]   
\end{corollary}

This corollary is proved easily due to all eigenspaces being one-dimensional, in which the local unitaries are local phase rotations.

Theorem~\ref{theorem:deg_mixed} and Corollary~\ref{Coro:nond_mixed} reveal a very interesting property of the charging distance, that it does not depend on the eigenvalues, as long as the rank and the eigenspaces remain the same. In other words, the distance does not
change when eigenvalues are continuously changed, as long as the corresponding eigenspaces remain the same. However, if two previously unequal eigenvalues become equal, then the rank of the density matrix reduces. As a result, also the distance might discontinuously decrease~\footnote{This behavior could be already observed in previous literature, see Fig.~3 in Supplemental Material of~\cite{PhysRevLett.120.060409}}.
This is because the optimization suddenly goes over fewer larger eigenspaces corresponding to fewer unitary operators $U_i$ that each covers the same or a bigger subspace than before. Conversely, if the number of eigenspaces increases due to eigenvalues that used to be the same being no longer the same, also the charging distance might discontinuously jump to a higher value. 

This independence of the continuously varied eigenvalues clearly separates the charging distance from other quantities designed for similar purposes appearing in the QSL literature~\cite{PhysRevLett.120.060409, VittorioGiovannetti_2004, Deffner.2017}, such as the Bures angle. It further allows us to tighten the lower bound on this quantity, as we will show next.

\begin{figure}[t]
\begin{center}
\includegraphics[width=0.9\hsize]{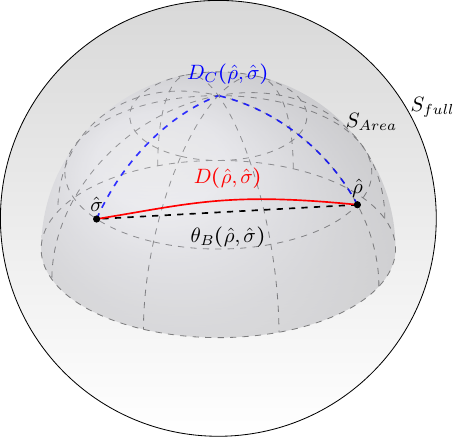}\\

\caption
{The quantum charging distance $D(\R,\S)$ is the length of the geodesic in the space given by a fixed spectrum of the density matrix. It measures the minimal time it takes to evolve one state to another by a unitary evolution generated by a driving Hamiltonian $\V$ with a fixed operator norm $\norm{\V}=1$. Bures angle, on the other hand, is always smaller than the charging distance. This is because the Bures angle $\theta_B(\R,\S)$ is equal to the quantum charging distance in the purification space \cite{gyhm2023to_appear}. In other words, if one is allowed to apply a unitary operator in an extended Hilbert space in which both $\R$ and $\S$ are pure, even a faster charging speed can be achieved. 
However, this is impossible without access to these unknown degrees of freedom. The unitary operation in the purification space corresponds to a non-unitary operation in the original Hilbert space, creating an additional shortcut for charging.  Classical charging allows only local unitary operations applied on the product states, 
thus, it does not create any entanglement in the charging process. As a result, classical charging is even more limited than quantum charging and the time to charge, as measured by the classical charging distance $D_C(\R,\S)$, is longer.
}\label{Fig:quantum_charging}
\end{center}
\end{figure}

\section{Bounds on the quantum charging distance}
\label{sec:bounds}

So far, we have only provided an analytic expression for the computation of the charging distance in the case of pure states, while for mixed states, numerical optimization methods may be required. To address this limitation, we also offer readily calculable upper and lower bounds for this quantity, applicable in all situations.




\begin{theorem}\label{theorem:bound_of_Dis}
    (Upper bound) For any $d$-dimensional density matrices $\R$ and $\S$ with the same spectra, the distance is bounded by
    \[\label{eq:uppermixed}
    D(\R,\S)\leq \pi\left(1-\frac{1}{d}\right).
    \]
\end{theorem}
\begin{proof} See Appendix~\ref{app:5}.
\end{proof}

The theorem guarantees that the distance is bounded by $\pi$ for even an infinite dimension of the Hilbert space. This reveals an interesting observation: while for pure states, the maximal charging distance is given by $\pi/2$ as per Theorem~\ref{C_U_pure}, for mixed states, this time is doubled.

We also found general lower bounds on the charging distance, related to the Bures angle. Defining the Uhlmann's fidelity~\cite{jozsa1994fidelity} between two mixed states as $\mathcal{F}(\R,\S)=\tr(\sqrt{\sqrt{\R}\S\sqrt{\R}})$, the Bures angle~\cite{nielsen2010quantum} is defined as
\[
\theta_B(\R,\S)=\arccos[\mathcal{F}(\R,\S)].
\]

The lower bounds follow.

\begin{theorem}\label{theorem:relation_QCD_Bures}(Lower bound) The charging distance is always larger than or equal to the Bures angle between two states with the same spectra, i.e.
\[\label{eq:lower_bound_1}
\theta_B(\R,\S)\leq D(\R,\S).
\]
\end{theorem}
\begin{proof}See Appendix~\ref{app:6}.
\end{proof}

See Fig.~\ref{Fig:quantum_charging} for illustration.
Note that Theorem~\ref{thm:pure_states} shows that for pure states, the charging distance and the Bures angle are equal, $D(\ket{\psi},\ket{\phi})=\theta_B(\ket{\psi},\ket{\phi})$.


The fact that the charging distance does not depend on the eigenvalues, as long as the corresponding eigenspaces do not change, allows us to make the lower bound above even tighter, by maximizing over these eigenvalues. We can formalize this as follows.




\begin{corollary}   
\label{theorem:relation_QCD_Bures_genelized2}
    (Tighter lower bound) Consider two density matrices with spectral decompositions given by Eq.~\eqref{eq:spectral_dec}. The charging distance is lower bounded as
    \[\label{eq:tigthermixed}
    \max_{i}\,\theta_B(\R_i,\S_i)\leq D(\R,\S).
    \]
    There, we denoted the maximally mixed states on the subspaces-eigenspaces as
    \[\R_i=\frac{1}{n_i}\sum_{k=1}^{n_i}\pro{r_i^k}{r_i^k},\quad \S_i=\frac{1}{n_i}\sum_{k=1}^{n_i}\pro{s_i^k}{s_i^k}.
    \]
\end{corollary}

\begin{proof} See Appendix~\ref{app:cor2}.
\end{proof}


This means that the minimal time it takes to charge a quantum system is at least as large as the slowest element in the chain: eigenspace of $\R$, represented by its maximally mixed state $\R_i$, that takes the longest time to transform into the corresponding eigenspace of $\S$ with the unitary transformation, lower bounds the total charging distance and thus also the total charging time.

\section{Examples}
\label{sec:examples}

Let us now discuss a couple of simple examples to make the reader acquainted with the machinery just developed.

\begin{figure}[t]
\begin{center}
\includegraphics[width=0.8\hsize]{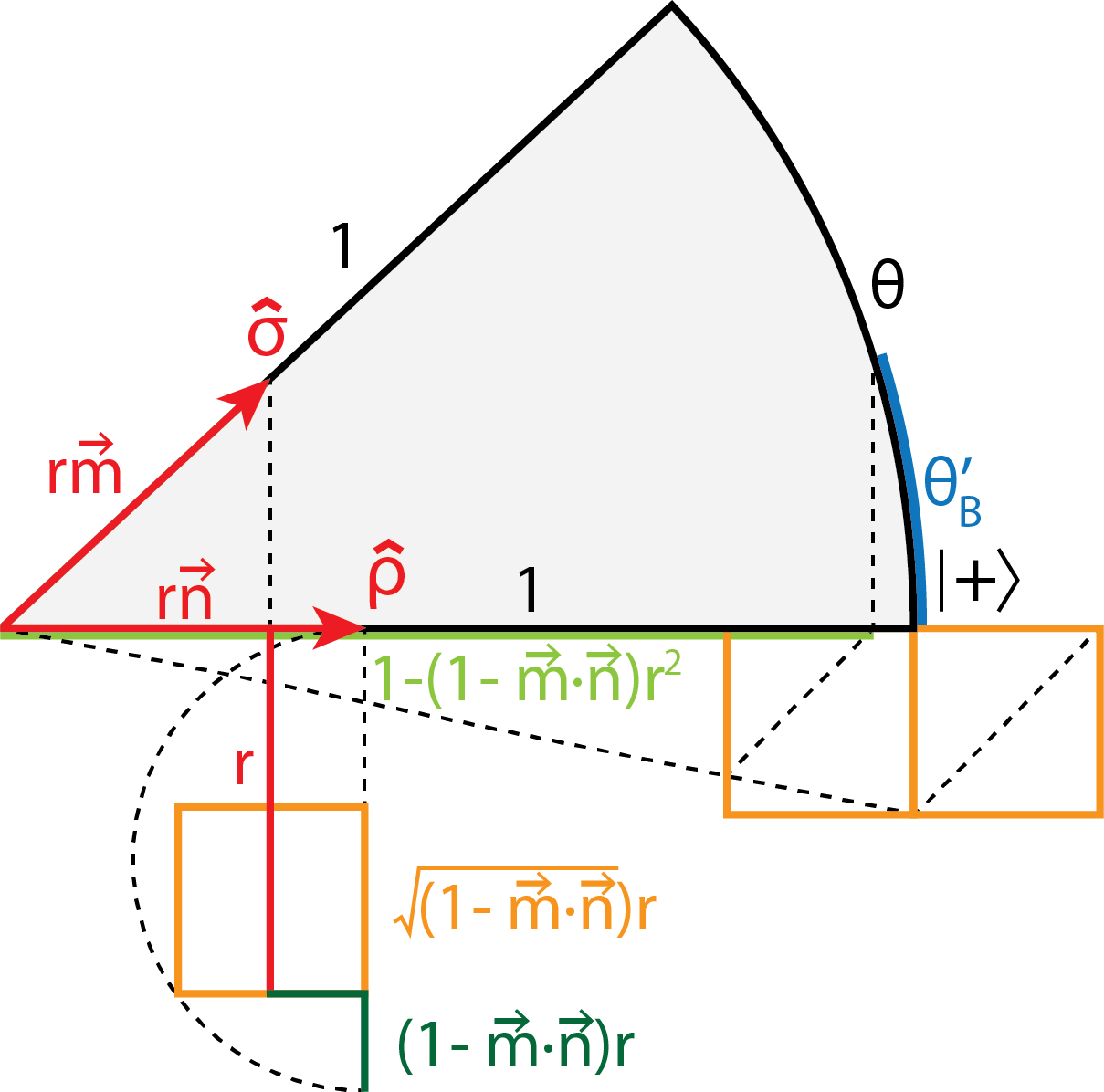}\\
\caption
{Cut through the Bloch sphere to illustrate the charging distance between two qubits. The charging distance is given by the curve's length on the Bloch sphere's surface, $D(\R,\S)=\theta/2$, connecting the two states. The Bures angle $\theta_B(\R,\S)=\theta_B'/2$ is always smaller and it depends on $r$. For $r=1$, they are both equal, while for $r\rightarrow 0$, $\theta_B\rightarrow 0$ while the charging distance stays constant. Objects below the triangle and the dashed lines are for the geometric construction of $\theta_B'$.
}
\label{Fig:angles}
\end{center}
\end{figure}

\subsection{Two level system}

Any two mixed states of a qubit with the same spectra can be written as $\R=(\I+r\vec{n}\cdot\vec{\sigma})/2$ and $\S=(\I+r\vec{m}\cdot\vec{\sigma})/2$. $\vec{n}$ and $\vec{m}$ are the unit vectors, $0\leq r\leq 1$, and $\vec{\sigma}$ denote the vector of Pauli matrices.

Using Corollary~\ref{Coro:nond_mixed}, for $r>0$ the charging distance is given by
\[
D(\R,\S)=\min_{\phi_1,\phi_2}\left\|i\ln(e^{i\phi_1}\pro{s_1}{r_1}+e^{i\phi_2}\pro{s_2}{r_2})\right\|,
\]
where $\{\ket{r_i}\}$ and $\{\ket{s_i}\}$ are eigenvectors of $\R$ and $\S$, respectively. This problem is mathematically equivalent to the case of pure states, see Theorem~\ref{thm:pure_states} and Appendix~\ref{app:3}. From this we obtain $D(\R,\S)=\arccos\abs{\braket{r_1}{s_1}}$. Diagonalizing the density matrices and inserting the eigenvectors into the formula yields 
\[
D(\R,\S)=\frac{1}{2}\theta=\frac{1}{2}\arccos\left(\vec{n}\cdot\vec{m}\right)
\]
for $r>0$, where $\theta$ is the angle between $\vec{n}$ and $\vec{m}$. For $r=0$, the charging distance is zero, which follows from Theorem~\ref{theorem:deg_mixed}.

In contrast, the Bures angle between the two qubits is given by~\cite{HUBNER1992explicit,jozsa1994fidelity} 
\[\begin{split}
\theta_B(\R,\S)&=\arccos(\sqrt{\tr[\R\S]+2\sqrt{\mathrm{det}(\R\S)}})\\
&=\frac{1}{2}\arccos(1-(1-\vec{n}\cdot\vec{m})r^2).
\end{split}
\]
When $r\rightarrow 0$, the charging distance conserves its value while the Bures angle converges to 0. We can easily derive $D\geq \theta_B$, corresponding to the result of Theorem~\ref{theorem:relation_QCD_Bures}. See Fig.~\ref{Fig:angles} for a geometric illustration.

\subsection{Three level system}

\begin{figure}[t]
\begin{center}
\includegraphics[width=1\hsize]{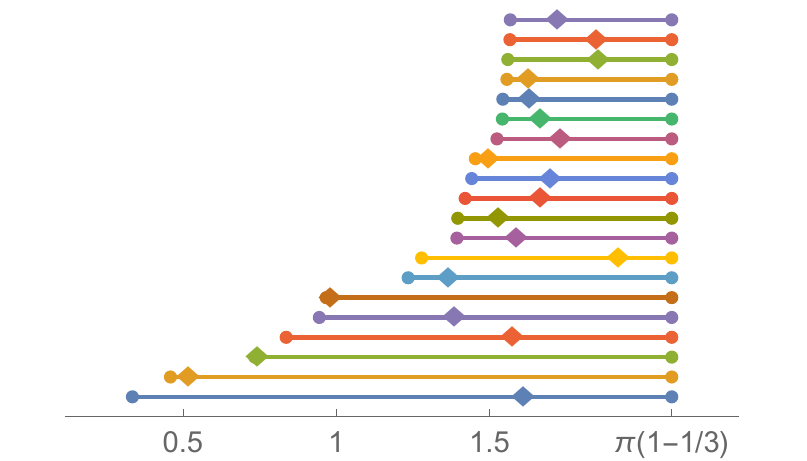}\\
\caption
{The quantum charging distance (diamonds), for twenty randomly generated couples $\R$ and $\S$ shown from top to bottom, in a three-dimensional Hilbert space was obtained numerically using Corollary~\ref{Coro:nond_mixed}. For each realization, we also plot the interval of the lower and the upper bounds $[\max_{i}\theta_B(\R_i,\S_i),\pi(1-1/3)]$, obtained from  Corollary~\ref{theorem:relation_QCD_Bures_genelized2} and Theorem~\ref{theorem:bound_of_Dis}, respectively.
}
\label{Fig:three}
\end{center}
\end{figure}

As a second example, we numerically compute the quantum charging distance for twenty randomly generated couples $\R$ and $\S$, in a three-dimensional system, together with upper and lower bounds. See Fig.~\ref{Fig:three}.

\section{Application to quantum batteries}
\label{sec:quantum_batteries}

The very notion of quantum charging distance has been inspired by the recent literature studying the charging power of quantum batteries (see \cite{campaioli2023colloquium} and references therein for an overview). As such, we are going to describe how the quantum charging distance relates to the problem of computing the charging power of a many-body quantum battery, in Subsection~\ref{subsec:charging_power}. After that, in Subsection~\ref{subsec:classical_vs_quantum}, we show how it allows for a quantitative description of the ``shortcuts in the space of quantum states'' which are at the heart of the quantum charging advantage of many-body quantum batteries.

\subsection{Relation to the charging power}
\label{subsec:charging_power}

Consider a battery Hamiltonian $\h$ which defines the natural energy levels of the system when it is not charged. The mean power of the quantum charging is defined as the difference between the initial and the final energy, divided by the time of charging,
\[
P=\frac{\left|\mean{\h}_\R-\mean{\h}_{\S}\right|}{
T}.
\]
Here, the mean energy is defined as 
$\mean{\h}_\R=\tr[\h \R]$.
A well-known and easy-to-derive bound states that the mean power is bounded by the product of the norms of the battery and the driving Hamiltonians as~\cite{PhysRevLett.118.150601, PhysRevLett.128.140501}
\[\label{eq:former_bound}
P\leq 2\|\h\| \|\V\|.
\]

We will show that the quantum charging distance provides a tighter bound.

The maximum energy difference between the two states is bounded by the trace distance. Trace distance, defined as $D_{\Tr}(\R,\S)=\frac{1}{2}\norm{\R-\S}_{1}$, can be equivalently written as~\cite{Rastegin_2007}
\[
D_{\Tr}(\R,\S)=\frac{1}{2}\max_{\hat{O}:\norm{\hat{O}}=1}\left(\mean{\hat{O}}_\R-\mean{\hat{O}}_{\S}\right).
\]
This gives the bound on the energy difference between the two states as
\[
\left|\mean{\h}_\R\!-\mean{\h}_{\S}\right|=\|\h\|\left|\mean{\frac{\h}{\|\h\|}}_{\!\R}\!\!\!-\mean{\frac{\h}{\|\h\|}}_{\!\S}\right|\leq 2\|\h\|D_{\Tr}(\R,\S).
\]

Realizing that the minimal time of charging is given by the quantum charging distance, $T\geq D(\R,\S)/\norm{\V}$, this gives a bound on the mean power of the charging as
\[\label{bound_of_power}
P\leq 2\|\h\| \|\V\| \frac{D_{\Tr}(\R,\S)}{D(\R,\S)}.
\]
The right-hand side is the largest mean charging power between $\R$ and $\S$ for the battery Hamiltonian $\h$ and the driving Hamiltonian $\V$, which are restricted by their operator norms, $\|\h\|$ and $\|\V\|$.

Using Theorem~\ref{theorem:relation_QCD_Bures} and a bound on the trace distance in terms of fidelity~\cite{PhysRevA.79.024302}, 
we derive
\[
\frac{D_{\Tr}(\R,\S)}{D(\R,\S)}\leq\frac{\sqrt{1-\mathcal{F}(\R,\S)^2}}{\arccos(\mathcal{F}(\R,\S))}\leq 1.
\]
The second inequality is saturated only when $\R=\S$. It follows from $\arccos{x}=\arcsin\sqrt{1-x^2}$ and $\sin x\leq x$ for $0\leq x \leq 1$, and from $\mathcal{F}(\R,\S)=1$ iff $\R=\S$. 
Thus, our newly derived bound is always tighter than the formerly known bound, Eq.~\eqref{eq:former_bound}.

Moreover, Eq.~\eqref{bound_of_power} is achievable due to the existence of the optimal battery Hamiltonian, $\h$, and the driving Hamiltonian, $\V_t$, which saturate the bound by definition.

\begin{figure}[t]
\begin{center}
\includegraphics[width=1\hsize]{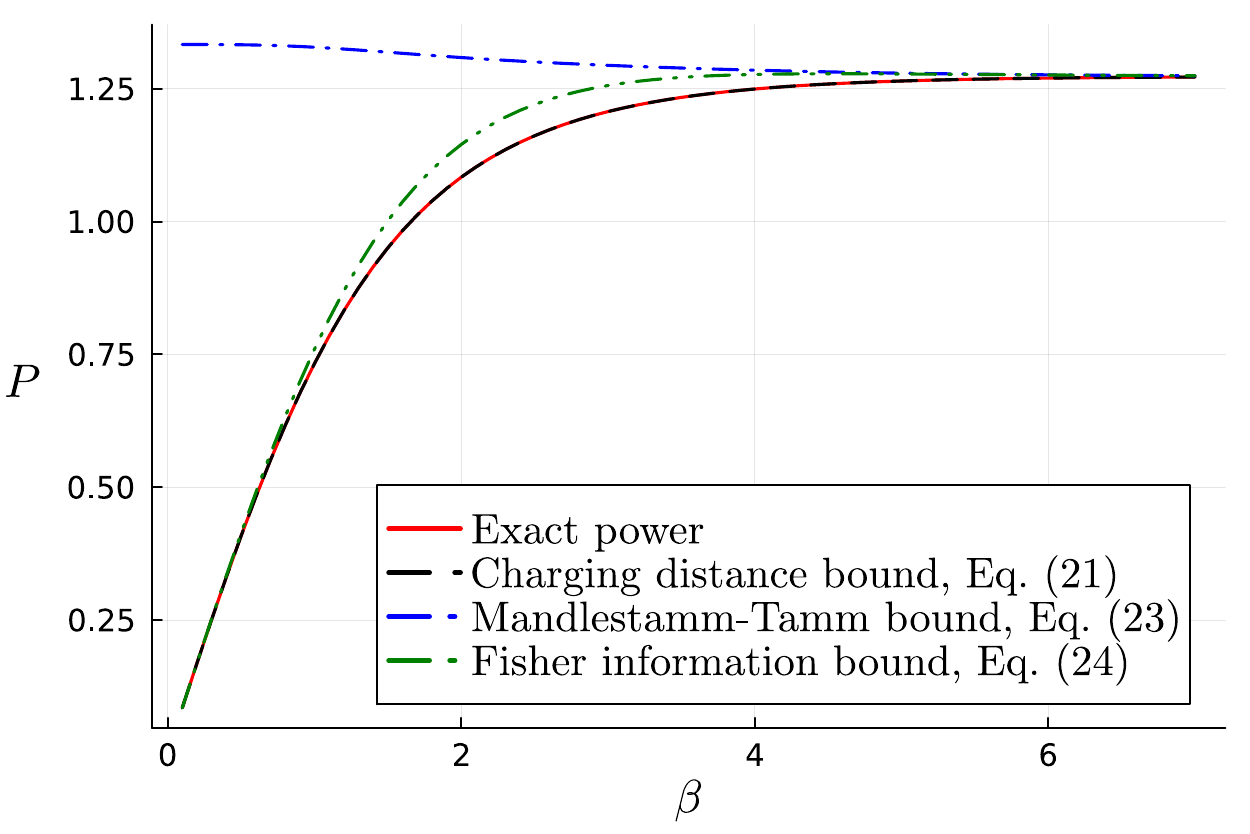}\\
\caption
{Exact power of a three-level mixed state charged with an optimal Hamiltonian and three bounds on power, with units $h=V=1$. The charging distance power bound is always tight for the optimal charging protocol when charging from the passive to the maximally charged state. For non-optimal charging protocols, other bounds might be tighter.
}
\label{Fig:power_bound}
\end{center}
\end{figure}

\subsubsection{Example and comparison}

We provide an example to show achievability of the charging distance bound Eq.~\eqref{bound_of_power} in comparison with previously derived bounds, coming from the Mandelstamm-Tamm QSL~\cite{PhysRevLett.118.150601} and from the Fisher information~\cite{PhysRevResearch.2.023113}. 
The first of the known bounds is tight for pure states, while the second is tight for both pure states and two-level systems. However, they do not provide tight bounds for a three-level system with a mixed state.

The Mandelstamm-Tamm power bound comes from the uncertainty relation between time and energy~\cite{PhysRevLett.118.150601}, so that
\[\label{bound_MT}
P\leq \dfrac{2}{T}\int_0^T \sqrt{\mean{\Delta \h^2}\mean{\Delta \V^2}}\ dt,
\]
which is tighter than the trivial bound, Eq.~\eqref{eq:former_bound}, due to the variance being always smaller than the operator norm. A tighter bound was derived using the Fisher information as~\cite{PhysRevResearch.2.023113},
\[\label{bound_prr}
P\leq \dfrac{1}{T}\int_0^T \sqrt {\mean{\Delta \h^2}I_E}\ dt,
\]
where the Fisher infromation $I_E$ is defined as
\[
I_E=2\sum_{i=1}^d \frac{(dp_{h_i}/dt)^2}{p_{h_i}}.
\]
$p_{h_i}$ is the probability of measuring the system to have energy $h_i$, given by the spectral decomposition of the battery Hamiltonian $\h=\sum_{i=1}^d h_i \pro{h_i}{h_i}$, i.e., $p_{h_i}=\bra{h_i}\R_t\ket{h_i}$.

We test the three-level system with a battery Hamiltonian with equidistant energy spacing,
\[
\h=-h\pro{-h}{-h}+0\pro{0}{0}+h\pro{h}{h}.
\]
We assume the initial state is prepared in the Gibbs state $\R=e^{-\beta \h}/Z $, where $Z$ is the partition function and $\beta$ is an inverse temperature. The charged state is $\S=e^{\beta \h}/Z$, with the maximum expectation energy of $\h$.

We find the optimal charging protocol using Corollary~\ref{Coro:nond_mixed}. The unitary operator that connects the depleted state, $\R$, with the charged state $\S$, is given by
\[
U=e^{i\phi_1}\pro{h}{-h}+e^{i\phi_2}\pro{0}{0}+e^{i\phi_3}\pro{-h}{h}.
\]
By minimizing over $\phi_1$, $\phi_2$ and $\phi_3$, we obtain the optimal driving Hamiltonian $\V$ through Theorem~\ref{theorem:D_equal_lnU} as
\[\label{op_V}
\V=V(\pro{h}{-h}+\pro{-h}{h}),
\]
independent of $\beta$.

Fig.~\ref{Fig:power_bound} illustrates the exact power when charging by the optimal driving Hamiltonian, Eq.~\eqref{op_V}. Further, we plot the charging distance bound, Eq.~\eqref{bound_of_power}, the Mandelstamm-Tamm bound, Eq.~\eqref{bound_MT} and the Fisher information bound, Eq.~\eqref{bound_prr}. 
The charging distance bound saturates to exact power, while other bounds do not. This tightness comes from our choice of $\V$, since the charging distance bound is saturated only when using the optimal protocol while transforming a passive state to the maximally charged state (with the same spectrum). This is exactly the situation that is the most relevant to quantum batteries, because charging from a passive state to the maximally charged state maximizes the capacity of the battery. In other words, when using the optimal protocol for charging the battery to its maximal capacity, the charging distance bound is always tight, and thus tighter than the other two bounds~\footnote{Note that in the charging power bound, Eq.~\eqref{bound_of_power}, we also assume that the battery Hamiltonian $\ham$ is centered around zero, meaning that the absolute values of the maximal and minimal eigenvalues are the same. If this does not hold, we can make the bound tighter by centering it by an additive constant.}.

Conversely, when non-optimal charging Hamiltonian $\V$ is employed, the other bounds may perform better.

\subsection{Quantum versus classical charging}
\label{subsec:classical_vs_quantum}

Consider two pure states, depleted state $\ket{00}$ and a charged state $\ket{11}$.
The optimal charging unitary is derived from Eq.~\eqref{eq:optimal_U} as
\[
U_{\mathrm{opt}}=\pro{00}{11}-\pro{11}{00}+\pro{10}{10}+\pro{01}{01}.
\]
This corresponds to the optimal Hamiltonian $\V_{\mathrm{opt}}=i(\pro{00}{11}-\pro{11}{00})$ and the minimal charging time  is $D(\ket{00},\ket{11})=\pi/2$. During this optimal evolution, the state passes through a maximally entangled state $\ket{\Phi^-}=(\ket{00}-\ket{11})/\sqrt{2}$ at time $t=\pi/4$.

Consider two product states $\R=\R_1\otimes\cdots\otimes \R_m$ and $\S=\S_1\otimes\cdots\otimes \S_m$, where $\R_i$ and $\S_i$ have the same spectra. We define the classical charging distance as the minimal time it takes to evolve $\R\rightarrow\S$ using only the local evolution,
\[
\V=\V_1\!\otimes\!\I\!\cdots\!\otimes\! \I\,+\,\cdots \,+\, \I\!\otimes\!\cdots \otimes\!\I\!\otimes\!\V_m,
\]
assumming $\norm{\V}=1$. (We allow time-dependent $\V$.)
The corresponding unitary $U_1\otimes\cdots \otimes U_m$ acts only on the local subsystems and consequently does not create any entanglement.

Taking the previous example, the classical charging distance is equal to $D_C(\ket{00},\ket{11})=\pi$. This can be achieved through different driving Hamiltonians, for example, by charging both qubits at the same time at half the power,
\[
\V=\frac{i}{2}\big[(\pro{0}{1}-\pro{1}{0})\otimes \I+\I\otimes(\pro{0}{1}-\pro{1}{0})\big],
\]
or by charging them sequentially with the full power,
\[
\V_t=\begin{cases}
i[(\pro{0}{1}-\pro{1}{0})\otimes \I], & 0\leq t< \pi/2,\\
i[\I\otimes(\pro{0}{1}-\pro{1}{0})],  & \pi/2\leq  t\leq \pi.
\end{cases}
\]

Clearly, both classical and quantum charging reach the target state. However, quantum charging utilized a shortcut by going through a maximally entangled state, leading to a shorter charging time. See Fig.~\ref{Fig:quantum_charging} for illustration.

This can be straightforwardly generalized to $N$ qubits as
$D_C(\ket{0\cdots 0},\ket{1\cdots 1})=N\pi/2$ and $D(\ket{0\cdots 0},\ket{1\cdots 1})=\pi/2$, showing that quantum charging provides $N$-fold advantage. The optimal charging Hamiltonian is given by $\V_{\mathrm{opt}}=i(\pro{0\cdots 0}{1\cdots 1}-\pro{1\cdots 1}{0\cdots 0})$ and the state passes through a GHZ state $(\ket{0\cdots 0}-\ket{1\cdots 1})/\sqrt{2}$ during the charging.

This offers a geometric interpretation, quantitatively formalizing the quantum charging advantage---a subject of extensive discussion within quantum batteries literature~\cite{Binder_2015, PhysRevLett.128.140501, andolina2019quantum, PhysRevResearch.2.023113, Campaioli2018, PhysRevLett.125.236402,  PhysRevLett.120.117702}.

\section{A Tight Quantum Speed Limit for Unitary Evolution}
\label{sec:QSL}

The charging distance introduced in previous sections can be used to define a tight quantum speed limit (QSL) for unitary evolution.

On general grounds, QSL is defined as the minimum time to evolve from one quantum state to another~\cite{PhysRevLett.65.1697, Deffner_2017}.
Mandelstam and Tamm proposed a first approach based on the time-energy uncertainty relation, suggesting that the minimum time is related to the standard deviation of energy~\cite{Mandelstam1991}. Margolus and Levitin later improved the QSL of Mandelstam and Tamm to make it tighter~\cite{MARGOLUS1998188}. In the following years, the concept has been refined and extended in many ways to include non-orthogonal states~\cite{PhysRevLett.103.160502, PhysRevLett.70.3365, pati2023exact}, non-unitary evolution~\cite{Funo_2019, PhysRevLett.110.050403, PhysRevLett.127.100404, PhysRevA.104.052620, Meng2015, Wu2020, Mohan2022}, and mixed states~\cite{UHLMANN1992329, Deffner_2013, PhysRevLett.72.3439, MONDAL2016689, PhysRevA.51.1820, PhysRevLett.120.060409}. 

A common feature of all these approaches is that the resulting QSLs are dependent on the spectra of the two states, $\rho$ and $\sigma$, under investigation.
This feature is desirable when dealing with two generic states which do not share the same spectra, and so they must be connected by some non-unitary evolution. On the other hand, when restricting to states which are unitarily connected, such a dependence on the spectra is redundant and, more in detail, it leads in general to loose bounds~\cite{PhysRevLett.120.060409}.

Therefore, it is an interesting problem to find QSLs that are more suitable for the peculiarities of unitary evolution, \textit{i.e.} which are independent of the eigenvalues and which give as tight as possible estimates of the minimum time required to unitarily evolve one state into another state.

This problem has been addressed and investigated in \cite{PhysRevLett.120.060409}. The authors defined the \emph{generalized Bloch angle} for $d$-dimensional mixed states, 
\[
\label{eq:generalized_phi}
\Phi(\R,\S)=\arccos(\frac{\tr(\R\S)-1/d}{\tr(\R^2)-1/d}) \ ,
\] 
which reduces to the standard notion of Bloch angle when dealing with 2-level systems.
From \eqref{eq:generalized_phi} the QSL follows
\[\label{eq:QSL_GB}
\begin{split}
    t_{GB}&=\frac{\Phi(\R,\S)}{v_{GB}},\quad \mathrm{with}\\
v_{GB}&=\frac{1}{T} \int_0^T \sqrt{\frac{2\tr(V_t^2 \R_t^2 - (V_t\R_t)^2)}{\tr(\R_t^2-1/d^2)}}dt.
\end{split}
\]
In \cite{PhysRevLett.120.060409}, it is shown that eq~\eqref{eq:QSL_GB} gives, for many but not all cases, a tighter bound than the Mandelstam-Tamm bound. On the other hand, it still depends explicitly on the spectra except for 2-level systems, and, in consequence, it is not always tight.

From its very definition, Eq.\eqref{Def:D_Her}, $D(\hat{\rho}, \hat{\sigma})$, gives a notion of QSL tailored towards unitary evolution only.
Also, from the results of the previous section, $D(\hat{\rho}, \hat{\sigma})$ is by construction a limit which is \textit{achievable}, since it is constructed by finding the best driving Hamiltonian connecting the two states $\hat{\rho}$ and $\hat{\sigma}$.
At the same time, as discussed at length, it is independent of the spectra of the two states $\hat{\rho}$ and $\hat{\sigma}$.

On the other hand, $D(\hat{\rho}, \hat{\sigma})$ has been obtained by finding the best driving protocol among the protocols satisfying the constraint $\norm{\hat{V}_t} = 1$, and this constraint is not necessarily satisfied by a generic driving $\hat{V}_t$.
However, this problem can be easily addressed by first defining a QSL as
\[\label{Eq:QSL_from_QCD}
t_{C\!D}=\frac{D(\R,\S)}{v_{C\!D}}.
\]
with \emph{charging distance evolution speed} $v_{CD}=\frac{1}{T}\int_0^T \|\V_t\|$, where $\V_t$ is the instantaneous driving Hamiltonian. We stress that $v_{CD}$, by construction, is only determined by the operator norm of the quench Hamiltonian, and so it is independent of the instantaneous state of the system. 

By definition, $t_{C\!D}$ is achievable, \textit{i.e.} it is always possible to find a protocol $\V_t$ that saturates the bound set by $t_{C\!D}$, and this protocol is then optimal. 

On the other hand, there could be some suboptimal protocols, for which this bound becomes unnecessarily loose, and better (higher) bounds can be found. Specifically, $v_{CD}=\frac{1}{T}\int_0^T \|\V_t\|$ can contain the some contributions that do not contribute to the evolution of the state, but they do contribute to the operator norm~\cite{physRevA.100.062328}. Thus, to find less loose bound for these non-optimal driving protocols, we can remove these superfluous contributions to improve the QSL. 


The source of these superfluous contributions can be understood by noticing that $\V_t$ can contain terms which \textit{commute} with the instantaneous state $\R_t$~\cite{campaioli2019algorithm}. These terms, of course, do not contribute to the dynamics, but they do contribute in the evaluation of $v_{CD}$. As result, these terms make the bounds looser.

To solve this issue, we define the following \emph{modified evolution speed} $v_{mCD}$
\[\label{theorem:mQSL}
v_{mCD}=\frac{1}{T}\min_{\hat{D_t}:[\hat{D_t},\R_t]=0}\int_0^T \|\V_t+\hat{D_t}\|,
\]
and the QSL that is tighter for non-optimal driving reads as
\[
\label{eq:modified_QSL}
    t_{mC\!D}=\frac{D(\R,\S)}{v_{\mathrm{mCD}}}.
\]
The detailed proof of Eq.~\eqref{theorem:mQSL} is in Appendix~\ref{app:QSL}. 

It should be noticed that, by construction, $v_{mCD}$ depends on the instantaneous state, but it is still independent of the eigenvalues of the instantaneous state.
Also, it is easy to see that $t_{mCD}$ gives a tighter bound than $t_{CD}$, although it requires larger resources and time to be evaluated. 

All in all, the QSL in Eq.~\eqref{eq:modified_QSL}, together with the results of Section \ref{sec:alternative_expressions}, allows to reduce the problem of finding the QSL between two states to a functional minimization problem living on smaller functional spaces (Theorem~\ref{theorem:deg_mixed}) thus simplifying significantly the computational complexity of the problem. In the particular case of $\R$ and $\S$ being non-degenerate and $d$-dimensional, the problem is dramatically simplified to a \textit{$d$-dimensional} minimization problem (Corollary~\ref{Coro:nond_mixed}).

\begin{table*}
    \centering
    \begin{tabular}{|l|l|l|l|}
    \hline
    \multicolumn{4}{|c|}{Mathematical relations}\\
    \hline
        Charging distance & Formulas & Bounds & Ref. \\
        \hline
        Pure states & $D(\R,\S)=\theta_B(\R,\S)$ & $0\leq D(\R,\S)\leq \pi/2$  & \eqref{C_U_pure} \\
        \hline
        Mixed states & ${\displaystyle D(\R,\S)=\min_{U_1,U_2,\cdots, U_m}\Big\|i\ln\Big(\bigoplus_{i=1}^m U_i \!\!\!\sum_{i=1,k=1}^{m,n_i}\!\!\!\pro{s_i^k}{r_i^k}\Big)\Big\|}$   & {$\displaystyle \theta_B(\R,\S)\leq \max_{i}\,\theta_B(\R_i,\S_i) \leq D(\R,\S)\leq \pi\big(1-\tfrac{1}{d}\big)$} & $\begin{matrix}
\eqref{eq:lower_bound_1},\!\eqref{eq:tigthermixed}\\
\eqref{eq:formula_mixed_states},\!\eqref{eq:uppermixed} 
        \end{matrix}$\\
                \hline
    \multicolumn{4}{|c|}{Applications}\\
    \hline
    Mean charging power &  \multicolumn{2}{l|}{{$\displaystyle P\leq 2\|\h\| \|\V\| D_{\Tr}(\R,\S)/D(\R,\S)$}} & \eqref{bound_of_power} \\
    \hline
    Quantum speed limit &  \multicolumn{2}{l|}{{$\displaystyle 
t_{C\!D}=D(\R,\S)/v_{C\!D}
$}} & \eqref{Eq:QSL_from_QCD} \\
    \hline
        Classical vs &  \multicolumn{2}{l|}{{$\displaystyle 
D_C(\ket{0\cdots 0},\ket{1\cdots 1})=N\pi/2$}} & Sec. \\
        quantum charging &  \multicolumn{2}{l|}{{$\displaystyle 
D(\ket{0\cdots 0},\ket{1\cdots 1})=\pi/2
$}} & \ref{subsec:classical_vs_quantum} \\
    \hline
    \end{tabular}
    \caption{Main results.}
    \label{tab:results}
\end{table*}

\section{Conclusion and future directions}
\label{sec:conclusions}

Quantum batteries promise a potentially large speed-up in charging. The basic principle behind this is by means of shortcuts through the Hilbert space, using entangled states in between. This leads to a natural definition of a distance: We define the quantum charging distance between two given states, as the minimal time it takes for one of the two states to evolve into the other, among all the possible driving Hamiltonians satisfying a constraint in their operator norm. The main results of this paper are summarized in Table~\ref{tab:results}.

By definition, this notion of distance is hard to compute, since it requires an optimization process during the calculation. 
This difficulty is mitigated by the theory developed in Section \ref{sec:alternative_expressions}, in which it is first shown that the minimization can be carried on \textit{time-independent} driving Hamiltonian only (Theorem~\ref{theorem:D_equal_lnU}).
For the special case of $\R$ and $\S$ being pure states, the charging distance reduces to the well-known Bures angle.
Moreover -- when the two states under investigation are non-degenerate and of maximal rank -- the minimization process can be further reduced to a \textit{d-dimensional} minimization problem, thus reducing significantly the computational complexity. 

Furthermore, we have found general upper and lower bounds satisfied by the charging distance. The distance is lower bounded by the Bures angle and upper bounded by $\pi$ for any dimensional Hilbert space and for any pair of states. When the states are pure, the upper bound is cut by half to $\pi/2$.

These results have been then applied to two problems of interest in recent literature: the maximal charging power of many-body quantum batteries, and the estimate of QSL for unitary evolution.

As for quantum batteries and their charging power, the quantum charging distance gives a bound on the power of quantum charging in terms of the operator norms of the battery Hamiltonian and the driving Hamiltonian. We proved that this new bound is always tighter than the trivial bound on the power, $P\leq 2\norm{\V}\norm{\h}$~\cite{PhysRevLett.118.150601, PhysRevLett.128.140501}. By comparing the quantum charging distance with classical charging, 
we have confirmed that the maximum quantum charging advantage scales linearly with the system size, thus providing a genuine geometric interpretation of the notion of shortcuts in the many-body Hilbert space. Furthermore, the shortcut that connects the depleted with the charged product states always goes through an entangled state. In comparison, classical charging, which forces the state to remain separable throughout the evolution, always leads to a longer route. Any two pure orthogonal states have the quantum charging distance equal to $\pi/2$, which follows from Theorem~\ref{thm:pure_states}, independent of the size of the system, the number of qubits in particular. Thus, ten thousand qubits can be charged as quickly as a single one if a sufficiently entangling driving Hamiltonian is applied.


Outside of the charging distance, we defined a new quantum speed limit -- tailored towards \textit{unitary evolution}. This QSL is automatically defined from the charging distance, up to a rescaling that considers the instantaneous operator norm of the driving Hamiltonian.
By construction, this QSL is achievable and independent of the spectrum of the instantaneous state. Thus, it solves the open issues for unitary evolution from previous works \cite{PhysRevLett.120.060409}.


One of the drawbacks of the charging distance introduced in this paper is that it is not formulated by an explicit analytic formula, except for the case of $\R$ and $\S$ being pure states.
For the general case of $\R$ and $\S$ being mixed, we reduced the problem to an optimization task. We confirmed that for small systems, such an optimization task takes a reasonable time to compute.
An immediate question that would be interesting to address is finding efficient numerical strategies to solve this optimization problem, since our numerical checks have been quite preliminary and restricted to small systems only. Furthermore, it would be remarkable to find a closed analytic expression for the charging distance in the case of mixed states.

Clearly, the time necessary to evolve one state into another depends heavily on the type of restrictions put on the driving Hamiltonian. In this paper, we used the operator norm for the normalization of the driving Hamiltonian, inspired by the quantum batteries literature. However, other norms, such as the trace and square norms, are also possible, leading to different charging geometries. The dependency of the minimum time on the normalization condition is an interesting question that we plan to address in the near future.

\section*{Acknowledgements}
DR and D\v{S} acknowledge financial support from the Institute for Basic Science (IBS) in the Republic of Korea through the project IBS-R024-D1. JG acknowledges financial support in part by the National Research Foundation of Korea Grant funded by the Korean Government (NRF-2020R1C1C1014436) and thanks Yongjoo Baek for academic support. 

\appendix

\section{Proof of Theorem~\ref{thm1}}\label{app:1}

There exists a driving Hamiltonian, $\V_t$, which connects two mixed states with the same spectra. To show that, let us assume that $\R=\sum_i r_i\ket{r_i}\bra{r_i}$ and $\S=\sum_i r_i \ket{s_i}\bra{s_i}$. We can define a unitary operator $\U=\sum_i \ket{s_i}\bra{r_i}$, that connects the two states. Each unitary operator can be represented as an exponential using lie algebra $\U=\exp(-iV)$. The time-independent Hamiltonian, $\V_t=\V/\|\V\|$, follow the normalization condition and the time of evolution $T$ is given as $\|\V\|$. Hence, it takes a finite time to evolve from $\R$ to $\S$ which guarantees the existence of the minimum time to evolve. The quantum charging distance is thus well-defined.

The proofs of other properties of distance follow.
1) (symmetry) $D(\R_1,\R_2)=D(\R_2,\R_1)$ is satisfied, because one can choose $-\V$ to go from $\R_2$ to $\R_1$ in the same time.\\
2) (positivity) The time is clearly positive. $D(\R_1,\R_2)=0 \Leftrightarrow \R_1=\R_2$ holds because it will always take non-zero time to go to a state that is not the same state.\\
3) (triangle inequality) $D(\R_1,\R_2)\leq D(\R_1,\R_3)+D(\R_3,\R_2)$ holds because one can choose a piece-wise function
\[
\V_t=\begin{cases}
\V^{13}_t, & t\in [0,T_{13}) \\
\V^{32}_{t-T_{13}}, & t\in [T_{13},T_{13}+T_{32}] \\
\end{cases}
\]
where $\V^{13}_t$ is the optimal driving Hamiltonian which takes state $\R_1$ to state $\R_3$ in time $T_{13}$ and similar with $\V^{32}_{t}$ and time $T_{32}$. This $\V_t$ is part of the set over which the time to reach $\R_2$ from $\R_1$ is optimized, so the optimum must be given by time $T$ equal or lower to the time it takes for the $\V_t$ defined above.

\section{Proof of Theorem~\ref{theorem:D_equal_lnU}}\label{app:2}

Here we prove an alternate formula for general states it terms of the charging unitary operator.

\emph{Sketch.}
The proof uses the perturbation method on the eigenvalues of $i\ln(U)$ expanded in the time differential as $i\ln((1-i\V_t dt)U)$. The time evolution rate of eigenvalues is bounded by the operator norm of $\V_t$. From this, it follows that the time it takes to evolve $\R$ to $\S$ is bounded below as $T(\R,\S)\geq\|i\ln(U)\|$. By definition, the distance is equal to or greater than $\min_{U:U\R U^\dag=\S} \|i \ln(U)\|$ and it is possible to always find a time-independent Hamiltonian $\V_t$ which saturates the bound. 

\begin{proof}(Theorem~\ref{theorem:D_equal_lnU})
    Any unitary operator $U$ can be written as
    \[\label{eq:first_U}
    \begin{split}
    U&=\mathrm{Texp}\left(-i\int_0^t \V_\tau\, d\tau\right)=\mathrm{exp}(-i\hat{W}(t))\\
        &=\sum_i \pro{i}{i}\mathrm{exp}(-i\w_i),
    \end{split}
    \]
    where $\ket{i}$ and $\w_i$ are eigenbasis and eigenvalues of $\hat{W}$. For an infinitesimal $dt$, we can expand $U$ as
    \[\label{Unitary_dt}
    \begin{split}
        U'&=\exp(-i \hat{W}(t+dt))=\mathrm{Texp}\left(-i\int_0^{t+dt} \V_\tau\, d\tau\right)
    \\
    &=\mathrm{Texp}\bigg(-i\int_t^{t+dt}\V_\tau d\tau\bigg)\mathrm{Texp}\left(-i\int_0^t \V_\tau\, d\tau\right)\\
    &=\mathrm{Texp}\bigg(-i\int_t^{t+dt}\V_\tau d\tau\bigg)\mathrm{exp}(-i\hat{W}(t))\\
    &=\sum_i \pro{i'}{i'}\mathrm{exp}(-i\w_i').
    \end{split}
    \]

    Additionally, we Taylor expand $\V_\tau$ at point $t$ as $\V_{t+dt}=\V_t+\frac{d\V_t}{dt}|_t dt+\mathcal{O}(dt^2)$ which yields
    \[
    \mathrm{Texp}\bigg(-i\int_t^{t+dt}\V_\tau d\tau\bigg)=\mathrm{exp}(-i\V_t dt+\mathcal{O}(dt^2))
    \]

    Similarlly, we expand $\ket{i'}$ and $\w_i'$ as $\ket{i'}=\ket{i^{(0)}}+\ket{i^{(1)}}dt+\ket{i^{(2)}}dt^2+\cdots$ and $\w_i'=\w_i^{(0)}+\w_i^{(1)}dt+\w_i^{(2)}dt^2+\cdots$ for $\ket{i^{(0)}}=\ket{i}$ and $\w_i^{(0)}=\w_i$. The last equality in Eq.~\eqref{Unitary_dt}, while inserting Eq.~\eqref{eq:first_U}, is rewritten as
    \[
    \begin{split}
            &(1-i\V_t dt+\mathcal{O}(dt^2))\sum_i\pro{i}{i}\mathrm{exp}(-i\w_i)\\
            &=\sum_i (\ket{i^{(0)}}+\ket{i^{(1)}}dt+\mathcal{O}(dt^2))\\
            &\times (\bra{i^{(0)}}+\bra{i^{(1)}}dt+\mathcal{O}(dt^2))\\
            &\times\exp(-i(\w_i^{(0)}+\w_i^{(1)}dt+\mathcal{O}(dt^2)))
    \end{split}
    \]
    By applying $\bra{i}$ and $\ket{i}$ on both sides and extracting the first order of $dt$, we obtain
    \[
    \begin{split}
           &\bra{i}(-i\V_t)\ket{i}\mathrm{exp}(-i\w_i)\\
           &=-i\w_i^{(1)}\exp{-i\w_i}+\left(\bra{i^{(1)}}\ket{i}+\bra{i}\ket{i^{(1)}}\right)\exp{-i\w_i}.
    \end{split}
    \]
    
    Due to normalization condition, $1=\bra{i'}\ket{i'}=\bra{i}\ket{i}+\left(\bra{i^{(1)}}\ket{i}+\bra{i}\ket{i^{(1)}}\right) dt +\mathcal{O}(dt^2)$, it follows that $\bra{i^{(1)}}\ket{i}+\bra{i}\ket{i^{(1)}}=0$. Hence, from the equation above, we obtain
    \[
    \bra{i}\V_t\ket{i}=\w_i^{(1)}.
    \]
    At the same time, $\norm{\V_t}= 1$. This bounds all the diagonal elements of $\V_t$, thus we have
    \[
    \abs{\w_i^{(1)}}=\abs{\bra{i}\V_t\ket{i}}\leq \norm{\V_t}= 1.
    \]
    The above equation holds for any time $t\in [0,T)$.

    We are going to use this inequality to bound the first-order coefficient in $dt$ as
    \[
    \begin{split}
    \frac{d}{dt}\norm{\hat{W}(t)}&=\frac{d}{dt}\max_i \abs{\w_i(t)}\leq\max_i \frac{d}{dt}\abs{\w_i(t)}\\
    &=\max_i \abs{\w_i^{(1)}(t)}\leq 1
    \end{split}
    \]
    where the bound on the right-hand side holds for any $t\in [0,T)$.

    From this, we derive
    \[
    \norm{\hat W(T)}=\int_0^T\frac{d}{dt}\norm{\hat W(t)} dt \leq \int_0^T 1 dt= T.
    \]
    $\hat W(T)$ can be equivalently written as the matrix logarithm of the unitary $U$, and the above equation can be thus rewritten as
    \[
    T\geq \norm{\hat W(T)} = \norm{i \ln(U)}.
    \]
    This equation holds for any kind of driving process, so, optimizing over all drivings that each lead to a unitary $U$ that connect $\R$ with $\S$, we obtain the lower bound on the minimal time that it takes to achieve this transformation, obtaining
    \[\label{ineq_Unitary}
    D(\R,\S)\geq \min_{U:U\R U^\dag=\S} \|i \ln(U)\|.
    \]

    Next, we show that this bound is achievable. The strategy is as follows. Because the inequality is always satisfied, to achieve the lowest possible left-hand side, we first minimize the right-hand side. Let us define $U_{\mathrm{opt}}$ as a unitary operator that achieves the minimum for the right-hand side (the minimum always exists because it is a continuous function on a compact set of unitary operators). For any optimal unitary operator $U_{\mathrm{opt}}$, there is a time-independent Hamiltonian $\V_{\mathrm{opt}}:=i \ln(U_{\mathrm{opt}})/D(\R,\S)$, which satisfies $U_{\mathrm{opt}}=\mathrm{exp}(-i \V_{\mathrm{opt}} D(\R,\S))$, saturating the bound. This proves the theorem and shows that there is always a time-independent Hamiltonian $\V$ which satisfies the equality condition~\eqref{ineq_Unitary} and
    \[\label{eq_Unitary_app}
    D(\R,\S)= \min_{U:U\R U^{\dag}=\S} \|i \ln(U)\|.
    \]
    The optimal driving Hamiltonian is obtained from the definition, $\V_{\mathrm{opt}}=i \ln(U_{\mathrm{opt}})/D(\R,\S)$.
    \end{proof}
    
\section{Proof of Theorem~\ref{thm:pure_states}}\label{app:3}

Here we prove an alternate formula for the charging distance for pure states.

\emph{Sketch.}
The theorem is proved by separating the Hilbert space dimension into two relevant and other irrelevant dimensions. The relevant dimension is given by the plane spanned by $\ket{\psi}$ and $\ket{\varphi}$. The optimal unitary transforms one of the states into the other by rotating it in this plane. Thus, the total unitary $U$ can be expressed as a direct sum of unitary operators $U=U_2\oplus U_{d-2}$. $U_{d-2}$ turns out to be irrelevant for the distance, and $U_2$ is a two-dimensional unitary that is easily parameterized. Analytical optimization in Theorem~\ref{theorem:D_equal_lnU} yields $\arccos(\abs{\braket{\psi}{\varphi}})$.

\begin{proof}(Theorem~\ref{thm:pure_states})
    The dimensions except for the two dimensions of a plane where states live, can not contribute to evolution by the optimal time-independent driving Hamiltonian. Terms of these additional dimensions in $U$ just make $\|i\ln(U)\|$ increase without changing the state, thus, they will be irrelevant for the optimal charging unitary.

    Generally, unitary operator $\ket{\varphi_\perp}$ lives in the plane defined by both $\ket{\psi}$ and $\ket{\varphi}$. Using the Gram-Schmidt process, we derive the orthogonal vector to $\ket{\psi}$ in this plane as
    \[
    \ket{\psi_\perp}=\frac{\ket{\varphi}-\braket{\psi}{\varphi}\ket{\psi}}{\sqrt{1-\abs{\braket{\psi}{\varphi}}^2}}e^{-i\phi'},
    \]
    except for the case $\abs{\braket{\psi}{\varphi}}=1$, which is trivial since $D(\ket{\psi},\ket{\varphi})$ is clearly zero.     $e^{-i\phi}$ is a phase that can be chosen arbitrarily, but note that it appears in the parametrization below.

    $U$ that connects the two pure states can be expressed as a sum of three terms,
    \[\label{eq:U_pure}
    U= \pro{\varphi}{\psi}e^{i\phi_1}+\pro{\varphi_\perp}{\psi_\perp}e^{i\phi_2}+U',
    \]
    where $U'$ is an arbitrary unitary operator that lives in the irrelevant dimensions orthogonal to $\ket{\psi}$ and $\ket{\varphi}$.
    $U$ can be expressed as a direct sum,
    \[\label{U_pure}
       U=\begin{pmatrix}
\cos(\theta)e^{i(\phi_1+\phi)} & -\sin(\theta)e^{i(\phi_2+\phi)}\\
\sin(\theta)e^{i(\phi_1+\phi')} & \cos(\theta)e^{i(\phi_2+\phi')}
\end{pmatrix}\oplus U',
    \]
    in the basis $\{\ket{\psi},\ket{\psi_\perp}\}$. We expressed the other states as $\ket{\varphi}=(\cos(\theta)e^{i\phi},\sin(\theta)e^{i\phi'})^T$ and $\ket{\varphi_\perp}=(-\sin(\theta)e^{i\phi},\cos(\theta)e^{i\phi'})^T$ in that basis, and we assume $\theta \in [0,\pi/2)$, $\phi,\phi' \in [0,2\pi)$. Note that in this parametrization,
    \[\label{eq:parameter_prescription}
    \begin{split}
    \theta&=\arccos(\abs{\braket{\psi}{\varphi}}),\\
    e^{i\phi}&=\frac{\braket{\psi}{\varphi}}{\abs{\braket{\psi}{\varphi}}},\quad 
    e^{i\phi'}=\frac{\braket{\psi_\perp}{\varphi_\perp}}{\abs{\braket{\psi_\perp}{\varphi_\perp}}},
    \end{split}
    \]
    as long as the denominator is nonzero. This excludes orthogonal states. If the denominator is zero, then
    \[
    e^{i\phi}=-\braket{\psi}{\varphi_\perp}
    ,\quad e^{i\phi'}=\braket{\psi_\perp}{\varphi}.
    \]

    $\|i\ln(U)\|$ is function of $\phi_1$ and $\phi_2$ for a given $\theta$, $\phi$ and $\phi'$ as
    \[\label{D_for_pure} 
    \begin{split}
        &\|i\ln(U)\|=\max\bigg\{\|i\ln(U')\| ,\\
        &\bigg|\arccos(\cos(\theta)\cos(\frac{\phi_1\!+\!\phi\!-\!\phi_2\!-\!\phi'}{2}))+i\frac{\phi_1\!+\!\phi\!+\!\phi_2\!+\!\phi'}{2}\bigg|,\\
        &\bigg|\arccos(\cos(\theta)\cos(\frac{\phi_1\!+\!\phi\!-\!\phi_2\!-\!\phi'}{2}))-i\frac{\phi_1\!+\!\phi\!+\!\phi_2\!+\!\phi'}{2}\bigg|\,\bigg\}
    \end{split}
    \]
    which is minimized to $\theta$ 
    when $\phi_1=-\phi$ and $\phi_2=-\phi'$ and when $\|i\ln(U')\|$ is smaller than the other two terms.
    By Theorem~\ref{theorem:D_equal_lnU}, $D(\ket{\psi},\ket{\varphi})$ is equal to $\theta=\arccos(\abs{\braket{\psi}{\varphi}})$.

    The prescription for $\U_{\mathrm{opt}}$ is obtained by inserting $\phi_1=-\phi$ and $\phi_2=-\phi'$ into equations \eqref{eq:parameter_prescription}, while considering Eq.~\eqref{eq:U_pure}. We can also choose $U'=\hat{I}_{d-2}$ as an identity on the orthogonal dimensions, since for such $U'$, $\|i\ln(U')\|=0$. The optimal Hamiltonian is then obtained as $\V_{\mathrm{opt}}=i \ln(U_{\mathrm{opt}})/D(\R,\S)$, as per Theorem~\ref{theorem:D_equal_lnU}.

\section{Pure state charging distance from Mandelstam-Tamm bound}\label{app:PurefromMT} 

We show that the result for the quantum charging distance for pure states, Eq.~\eqref{C_U_pure} can be derived from the Mandelstam-Tamm bound~\cite{uffink1993},
\[
\abs{\braket{\psi_0}{\psi_t}}^2\geq \cos^2\frac{\Delta \V_t\ t}{\hbar}.
\]
using that it is tight for pure states. In the above, we set $\hbar=1$ as in the rest of our paper, and the standard deviation is $\Delta \V_t=\sqrt{\tr[\V_t^2\R]-\tr[\V_t\R]^2}$.

It can be trivially proved that
\[
\Delta \V_t\leq \norm{\V_t}.
\]
Rewriting the bound, we have
\[
\frac{\arccos\abs{\braket{\psi_0}{\psi_t}}}{\Delta \V_t}\leq \ t.
\]
Minimizing the left-hand side over all driving Hamiltonians gives
\[
\frac{\arccos\abs{\braket{\psi_0}{\psi_t}}}{\norm{\V_t}}\leq \ t
\]
Considering that for the charging distance, we require $\norm{\V_t}=1$ and that the Mandlestam-Tamm bound is tight, we obtain Eq.~\eqref{C_U_pure}.



\end{proof}

\section{Proof of Theorem~\ref{theorem:deg_mixed}}\label{app:4}

Here we prove an alternate formula for the charging distance for mixed states.

\emph{Sketch.} We will show that condition $U \R U^{\dag}=\S$ implies the form $U=\bigoplus_{i=1}^m U_i (\sum_{i=1}^m \sum_{k=1}^{n_i}\pro{s_i^k}{r_i^k})$.


\begin{proof}(Theorem~\ref{theorem:deg_mixed})
Any unitary operator can be written as a product of two unitary operators,
\[
U=U' U''.
\]
$U''$ can be chosen arbitrarily and $U'$ is determined by $U$ and $U''$. We set $U''=\sum_{i=1}^m \sum_{k=1}^{n_i}\pro{s_i^k}{r_i^k}$. Then
\[
U=U' \sum_{i=1}^m \sum_{k=1}^{n_i}\pro{s_i^k}{r_i^k}.
\]
$U$ has to satisfy 
\[
U \R U^{\dag}=U' \S U'^{\dag}=\S,
\]
which can be rewritten as 
\[
[U', \S]=0.
\]
We use a general representation of a matrix to express $U'$ as
\[
U'=\sum_{i=1}^m \sum_{k=1}^{n_i} \sum_{j=1}^m \sum_{l=1}^{n_i} u_{ikjl}\pro{s_i^k}{s_j^l}.
\]
We compute the commutator,
\[\label{commut_U_S}
[U', \S]=\sum_{i=1}^m \sum_{k=1}^{n_i} \sum_{j=1}^m \sum_{l=1}^{n_i} (r_j-r_i)u_{ikjl}\pro{s_i^k}{s_j^l}.
\]
To make \eqref{commut_U_S} zero, $u_{imjl}$ should be zero for all $i\neq j$, $k$ and $l$. Hence we have
\[
U'=\sum_{i=1}^m \sum_{k=1}^{n_i}\sum_{l=1}^{n_i} u_{ikil}\pro{s_i^k}{s_i^l}=\bigoplus_{i=1}^m U_i,
\]
where $U_i$ is an arbitrary unitary operator that acts only on the subspace $\HS_i$ of the Hilbert space spanned by $\ket{s_i^k}$ for $1\leq k\leq n_i$.
\end{proof}

\section{Proof of Theorem~\ref{theorem:bound_of_Dis}}\label{app:5}

Here we prove the upper bound on the charging distance.

\emph{Sketch.}  The proof comes from Theorem~\ref{theorem:D_equal_lnU} and the fact that there is freedom for the global phase of unitary operators without breaking the condition $U\R U^\dag=\S$. The dependence on dimension $d$ 
is derived from the fact that the number of eigenvalues is equal to the dimension of the Hilbert space, $d$.


\begin{figure*}[t]
\begin{center}
\includegraphics[width=0.8\hsize]{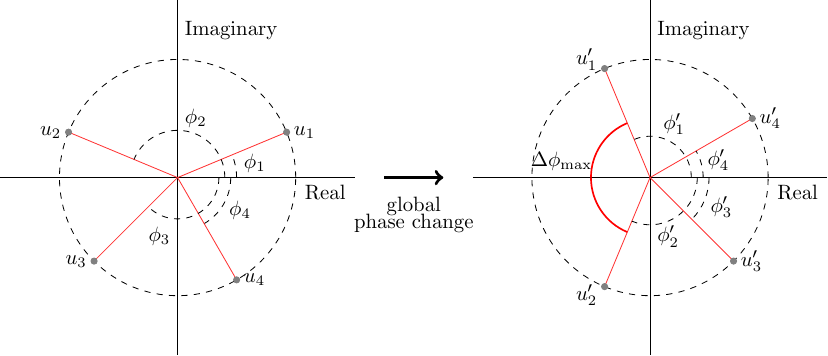}\\
\caption
{The schematic illustration of eigenvalues living in a complex space for $d=4$. By the changing global phase, eigenvalues rotate around zero conserving their relative angle. We always minimize $\|i\ln(U)\|$ to $(\pi-\Delta\phi_{\max}/2)$ by this method.}
\label{Fig:distribution_zeros}
\end{center}
\end{figure*}
\begin{proof}(Theorem~\ref{theorem:bound_of_Dis})
Unitary operators have eigenvalues $u_1,\dots,u_d$ that are complex numbers with norm one. Thus, they lie in a circle of radius one in the complex plane. See Fig~\ref{Fig:distribution_zeros} for their depiction. The eigenvalues of $i\ln( U)$ are equal to the corresponding angles, $\phi_1$, ..., $\phi_d$. The operator norm of $i\ln( U)$ is equal to $\max(\phi_1, ..., \phi_d)$. 

To find the quantum charging distance, according to Theorem~\ref{theorem:D_equal_lnU}, we need to find a unitary operator that satisfies $U\R U^\dag=\S$ and at the same time minimizes $\max(\phi_1, \cdots, \phi_d)$. Consider some operator $U$ that satisfies $U\R U^\dag=\S$. Then also $U'=e^{i\phi}U$ satisfies the same condition. This global phase change $e^{i\phi}$ only rotates the eigenvalues on the circle but does not change their relative angles. Thus, there is a degree of freedom $\phi$ of the global phase change that we can freely choose, but which at the same time has the potential make the minimum of $\max(\phi'_1,\cdots,\phi'_d)=\max(\phi_1+\phi, \cdots, \phi_d+\phi)$ smaller. Optimizing (minimizing) over $\phi$, the maximum angle is equal to $\max(\phi'_1,\cdots,\phi'_d)=\pi-\Delta\phi_{\max}/2$, where $\Delta\phi_{\max}$ is the maximum relative angle; see the right side of Fig~\ref{Fig:distribution_zeros} for illustration. At the same time, $\Delta\phi_{\max}$ is greater than or equal to $2\pi/d$. Thus, 
\[
\|i \ln(U')\|=\max(\phi'_1,\cdots,\phi'_d)\leq \pi-\pi/d.
\]
This holds for every $U'$ that connects $\R$ with $\S$, thus, it must also hold for the optimal $U'$ that achieves the minimal charging distance. Therefore, $D(\R,\S)\leq \pi-\pi/d$, which proves the theorem.


At last, we show that the bound can be saturated with the following example. Consider two states that are given by $\R=\sum_{i=1}^d r_i\ket{r_i}\bra{r_i}$ and $\S=\sum_{i=1}^d r_i\ket{r_{i+1}}\bra{r_{i+1}}$, where we have a cycling order, so that $\ket{r_{d+1}}=\ket{r_1}$. The distance between these two states equals $(1-1/d)\pi$.
\end{proof}

\section{Proof of Theorem~\ref{theorem:relation_QCD_Bures}}\label{app:6}

Here we prove that the charging distance is lower bounded by the Bures angle.

\emph{Sketch.}
The Bures angle is equal to the quantum charging distance at the purification space with the optimal choice of driving Hamiltonian using a larger dimension of purification space than the original space. It is obvious that we can not use all of the unitary evolution belonging to purification space; we only use the sub-set of them which means the minimum time is larger than the minimum time in the full space. 


\begin{proof}(Theorem~\ref{theorem:relation_QCD_Bures})
    Where the purified states, $\ket{\psi_\rho}$ and $\ket{\psi_\sigma}$, are defined by
    \[\begin{split}
        \ket{\psi_\rho}=\sum_{i=1}^d  r_i\ket{r_i}\otimes \ket{e_i}\\
        \ket{\psi_\sigma}=\sum_{i=1}^d r_i\ket{s_i}\otimes \ket{e_i'},
    \end{split}
    \]
    By the theorem~\ref{C_U_pure}, the $D(\ket{\psi_\rho},\ket{\psi_\rho})$ is equal to the $\arccos(|\braket{\psi_\rho}{\psi_\sigma}|)$. 
    Since the maximum overlap between purified states is equal to fidelity between two mixed states, the minimum $\arccos(|\braket{\psi_\rho}{\psi_\sigma}|)$ is equal to the Bures angle, such that
    \[\theta_B=\min_{\ket{e_i},\ket{e'_i}}\min_{\V_t:\|\V_t\|=1}T(\ket{\psi_\rho},\ket{\psi_\sigma}).
    \]
    At the same time, there is an optimal $\V_\mathrm{optimal}$ which evolution time is equal to $D(\R,\S)$ and which give the driving Hamiltonian $\V=\V_\mathrm{optimal}\otimes \hat{I}$. $\V$ has operator norm as one, and it connect to two purified states, $\ket{\psi_\rho}=\sum_{i=1}^d \sqrt{r_i}\ket{r_i}\otimes \ket{e_i}$ and $\ket{\psi_\sigma}=\sum_{i=1}^d \sqrt{r_i}\ket{s_i}\otimes \ket{e_i}$ with same purification basis $\ket{e_i}$'s. The evolution time by $\V=\V_\mathrm{optimal}\otimes \hat{I}$ is equal to the charging distance between origin mixed states, i.e., $T(\ket{\psi_\rho},\ket{\psi_\sigma})=D(\R,\S)$, such that
    \[
    \begin{split}
        \theta_B&=\min_{\ket{e_i},\ket{e'_i}}\min_{\V_t:\|\V_t\|=1}T(\ket{\psi_\rho},\ket{\psi_\sigma})\\
        &\leq T(\ket{\psi_\rho},\ket{\psi_\sigma})_{\V_\mathrm{optimal}\otimes \hat{I},\;\ket{e_i}=\ket{e'_i}}
    =D(\R,\S),
    \end{split}
    \]
    which proves the statement.

    Fig~\ref{Fig:quantum_charging} helps an intuitive understanding of the relation between the Bures angle and Quantum charging Distance. By using optimal Hamiltonian in the purification, the spectrum of $\R_t$ is able to be different from $\R$ and $\S$, which is not allowed by unitary evolution. The Bures angle can have less quantity than the quantum charging distance since it uses the larger freedom to choose a path between states.
\end{proof}

\section{Proof of Corollary~\ref{theorem:relation_QCD_Bures_genelized2}}\label{app:cor2}

Here we prove a tighter lower bound on the charging distance.

\emph{Sketch.}
We maximize left-hand side of Eq.~\eqref{eq:lower_bound_1} over all eigenvalues, and using the joint concavity of fidelity we evaluate this maximum. According to Theorem~\ref{theorem:deg_mixed}, the charging distance does not depend on the eigenvalues. Thus, the maximum must still lower bound the charging distance. 

\begin{proof}(Corollary~\ref{theorem:relation_QCD_Bures_genelized2})
Notice that we can rewrite the density matrices as
\[
\R=\sum_{i=1}^{m}r_in_i\R_i,\quad  \S=\sum_{i=1}^{m}r_in_i\S_i.
\]
The joint concavity of fidelity~\cite{nielsen2010quantum} implies 
\[
\mathcal{F}(\R,\S)\geq \sum_ir_i n_i\mathcal{F}(\R_i,\S_i).
\]
Considering that $\arccos$ is a decreasing function, we have
\[
\theta_B(\R,\S)\leq \sum_ir_i n_i\theta_B(\R_i,\S_i).
\]
Taking the maximum gives
\[
\max_{r_1,\cdots,r_m}\theta_B(\R,\S)\leq \max_{r_1,\cdots,r_m}\sum_ir_in_i\theta_B(\R_i,\S_i).
\]
This bound saturates for a specific $r_i$, which maximizes the right-hand side (while satisfying $r_in_i=1$ and  $r_{i'}n_{i'}=0$ for $i'\neq i$). This gives
\[
\max_{r_1,\cdots,r_m}\theta_B(\R,\S)=\max_{i}\, \theta_B(\R_i,\S_i).
\]
According to Theorem~\ref{theorem:deg_mixed}, the charging distance does not depend on the eigenvalues $r_i$. Thus, the maximum must be still lower than the charging distance,
\[
\max_{i}\,\theta_B(\R_i,\S_i)\leq D(\R,\S).
\]
\end{proof}

\section{Quantum speed limits}\label{app:QSL}

\begin{theorem}
    QSL from $D(\R,\S)$ is given as
    \[
    t_{mCD}=\frac{D(\R,\S)}{v_{mCD}},
    \]
    where
    \[
    v_{mCD}=\frac{1}{T} \min_{\hat{D}_t:[\hat{D}_t,\R_t]=0}\int_0^T \| \V_t+\hat{D}_t\|.
    \]
\end{theorem}
\begin{proof}
    The proof follow the conventional way for the proof of other QSL. It comes from the $D(\R_t,\R_{t+dt})$ with infinitesimal time $dt$. By the theorem~\ref{theorem:D_equal_lnU}, $D(\R_t,\R_{t+dt})$ is equal to $\min_{U:U\R_t U^\dagger=\R_{t+dt}}\|i\ln{U}\|$. By the condition, we obtain that 
    \[
    U\R_t U^\dagger =\R_{t+dt} =e^{-i \V_t dt} \R_t e^{i \V_t dt}.
    \]
    Then
    \[
    [e^{i \V_t dt} U,\R_t] = 0.
    \]
    Since $e^{i \V_t dt} U$ is also unitary operator, we can represent $e^{i \V_t dt} U$ as $e^{-i \hat{D}_t dt}$ when $[\hat{D}_t,\R_t]=0$.
    For infinitesimal time $dt$, we have
    \[
    U=e^{-i(\V_t+\hat{D}_t)dt}.
    \]
    It follows that
    \[
        D(\R_t,\R_{t+dt})=\!\!\!\min_{U:U\R_t U^\dagger=\R_{t+dt}}\!\!\!\|i\ln{U}\|=\!\!
        \min_{\hat{D}_t:[\hat{D}_t,\R_t]=0}\!\! \| \V_t+\hat{D}_t\|dt.
    \]
    Now we can derive an inequality
    \[
    D(\R,\S)\leq \int_0^T D(\R_t,\R_{t+dt})=\min_{\hat{D}_t:[\hat{D}_t,\R_t]=0} \int_0^T \| \V_t+\hat{D}_t\|dt,
    \]
    which yields $t_{mCD}$.

    Because $\min_{\hat{D}_t:[\hat{D}_t,\R_t]=0} \int_0^T \| \V_t+\hat{D}_t\|dt$ is always smaller or equal to $\int_0^T \| \V_t\|dt$, $t_{CD}$ also gives an achievable QSL but it is looser than $t_{mCD}$, $t_{mCD}\geq t_{CD}$.
\end{proof}

\bibliography{bibliography}

\begin{thebibliography}{106}%
\makeatletter
\providecommand \@ifxundefined [1]{%
 \@ifx{#1\undefined}
}%
\providecommand \@ifnum [1]{%
 \ifnum #1\expandafter \@firstoftwo
 \else \expandafter \@secondoftwo
 \fi
}%
\providecommand \@ifx [1]{%
 \ifx #1\expandafter \@firstoftwo
 \else \expandafter \@secondoftwo
 \fi
}%
\providecommand \natexlab [1]{#1}%
\providecommand \enquote  [1]{``#1''}%
\providecommand \bibnamefont  [1]{#1}%
\providecommand \bibfnamefont [1]{#1}%
\providecommand \citenamefont [1]{#1}%
\providecommand \href@noop [0]{\@secondoftwo}%
\providecommand \href [0]{\begingroup \@sanitize@url \@href}%
\providecommand \@href[1]{\@@startlink{#1}\@@href}%
\providecommand \@@href[1]{\endgroup#1\@@endlink}%
\providecommand \@sanitize@url [0]{\catcode `\\12\catcode `\$12\catcode
  `\&12\catcode `\#12\catcode `\^12\catcode `\_12\catcode `\%12\relax}%
\providecommand \@@startlink[1]{}%
\providecommand \@@endlink[0]{}%
\providecommand \url  [0]{\begingroup\@sanitize@url \@url }%
\providecommand \@url [1]{\endgroup\@href {#1}{\urlprefix }}%
\providecommand \urlprefix  [0]{URL }%
\providecommand \Eprint [0]{\href }%
\providecommand \doibase [0]{https://doi.org/}%
\providecommand \selectlanguage [0]{\@gobble}%
\providecommand \bibinfo  [0]{\@secondoftwo}%
\providecommand \bibfield  [0]{\@secondoftwo}%
\providecommand \translation [1]{[#1]}%
\providecommand \BibitemOpen [0]{}%
\providecommand \bibitemStop [0]{}%
\providecommand \bibitemNoStop [0]{.\EOS\space}%
\providecommand \EOS [0]{\spacefactor3000\relax}%
\providecommand \BibitemShut  [1]{\csname bibitem#1\endcsname}%
\let\auto@bib@innerbib\@empty
\bibitem [{\citenamefont {Deutsch}(2018)}]{Deutsch_2018}%
  \BibitemOpen
  \bibfield  {author} {\bibinfo {author} {\bibfnamefont {J.~M.}\ \bibnamefont
  {Deutsch}},\ }\bibfield  {title} {\bibinfo {title} {Eigenstate thermalization
  hypothesis},\ }\href {https://doi.org/10.1088/1361-6633/aac9f1} {\bibfield
  {journal} {\bibinfo  {journal} {Reports on Progress in Physics}\ }\textbf
  {\bibinfo {volume} {81}},\ \bibinfo {pages} {082001} (\bibinfo {year}
  {2018})}\BibitemShut {NoStop}%
\bibitem [{\citenamefont {Kim}\ \emph {et~al.}(2014)\citenamefont {Kim},
  \citenamefont {Ikeda},\ and\ \citenamefont {Huse}}]{PhysRevE.90.052105}%
  \BibitemOpen
  \bibfield  {author} {\bibinfo {author} {\bibfnamefont {H.}~\bibnamefont
  {Kim}}, \bibinfo {author} {\bibfnamefont {T.~N.}\ \bibnamefont {Ikeda}},\
  and\ \bibinfo {author} {\bibfnamefont {D.~A.}\ \bibnamefont {Huse}},\
  }\bibfield  {title} {\bibinfo {title} {Testing whether all eigenstates obey
  the eigenstate thermalization hypothesis},\ }\href
  {https://doi.org/10.1103/PhysRevE.90.052105} {\bibfield  {journal} {\bibinfo
  {journal} {Phys. Rev. E}\ }\textbf {\bibinfo {volume} {90}},\ \bibinfo
  {pages} {052105} (\bibinfo {year} {2014})}\BibitemShut {NoStop}%
\bibitem [{\citenamefont {Dymarsky}\ \emph {et~al.}(2018)\citenamefont
  {Dymarsky}, \citenamefont {Lashkari},\ and\ \citenamefont
  {Liu}}]{PhysRevE.97.012140}%
  \BibitemOpen
  \bibfield  {author} {\bibinfo {author} {\bibfnamefont {A.}~\bibnamefont
  {Dymarsky}}, \bibinfo {author} {\bibfnamefont {N.}~\bibnamefont {Lashkari}},\
  and\ \bibinfo {author} {\bibfnamefont {H.}~\bibnamefont {Liu}},\ }\bibfield
  {title} {\bibinfo {title} {Subsystem eigenstate thermalization hypothesis},\
  }\href {https://doi.org/10.1103/PhysRevE.97.012140} {\bibfield  {journal}
  {\bibinfo  {journal} {Phys. Rev. E}\ }\textbf {\bibinfo {volume} {97}},\
  \bibinfo {pages} {012140} (\bibinfo {year} {2018})}\BibitemShut {NoStop}%
\bibitem [{\citenamefont {Foini}\ and\ \citenamefont
  {Kurchan}(2019)}]{PhysRevE.99.042139}%
  \BibitemOpen
  \bibfield  {author} {\bibinfo {author} {\bibfnamefont {L.}~\bibnamefont
  {Foini}}\ and\ \bibinfo {author} {\bibfnamefont {J.}~\bibnamefont
  {Kurchan}},\ }\bibfield  {title} {\bibinfo {title} {Eigenstate thermalization
  hypothesis and out of time order correlators},\ }\href
  {https://doi.org/10.1103/PhysRevE.99.042139} {\bibfield  {journal} {\bibinfo
  {journal} {Phys. Rev. E}\ }\textbf {\bibinfo {volume} {99}},\ \bibinfo
  {pages} {042139} (\bibinfo {year} {2019})}\BibitemShut {NoStop}%
\bibitem [{\citenamefont {Murthy}\ and\ \citenamefont
  {Srednicki}(2019)}]{PhysRevLett.123.230606}%
  \BibitemOpen
  \bibfield  {author} {\bibinfo {author} {\bibfnamefont {C.}~\bibnamefont
  {Murthy}}\ and\ \bibinfo {author} {\bibfnamefont {M.}~\bibnamefont
  {Srednicki}},\ }\bibfield  {title} {\bibinfo {title} {Bounds on chaos from
  the eigenstate thermalization hypothesis},\ }\href
  {https://doi.org/10.1103/PhysRevLett.123.230606} {\bibfield  {journal}
  {\bibinfo  {journal} {Phys. Rev. Lett.}\ }\textbf {\bibinfo {volume} {123}},\
  \bibinfo {pages} {230606} (\bibinfo {year} {2019})}\BibitemShut {NoStop}%
\bibitem [{\citenamefont {Murthy}\ \emph {et~al.}(2023)\citenamefont {Murthy},
  \citenamefont {Babakhani}, \citenamefont {Iniguez}, \citenamefont
  {Srednicki},\ and\ \citenamefont {Yunger~Halpern}}]{PhysRevLett.130.140402}%
  \BibitemOpen
  \bibfield  {author} {\bibinfo {author} {\bibfnamefont {C.}~\bibnamefont
  {Murthy}}, \bibinfo {author} {\bibfnamefont {A.}~\bibnamefont {Babakhani}},
  \bibinfo {author} {\bibfnamefont {F.}~\bibnamefont {Iniguez}}, \bibinfo
  {author} {\bibfnamefont {M.}~\bibnamefont {Srednicki}},\ and\ \bibinfo
  {author} {\bibfnamefont {N.}~\bibnamefont {Yunger~Halpern}},\ }\bibfield
  {title} {\bibinfo {title} {Non-abelian eigenstate thermalization
  hypothesis},\ }\href {https://doi.org/10.1103/PhysRevLett.130.140402}
  {\bibfield  {journal} {\bibinfo  {journal} {Phys. Rev. Lett.}\ }\textbf
  {\bibinfo {volume} {130}},\ \bibinfo {pages} {140402} (\bibinfo {year}
  {2023})}\BibitemShut {NoStop}%
\bibitem [{\citenamefont {Mori}\ \emph {et~al.}(2018)\citenamefont {Mori},
  \citenamefont {Ikeda}, \citenamefont {Kaminishi},\ and\ \citenamefont
  {Ueda}}]{Mori_2018}%
  \BibitemOpen
  \bibfield  {author} {\bibinfo {author} {\bibfnamefont {T.}~\bibnamefont
  {Mori}}, \bibinfo {author} {\bibfnamefont {T.~N.}\ \bibnamefont {Ikeda}},
  \bibinfo {author} {\bibfnamefont {E.}~\bibnamefont {Kaminishi}},\ and\
  \bibinfo {author} {\bibfnamefont {M.}~\bibnamefont {Ueda}},\ }\bibfield
  {title} {\bibinfo {title} {Thermalization and prethermalization in isolated
  quantum systems: a theoretical overview},\ }\href
  {https://doi.org/10.1088/1361-6455/aabcdf} {\bibfield  {journal} {\bibinfo
  {journal} {Journal of Physics B: Atomic, Molecular and Optical Physics}\
  }\textbf {\bibinfo {volume} {51}},\ \bibinfo {pages} {112001} (\bibinfo
  {year} {2018})}\BibitemShut {NoStop}%
\bibitem [{\citenamefont {Luitz}\ and\ \citenamefont
  {Bar~Lev}(2017)}]{PhysRevB.96.020406}%
  \BibitemOpen
  \bibfield  {author} {\bibinfo {author} {\bibfnamefont {D.~J.}\ \bibnamefont
  {Luitz}}\ and\ \bibinfo {author} {\bibfnamefont {Y.}~\bibnamefont
  {Bar~Lev}},\ }\bibfield  {title} {\bibinfo {title} {Information propagation
  in isolated quantum systems},\ }\href
  {https://doi.org/10.1103/PhysRevB.96.020406} {\bibfield  {journal} {\bibinfo
  {journal} {Phys. Rev. B}\ }\textbf {\bibinfo {volume} {96}},\ \bibinfo
  {pages} {020406} (\bibinfo {year} {2017})}\BibitemShut {NoStop}%
\bibitem [{\citenamefont {Osborne}\ and\ \citenamefont
  {Linden}(2004)}]{PhysRevA.69.052315}%
  \BibitemOpen
  \bibfield  {author} {\bibinfo {author} {\bibfnamefont {T.~J.}\ \bibnamefont
  {Osborne}}\ and\ \bibinfo {author} {\bibfnamefont {N.}~\bibnamefont
  {Linden}},\ }\bibfield  {title} {\bibinfo {title} {Propagation of quantum
  information through a spin system},\ }\href
  {https://doi.org/10.1103/PhysRevA.69.052315} {\bibfield  {journal} {\bibinfo
  {journal} {Phys. Rev. A}\ }\textbf {\bibinfo {volume} {69}},\ \bibinfo
  {pages} {052315} (\bibinfo {year} {2004})}\BibitemShut {NoStop}%
\bibitem [{\citenamefont {Tanimura}\ and\ \citenamefont
  {Kubo}(1989)}]{1989101}%
  \BibitemOpen
  \bibfield  {author} {\bibinfo {author} {\bibfnamefont {Y.}~\bibnamefont
  {Tanimura}}\ and\ \bibinfo {author} {\bibfnamefont {R.}~\bibnamefont
  {Kubo}},\ }\bibfield  {title} {\bibinfo {title} {Time evolution of a quantum
  system in contact with a nearly gaussian-markoffian noise bath},\ }\href
  {https://doi.org/10.1143/JPSJ.58.101} {\bibfield  {journal} {\bibinfo
  {journal} {Journal of the Physical Society of Japan}\ }\textbf {\bibinfo
  {volume} {58}},\ \bibinfo {pages} {101} (\bibinfo {year} {1989})}\BibitemShut
  {NoStop}%
\bibitem [{\citenamefont {Johnson}\ and\ \citenamefont
  {Soff}(1985)}]{JOHNSON1985405}%
  \BibitemOpen
  \bibfield  {author} {\bibinfo {author} {\bibfnamefont {W.}~\bibnamefont
  {Johnson}}\ and\ \bibinfo {author} {\bibfnamefont {G.}~\bibnamefont {Soff}},\
  }\bibfield  {title} {\bibinfo {title} {The lamb shift in hydrogen-like atoms,
  $1 \leq z \leq 110$},\ }\href
  {https://doi.org/https://doi.org/10.1016/0092-640X(85)90010-5} {\bibfield
  {journal} {\bibinfo  {journal} {Atomic Data and Nuclear Data Tables}\
  }\textbf {\bibinfo {volume} {33}},\ \bibinfo {pages} {405} (\bibinfo {year}
  {1985})}\BibitemShut {NoStop}%
\bibitem [{\citenamefont {Rotter}\ and\ \citenamefont
  {Bird}(2015)}]{Rotter_2015}%
  \BibitemOpen
  \bibfield  {author} {\bibinfo {author} {\bibfnamefont {I.}~\bibnamefont
  {Rotter}}\ and\ \bibinfo {author} {\bibfnamefont {J.~P.}\ \bibnamefont
  {Bird}},\ }\bibfield  {title} {\bibinfo {title} {A review of progress in the
  physics of open quantum systems: theory and experiment},\ }\href
  {https://doi.org/10.1088/0034-4885/78/11/114001} {\bibfield  {journal}
  {\bibinfo  {journal} {Reports on Progress in Physics}\ }\textbf {\bibinfo
  {volume} {78}},\ \bibinfo {pages} {114001} (\bibinfo {year}
  {2015})}\BibitemShut {NoStop}%
\bibitem [{\citenamefont {REDFIELD}(1965)}]{REDFIELD19651}%
  \BibitemOpen
  \bibfield  {author} {\bibinfo {author} {\bibfnamefont {A.}~\bibnamefont
  {REDFIELD}},\ }\bibfield  {title} {\bibinfo {title} {The theory of relaxation
  processes* *this work was started while the author was at harvard university,
  and was then partially supported by joint services contract n5ori-76, project
  order i.},\ }in\ \href
  {https://doi.org/https://doi.org/10.1016/B978-1-4832-3114-3.50007-6} {\emph
  {\bibinfo {booktitle} {Advances in Magnetic Resonance}}},\ \bibinfo {series}
  {Advances in Magnetic and Optical Resonance}, Vol.~\bibinfo {volume} {1},\
  \bibinfo {editor} {edited by\ \bibinfo {editor} {\bibfnamefont {J.~S.}\
  \bibnamefont {Waugh}}}\ (\bibinfo  {publisher} {Academic Press},\ \bibinfo
  {year} {1965})\ pp.\ \bibinfo {pages} {1--32}\BibitemShut {NoStop}%
\bibitem [{\citenamefont {Wineland}\ and\ \citenamefont
  {Itano}(1979)}]{PhysRevA.20.1521}%
  \BibitemOpen
  \bibfield  {author} {\bibinfo {author} {\bibfnamefont {D.~J.}\ \bibnamefont
  {Wineland}}\ and\ \bibinfo {author} {\bibfnamefont {W.~M.}\ \bibnamefont
  {Itano}},\ }\bibfield  {title} {\bibinfo {title} {Laser cooling of atoms},\
  }\href {https://doi.org/10.1103/PhysRevA.20.1521} {\bibfield  {journal}
  {\bibinfo  {journal} {Phys. Rev. A}\ }\textbf {\bibinfo {volume} {20}},\
  \bibinfo {pages} {1521} (\bibinfo {year} {1979})}\BibitemShut {NoStop}%
\bibitem [{\citenamefont {Gong}\ \emph {et~al.}(2021)\citenamefont {Gong},
  \citenamefont {de~Moraes~Neto}, \citenamefont {Zha}, \citenamefont {Wu},
  \citenamefont {Rong}, \citenamefont {Ye}, \citenamefont {Li}, \citenamefont
  {Zhu}, \citenamefont {Wang}, \citenamefont {Zhao}, \citenamefont {Liang},
  \citenamefont {Lin}, \citenamefont {Xu}, \citenamefont {Peng}, \citenamefont
  {Deng}, \citenamefont {Bayat}, \citenamefont {Zhu},\ and\ \citenamefont
  {Pan}}]{PhysRevResearch.3.033043}%
  \BibitemOpen
  \bibfield  {author} {\bibinfo {author} {\bibfnamefont {M.}~\bibnamefont
  {Gong}}, \bibinfo {author} {\bibfnamefont {G.~D.}\ \bibnamefont
  {de~Moraes~Neto}}, \bibinfo {author} {\bibfnamefont {C.}~\bibnamefont {Zha}},
  \bibinfo {author} {\bibfnamefont {Y.}~\bibnamefont {Wu}}, \bibinfo {author}
  {\bibfnamefont {H.}~\bibnamefont {Rong}}, \bibinfo {author} {\bibfnamefont
  {Y.}~\bibnamefont {Ye}}, \bibinfo {author} {\bibfnamefont {S.}~\bibnamefont
  {Li}}, \bibinfo {author} {\bibfnamefont {Q.}~\bibnamefont {Zhu}}, \bibinfo
  {author} {\bibfnamefont {S.}~\bibnamefont {Wang}}, \bibinfo {author}
  {\bibfnamefont {Y.}~\bibnamefont {Zhao}}, \bibinfo {author} {\bibfnamefont
  {F.}~\bibnamefont {Liang}}, \bibinfo {author} {\bibfnamefont
  {J.}~\bibnamefont {Lin}}, \bibinfo {author} {\bibfnamefont {Y.}~\bibnamefont
  {Xu}}, \bibinfo {author} {\bibfnamefont {C.-Z.}\ \bibnamefont {Peng}},
  \bibinfo {author} {\bibfnamefont {H.}~\bibnamefont {Deng}}, \bibinfo {author}
  {\bibfnamefont {A.}~\bibnamefont {Bayat}}, \bibinfo {author} {\bibfnamefont
  {X.}~\bibnamefont {Zhu}},\ and\ \bibinfo {author} {\bibfnamefont {J.-W.}\
  \bibnamefont {Pan}},\ }\bibfield  {title} {\bibinfo {title} {Experimental
  characterization of the quantum many-body localization transition},\ }\href
  {https://doi.org/10.1103/PhysRevResearch.3.033043} {\bibfield  {journal}
  {\bibinfo  {journal} {Phys. Rev. Res.}\ }\textbf {\bibinfo {volume} {3}},\
  \bibinfo {pages} {033043} (\bibinfo {year} {2021})}\BibitemShut {NoStop}%
\bibitem [{\citenamefont {Bernon}\ \emph {et~al.}(2013)\citenamefont {Bernon},
  \citenamefont {Hattermann}, \citenamefont {Bothner}, \citenamefont
  {Knufinke}, \citenamefont {Weiss}, \citenamefont {Jessen}, \citenamefont
  {Cano}, \citenamefont {Kemmler}, \citenamefont {Kleiner}, \citenamefont
  {Koelle},\ and\ \citenamefont {Fort{\'a}gh}}]{Bernon2013}%
  \BibitemOpen
  \bibfield  {author} {\bibinfo {author} {\bibfnamefont {S.}~\bibnamefont
  {Bernon}}, \bibinfo {author} {\bibfnamefont {H.}~\bibnamefont {Hattermann}},
  \bibinfo {author} {\bibfnamefont {D.}~\bibnamefont {Bothner}}, \bibinfo
  {author} {\bibfnamefont {M.}~\bibnamefont {Knufinke}}, \bibinfo {author}
  {\bibfnamefont {P.}~\bibnamefont {Weiss}}, \bibinfo {author} {\bibfnamefont
  {F.}~\bibnamefont {Jessen}}, \bibinfo {author} {\bibfnamefont
  {D.}~\bibnamefont {Cano}}, \bibinfo {author} {\bibfnamefont {M.}~\bibnamefont
  {Kemmler}}, \bibinfo {author} {\bibfnamefont {R.}~\bibnamefont {Kleiner}},
  \bibinfo {author} {\bibfnamefont {D.}~\bibnamefont {Koelle}},\ and\ \bibinfo
  {author} {\bibfnamefont {J.}~\bibnamefont {Fort{\'a}gh}},\ }\bibfield
  {title} {\bibinfo {title} {Manipulation and coherence of ultra-cold atoms on
  a superconducting atom chip},\ }\href {https://doi.org/10.1038/ncomms3380}
  {\bibfield  {journal} {\bibinfo  {journal} {Nature Communications}\ }\textbf
  {\bibinfo {volume} {4}},\ \bibinfo {pages} {2380} (\bibinfo {year}
  {2013})}\BibitemShut {NoStop}%
\bibitem [{\citenamefont {L{\'e}onard}\ \emph {et~al.}(2023)\citenamefont
  {L{\'e}onard}, \citenamefont {Kim}, \citenamefont {Rispoli}, \citenamefont
  {Lukin}, \citenamefont {Schittko}, \citenamefont {Kwan}, \citenamefont
  {Demler}, \citenamefont {Sels},\ and\ \citenamefont
  {Greiner}}]{Léonard2023}%
  \BibitemOpen
  \bibfield  {author} {\bibinfo {author} {\bibfnamefont {J.}~\bibnamefont
  {L{\'e}onard}}, \bibinfo {author} {\bibfnamefont {S.}~\bibnamefont {Kim}},
  \bibinfo {author} {\bibfnamefont {M.}~\bibnamefont {Rispoli}}, \bibinfo
  {author} {\bibfnamefont {A.}~\bibnamefont {Lukin}}, \bibinfo {author}
  {\bibfnamefont {R.}~\bibnamefont {Schittko}}, \bibinfo {author}
  {\bibfnamefont {J.}~\bibnamefont {Kwan}}, \bibinfo {author} {\bibfnamefont
  {E.}~\bibnamefont {Demler}}, \bibinfo {author} {\bibfnamefont
  {D.}~\bibnamefont {Sels}},\ and\ \bibinfo {author} {\bibfnamefont
  {M.}~\bibnamefont {Greiner}},\ }\bibfield  {title} {\bibinfo {title} {Probing
  the onset of quantum avalanches in a many-body localized system},\ }\href
  {https://doi.org/10.1038/s41567-022-01887-3} {\bibfield  {journal} {\bibinfo
  {journal} {Nature Physics}\ }\textbf {\bibinfo {volume} {19}},\ \bibinfo
  {pages} {481} (\bibinfo {year} {2023})}\BibitemShut {NoStop}%
\bibitem [{\citenamefont {Dowling}\ and\ \citenamefont
  {Milburn}(2003)}]{dowling2003quantum}%
  \BibitemOpen
  \bibfield  {author} {\bibinfo {author} {\bibfnamefont {J.~P.}\ \bibnamefont
  {Dowling}}\ and\ \bibinfo {author} {\bibfnamefont {G.~J.}\ \bibnamefont
  {Milburn}},\ }\bibfield  {title} {\bibinfo {title} {Quantum technology: the
  second quantum revolution},\ }\href@noop {} {\bibfield  {journal} {\bibinfo
  {journal} {Philosophical Transactions of the Royal Society of London. Series
  A: Mathematical, Physical and Engineering Sciences}\ }\textbf {\bibinfo
  {volume} {361}},\ \bibinfo {pages} {1655} (\bibinfo {year}
  {2003})}\BibitemShut {NoStop}%
\bibitem [{\citenamefont {Daley}\ \emph {et~al.}(2022)\citenamefont {Daley},
  \citenamefont {Bloch}, \citenamefont {Kokail}, \citenamefont {Flannigan},
  \citenamefont {Pearson}, \citenamefont {Troyer},\ and\ \citenamefont
  {Zoller}}]{Daley2022}%
  \BibitemOpen
  \bibfield  {author} {\bibinfo {author} {\bibfnamefont {A.~J.}\ \bibnamefont
  {Daley}}, \bibinfo {author} {\bibfnamefont {I.}~\bibnamefont {Bloch}},
  \bibinfo {author} {\bibfnamefont {C.}~\bibnamefont {Kokail}}, \bibinfo
  {author} {\bibfnamefont {S.}~\bibnamefont {Flannigan}}, \bibinfo {author}
  {\bibfnamefont {N.}~\bibnamefont {Pearson}}, \bibinfo {author} {\bibfnamefont
  {M.}~\bibnamefont {Troyer}},\ and\ \bibinfo {author} {\bibfnamefont
  {P.}~\bibnamefont {Zoller}},\ }\bibfield  {title} {\bibinfo {title}
  {Practical quantum advantage in quantum simulation},\ }\href
  {https://doi.org/10.1038/s41586-022-04940-6} {\bibfield  {journal} {\bibinfo
  {journal} {Nature}\ }\textbf {\bibinfo {volume} {607}},\ \bibinfo {pages}
  {667} (\bibinfo {year} {2022})}\BibitemShut {NoStop}%
\bibitem [{\citenamefont {Shor}(1997)}]{doi:10.1137/S0097539795293172}%
  \BibitemOpen
  \bibfield  {author} {\bibinfo {author} {\bibfnamefont {P.~W.}\ \bibnamefont
  {Shor}},\ }\bibfield  {title} {\bibinfo {title} {Polynomial-time algorithms
  for prime factorization and discrete logarithms on a quantum computer},\
  }\href {https://doi.org/10.1137/S0097539795293172} {\bibfield  {journal}
  {\bibinfo  {journal} {SIAM Journal on Computing}\ }\textbf {\bibinfo {volume}
  {26}},\ \bibinfo {pages} {1484} (\bibinfo {year} {1997})},\ \Eprint
  {https://arxiv.org/abs/https://doi.org/10.1137/S0097539795293172}
  {https://doi.org/10.1137/S0097539795293172} \BibitemShut {NoStop}%
\bibitem [{\citenamefont {Aaronson}(2008)}]{aaronson2008limits}%
  \BibitemOpen
  \bibfield  {author} {\bibinfo {author} {\bibfnamefont {S.}~\bibnamefont
  {Aaronson}},\ }\bibfield  {title} {\bibinfo {title} {The limits of quantum
  computers},\ }\href
  {https://www.scientificamerican.com/article/the-limits-of-quantum-computers/}
  {\bibfield  {journal} {\bibinfo  {journal} {Scientific American}\ }\textbf
  {\bibinfo {volume} {298}},\ \bibinfo {pages} {62} (\bibinfo {year}
  {2008})}\BibitemShut {NoStop}%
\bibitem [{\citenamefont {{Cao}}\ \emph {et~al.}(2018)\citenamefont {{Cao}},
  \citenamefont {{Romero}}, \citenamefont {{Olson}}, \citenamefont
  {{Degroote}}, \citenamefont {{Johnson}}, \citenamefont {{Kieferov{\'a}}},
  \citenamefont {{Kivlichan}}, \citenamefont {{Menke}}, \citenamefont
  {{Peropadre}}, \citenamefont {{Sawaya}}, \citenamefont {{Sim}}, \citenamefont
  {{Veis}},\ and\ \citenamefont {{Aspuru-Guzik}}}]{cao2018quantum}%
  \BibitemOpen
  \bibfield  {author} {\bibinfo {author} {\bibfnamefont {Y.}~\bibnamefont
  {{Cao}}}, \bibinfo {author} {\bibfnamefont {J.}~\bibnamefont {{Romero}}},
  \bibinfo {author} {\bibfnamefont {J.~P.}\ \bibnamefont {{Olson}}}, \bibinfo
  {author} {\bibfnamefont {M.}~\bibnamefont {{Degroote}}}, \bibinfo {author}
  {\bibfnamefont {P.~D.}\ \bibnamefont {{Johnson}}}, \bibinfo {author}
  {\bibfnamefont {M.}~\bibnamefont {{Kieferov{\'a}}}}, \bibinfo {author}
  {\bibfnamefont {I.~D.}\ \bibnamefont {{Kivlichan}}}, \bibinfo {author}
  {\bibfnamefont {T.}~\bibnamefont {{Menke}}}, \bibinfo {author} {\bibfnamefont
  {B.}~\bibnamefont {{Peropadre}}}, \bibinfo {author} {\bibfnamefont
  {N.~P.~D.}\ \bibnamefont {{Sawaya}}}, \bibinfo {author} {\bibfnamefont
  {S.}~\bibnamefont {{Sim}}}, \bibinfo {author} {\bibfnamefont
  {L.}~\bibnamefont {{Veis}}},\ and\ \bibinfo {author} {\bibfnamefont
  {A.}~\bibnamefont {{Aspuru-Guzik}}},\ }\bibfield  {title} {\bibinfo {title}
  {{Quantum Chemistry in the Age of Quantum Computing}},\ }\href@noop {}
  {\bibfield  {journal} {\bibinfo  {journal} {arXiv e-prints}\ ,\ \bibinfo
  {eid} {arXiv:1812.09976}} (\bibinfo {year} {2018})},\ \Eprint
  {https://arxiv.org/abs/1812.09976} {arXiv:1812.09976 [quant-ph]} \BibitemShut
  {NoStop}%
\bibitem [{\citenamefont {{Menges}}\ \emph {et~al.}(2016)\citenamefont
  {{Menges}}, \citenamefont {{Mensch}}, \citenamefont {{Schmid}}, \citenamefont
  {{Riel}}, \citenamefont {{Stemmer}},\ and\ \citenamefont
  {{Gotsmann}}}]{menges2016temperature}%
  \BibitemOpen
  \bibfield  {author} {\bibinfo {author} {\bibfnamefont {F.}~\bibnamefont
  {{Menges}}}, \bibinfo {author} {\bibfnamefont {P.}~\bibnamefont {{Mensch}}},
  \bibinfo {author} {\bibfnamefont {H.}~\bibnamefont {{Schmid}}}, \bibinfo
  {author} {\bibfnamefont {H.}~\bibnamefont {{Riel}}}, \bibinfo {author}
  {\bibfnamefont {A.}~\bibnamefont {{Stemmer}}},\ and\ \bibinfo {author}
  {\bibfnamefont {B.}~\bibnamefont {{Gotsmann}}},\ }\bibfield  {title}
  {\bibinfo {title} {{Temperature mapping of operating nanoscale devices by
  scanning probe thermometry}},\ }\href {https://doi.org/10.1038/ncomms10874}
  {\bibfield  {journal} {\bibinfo  {journal} {Nature Communications}\ }\textbf
  {\bibinfo {volume} {7}},\ \bibinfo {eid} {10874} (\bibinfo {year}
  {2016})}\BibitemShut {NoStop}%
\bibitem [{\citenamefont {Santos}\ \emph {et~al.}(2016)\citenamefont {Santos},
  \citenamefont {Silva},\ and\ \citenamefont {Sarandy}}]{PhysRevA.93.012311}%
  \BibitemOpen
  \bibfield  {author} {\bibinfo {author} {\bibfnamefont {A.~C.}\ \bibnamefont
  {Santos}}, \bibinfo {author} {\bibfnamefont {R.~D.}\ \bibnamefont {Silva}},\
  and\ \bibinfo {author} {\bibfnamefont {M.~S.}\ \bibnamefont {Sarandy}},\
  }\bibfield  {title} {\bibinfo {title} {Shortcut to adiabatic gate
  teleportation},\ }\href {https://doi.org/10.1103/PhysRevA.93.012311}
  {\bibfield  {journal} {\bibinfo  {journal} {Phys. Rev. A}\ }\textbf {\bibinfo
  {volume} {93}},\ \bibinfo {pages} {012311} (\bibinfo {year}
  {2016})}\BibitemShut {NoStop}%
\bibitem [{\citenamefont {Georgescu}\ \emph {et~al.}(2014)\citenamefont
  {Georgescu}, \citenamefont {Ashhab},\ and\ \citenamefont
  {Nori}}]{RevModPhys.86.153}%
  \BibitemOpen
  \bibfield  {author} {\bibinfo {author} {\bibfnamefont {I.~M.}\ \bibnamefont
  {Georgescu}}, \bibinfo {author} {\bibfnamefont {S.}~\bibnamefont {Ashhab}},\
  and\ \bibinfo {author} {\bibfnamefont {F.}~\bibnamefont {Nori}},\ }\bibfield
  {title} {\bibinfo {title} {Quantum simulation},\ }\href
  {https://doi.org/10.1103/RevModPhys.86.153} {\bibfield  {journal} {\bibinfo
  {journal} {Rev. Mod. Phys.}\ }\textbf {\bibinfo {volume} {86}},\ \bibinfo
  {pages} {153} (\bibinfo {year} {2014})}\BibitemShut {NoStop}%
\bibitem [{\citenamefont {{Pirandola}}\ \emph {et~al.}(2020)\citenamefont
  {{Pirandola}}, \citenamefont {{Andersen}}, \citenamefont {{Banchi}},
  \citenamefont {{Berta}}, \citenamefont {{Bunandar}}, \citenamefont
  {{Colbeck}}, \citenamefont {{Englund}}, \citenamefont {{Gehring}},
  \citenamefont {{Lupo}}, \citenamefont {{Ottaviani}}, \citenamefont
  {{Pereira}}, \citenamefont {{Razavi}}, \citenamefont {{Shamsul Shaari}},
  \citenamefont {{Tomamichel}}, \citenamefont {{Usenko}}, \citenamefont
  {{Vallone}}, \citenamefont {{Villoresi}},\ and\ \citenamefont
  {{Wallden}}}]{pirandola2020advances}%
  \BibitemOpen
  \bibfield  {author} {\bibinfo {author} {\bibfnamefont {S.}~\bibnamefont
  {{Pirandola}}}, \bibinfo {author} {\bibfnamefont {U.~L.}\ \bibnamefont
  {{Andersen}}}, \bibinfo {author} {\bibfnamefont {L.}~\bibnamefont
  {{Banchi}}}, \bibinfo {author} {\bibfnamefont {M.}~\bibnamefont {{Berta}}},
  \bibinfo {author} {\bibfnamefont {D.}~\bibnamefont {{Bunandar}}}, \bibinfo
  {author} {\bibfnamefont {R.}~\bibnamefont {{Colbeck}}}, \bibinfo {author}
  {\bibfnamefont {D.}~\bibnamefont {{Englund}}}, \bibinfo {author}
  {\bibfnamefont {T.}~\bibnamefont {{Gehring}}}, \bibinfo {author}
  {\bibfnamefont {C.}~\bibnamefont {{Lupo}}}, \bibinfo {author} {\bibfnamefont
  {C.}~\bibnamefont {{Ottaviani}}}, \bibinfo {author} {\bibfnamefont {J.~L.}\
  \bibnamefont {{Pereira}}}, \bibinfo {author} {\bibfnamefont {M.}~\bibnamefont
  {{Razavi}}}, \bibinfo {author} {\bibfnamefont {J.}~\bibnamefont {{Shamsul
  Shaari}}}, \bibinfo {author} {\bibfnamefont {M.}~\bibnamefont
  {{Tomamichel}}}, \bibinfo {author} {\bibfnamefont {V.~C.}\ \bibnamefont
  {{Usenko}}}, \bibinfo {author} {\bibfnamefont {G.}~\bibnamefont {{Vallone}}},
  \bibinfo {author} {\bibfnamefont {P.}~\bibnamefont {{Villoresi}}},\ and\
  \bibinfo {author} {\bibfnamefont {P.}~\bibnamefont {{Wallden}}},\ }\bibfield
  {title} {\bibinfo {title} {{Advances in quantum cryptography}},\ }\href
  {https://doi.org/10.1364/AOP.361502} {\bibfield  {journal} {\bibinfo
  {journal} {Advances in Optics and Photonics}\ }\textbf {\bibinfo {volume}
  {12}},\ \bibinfo {pages} {1012} (\bibinfo {year} {2020})},\ \Eprint
  {https://arxiv.org/abs/1906.01645} {arXiv:1906.01645 [quant-ph]} \BibitemShut
  {NoStop}%
\bibitem [{\citenamefont {{Aasi et al.}}(2013)}]{LIGO2013}%
  \BibitemOpen
  \bibfield  {author} {\bibinfo {author} {\bibfnamefont {J.}~\bibnamefont
  {{Aasi et al.}}},\ }\bibfield  {title} {\bibinfo {title} {{Enhanced
  sensitivity of the LIGO gravitational wave detector by using squeezed states
  of light}},\ }\href {https://doi.org/10.1038/nphoton.2013.177} {\bibfield
  {journal} {\bibinfo  {journal} {Nature Photonics}\ }\textbf {\bibinfo
  {volume} {7}},\ \bibinfo {pages} {613} (\bibinfo {year} {2013})},\ \Eprint
  {https://arxiv.org/abs/1310.0383} {arXiv:1310.0383 [quant-ph]} \BibitemShut
  {NoStop}%
\bibitem [{\citenamefont {Degen}\ \emph {et~al.}(2017)\citenamefont {Degen},
  \citenamefont {Reinhard},\ and\ \citenamefont
  {Cappellaro}}]{degen2017quantum}%
  \BibitemOpen
  \bibfield  {author} {\bibinfo {author} {\bibfnamefont {C.~L.}\ \bibnamefont
  {Degen}}, \bibinfo {author} {\bibfnamefont {F.}~\bibnamefont {Reinhard}},\
  and\ \bibinfo {author} {\bibfnamefont {P.}~\bibnamefont {Cappellaro}},\
  }\bibfield  {title} {\bibinfo {title} {Quantum sensing},\ }\href
  {https://doi.org/10.1103/RevModPhys.89.035002} {\bibfield  {journal}
  {\bibinfo  {journal} {Rev. Mod. Phys.}\ }\textbf {\bibinfo {volume} {89}},\
  \bibinfo {pages} {035002} (\bibinfo {year} {2017})}\BibitemShut {NoStop}%
\bibitem [{\citenamefont {Cheiney}\ \emph {et~al.}(2018)\citenamefont
  {Cheiney}, \citenamefont {Fouch\'e}, \citenamefont {Templier}, \citenamefont
  {Napolitano}, \citenamefont {Battelier}, \citenamefont {Bouyer},\ and\
  \citenamefont {Barrett}}]{cheiney2018navigation}%
  \BibitemOpen
  \bibfield  {author} {\bibinfo {author} {\bibfnamefont {P.}~\bibnamefont
  {Cheiney}}, \bibinfo {author} {\bibfnamefont {L.}~\bibnamefont {Fouch\'e}},
  \bibinfo {author} {\bibfnamefont {S.}~\bibnamefont {Templier}}, \bibinfo
  {author} {\bibfnamefont {F.}~\bibnamefont {Napolitano}}, \bibinfo {author}
  {\bibfnamefont {B.}~\bibnamefont {Battelier}}, \bibinfo {author}
  {\bibfnamefont {P.}~\bibnamefont {Bouyer}},\ and\ \bibinfo {author}
  {\bibfnamefont {B.}~\bibnamefont {Barrett}},\ }\bibfield  {title} {\bibinfo
  {title} {Navigation-compatible hybrid quantum accelerometer using a {Kalman}
  filter},\ }\href {https://doi.org/10.1103/PhysRevApplied.10.034030}
  {\bibfield  {journal} {\bibinfo  {journal} {Phys. Rev. Appl.}\ }\textbf
  {\bibinfo {volume} {10}},\ \bibinfo {pages} {034030} (\bibinfo {year}
  {2018})}\BibitemShut {NoStop}%
\bibitem [{\citenamefont {Campaioli}\ \emph {et~al.}(2023)\citenamefont
  {Campaioli}, \citenamefont {Gherardini}, \citenamefont {Quach}, \citenamefont
  {Polini},\ and\ \citenamefont {Andolina}}]{campaioli2023colloquium}%
  \BibitemOpen
  \bibfield  {author} {\bibinfo {author} {\bibfnamefont {F.}~\bibnamefont
  {Campaioli}}, \bibinfo {author} {\bibfnamefont {S.}~\bibnamefont
  {Gherardini}}, \bibinfo {author} {\bibfnamefont {J.~Q.}\ \bibnamefont
  {Quach}}, \bibinfo {author} {\bibfnamefont {M.}~\bibnamefont {Polini}},\ and\
  \bibinfo {author} {\bibfnamefont {G.~M.}\ \bibnamefont {Andolina}},\
  }\href@noop {} {\bibinfo {title} {Colloquium: Quantum batteries}} (\bibinfo
  {year} {2023}),\ \Eprint {https://arxiv.org/abs/2308.02277} {arXiv:2308.02277
  [quant-ph]} \BibitemShut {NoStop}%
\bibitem [{\citenamefont {Binder}\ \emph {et~al.}(2015)\citenamefont {Binder},
  \citenamefont {Vinjanampathy}, \citenamefont {Modi},\ and\ \citenamefont
  {Goold}}]{Binder_2015}%
  \BibitemOpen
  \bibfield  {author} {\bibinfo {author} {\bibfnamefont {F.~C.}\ \bibnamefont
  {Binder}}, \bibinfo {author} {\bibfnamefont {S.}~\bibnamefont
  {Vinjanampathy}}, \bibinfo {author} {\bibfnamefont {K.}~\bibnamefont
  {Modi}},\ and\ \bibinfo {author} {\bibfnamefont {J.}~\bibnamefont {Goold}},\
  }\bibfield  {title} {\bibinfo {title} {Quantacell: powerful charging of
  quantum batteries},\ }\href {https://doi.org/10.1088/1367-2630/17/7/075015}
  {\bibfield  {journal} {\bibinfo  {journal} {New Journal of Physics}\ }\textbf
  {\bibinfo {volume} {17}},\ \bibinfo {pages} {075015} (\bibinfo {year}
  {2015})}\BibitemShut {NoStop}%
\bibitem [{\citenamefont {Alicki}\ and\ \citenamefont
  {Fannes}(2013)}]{PhysRevE.87.042123}%
  \BibitemOpen
  \bibfield  {author} {\bibinfo {author} {\bibfnamefont {R.}~\bibnamefont
  {Alicki}}\ and\ \bibinfo {author} {\bibfnamefont {M.}~\bibnamefont
  {Fannes}},\ }\bibfield  {title} {\bibinfo {title} {Entanglement boost for
  extractable work from ensembles of quantum batteries},\ }\href
  {https://doi.org/10.1103/PhysRevE.87.042123} {\bibfield  {journal} {\bibinfo
  {journal} {Phys. Rev. E}\ }\textbf {\bibinfo {volume} {87}},\ \bibinfo
  {pages} {042123} (\bibinfo {year} {2013})}\BibitemShut {NoStop}%
\bibitem [{\citenamefont {Hovhannisyan}\ \emph {et~al.}(2013)\citenamefont
  {Hovhannisyan}, \citenamefont {Perarnau-Llobet}, \citenamefont {Huber},\ and\
  \citenamefont {Ac\'{\i}n}}]{PhysRevLett.111.240401}%
  \BibitemOpen
  \bibfield  {author} {\bibinfo {author} {\bibfnamefont {K.~V.}\ \bibnamefont
  {Hovhannisyan}}, \bibinfo {author} {\bibfnamefont {M.}~\bibnamefont
  {Perarnau-Llobet}}, \bibinfo {author} {\bibfnamefont {M.}~\bibnamefont
  {Huber}},\ and\ \bibinfo {author} {\bibfnamefont {A.}~\bibnamefont
  {Ac\'{\i}n}},\ }\bibfield  {title} {\bibinfo {title} {Entanglement generation
  is not necessary for optimal work extraction},\ }\href
  {https://doi.org/10.1103/PhysRevLett.111.240401} {\bibfield  {journal}
  {\bibinfo  {journal} {Phys. Rev. Lett.}\ }\textbf {\bibinfo {volume} {111}},\
  \bibinfo {pages} {240401} (\bibinfo {year} {2013})}\BibitemShut {NoStop}%
\bibitem [{\citenamefont {Tirone}\ \emph {et~al.}(2023)\citenamefont {Tirone},
  \citenamefont {Salvia}, \citenamefont {Chessa},\ and\ \citenamefont
  {Giovannetti}}]{tirone2023work}%
  \BibitemOpen
  \bibfield  {author} {\bibinfo {author} {\bibfnamefont {S.}~\bibnamefont
  {Tirone}}, \bibinfo {author} {\bibfnamefont {R.}~\bibnamefont {Salvia}},
  \bibinfo {author} {\bibfnamefont {S.}~\bibnamefont {Chessa}},\ and\ \bibinfo
  {author} {\bibfnamefont {V.}~\bibnamefont {Giovannetti}},\ }\bibfield
  {title} {\bibinfo {title} {Work extraction processes from noisy quantum
  batteries: The role of nonlocal resources},\ }\href
  {https://doi.org/10.1103/PhysRevLett.131.060402} {\bibfield  {journal}
  {\bibinfo  {journal} {Phys. Rev. Lett.}\ }\textbf {\bibinfo {volume} {131}},\
  \bibinfo {pages} {060402} (\bibinfo {year} {2023})}\BibitemShut {NoStop}%
\bibitem [{\citenamefont {Zhang}\ \emph {et~al.}(2019)\citenamefont {Zhang},
  \citenamefont {Yang}, \citenamefont {Fu},\ and\ \citenamefont
  {Wang}}]{PhysRevE.99.052106}%
  \BibitemOpen
  \bibfield  {author} {\bibinfo {author} {\bibfnamefont {Y.-Y.}\ \bibnamefont
  {Zhang}}, \bibinfo {author} {\bibfnamefont {T.-R.}\ \bibnamefont {Yang}},
  \bibinfo {author} {\bibfnamefont {L.}~\bibnamefont {Fu}},\ and\ \bibinfo
  {author} {\bibfnamefont {X.}~\bibnamefont {Wang}},\ }\bibfield  {title}
  {\bibinfo {title} {Powerful harmonic charging in a quantum battery},\ }\href
  {https://doi.org/10.1103/PhysRevE.99.052106} {\bibfield  {journal} {\bibinfo
  {journal} {Phys. Rev. E}\ }\textbf {\bibinfo {volume} {99}},\ \bibinfo
  {pages} {052106} (\bibinfo {year} {2019})}\BibitemShut {NoStop}%
\bibitem [{\citenamefont {Caravelli}\ \emph {et~al.}(2020)\citenamefont
  {Caravelli}, \citenamefont {Coulter-De~Wit}, \citenamefont
  {Garc\'{\i}a-Pintos},\ and\ \citenamefont
  {Hamma}}]{PhysRevResearch.2.023095}%
  \BibitemOpen
  \bibfield  {author} {\bibinfo {author} {\bibfnamefont {F.}~\bibnamefont
  {Caravelli}}, \bibinfo {author} {\bibfnamefont {G.}~\bibnamefont
  {Coulter-De~Wit}}, \bibinfo {author} {\bibfnamefont {L.~P.}\ \bibnamefont
  {Garc\'{\i}a-Pintos}},\ and\ \bibinfo {author} {\bibfnamefont
  {A.}~\bibnamefont {Hamma}},\ }\bibfield  {title} {\bibinfo {title} {Random
  quantum batteries},\ }\href
  {https://doi.org/10.1103/PhysRevResearch.2.023095} {\bibfield  {journal}
  {\bibinfo  {journal} {Phys. Rev. Res.}\ }\textbf {\bibinfo {volume} {2}},\
  \bibinfo {pages} {023095} (\bibinfo {year} {2020})}\BibitemShut {NoStop}%
\bibitem [{\citenamefont {Quach}\ and\ \citenamefont
  {Munro}(2020)}]{PhysRevApplied.14.024092}%
  \BibitemOpen
  \bibfield  {author} {\bibinfo {author} {\bibfnamefont {J.~Q.}\ \bibnamefont
  {Quach}}\ and\ \bibinfo {author} {\bibfnamefont {W.~J.}\ \bibnamefont
  {Munro}},\ }\bibfield  {title} {\bibinfo {title} {Using dark states to charge
  and stabilize open quantum batteries},\ }\href
  {https://doi.org/10.1103/PhysRevApplied.14.024092} {\bibfield  {journal}
  {\bibinfo  {journal} {Phys. Rev. Appl.}\ }\textbf {\bibinfo {volume} {14}},\
  \bibinfo {pages} {024092} (\bibinfo {year} {2020})}\BibitemShut {NoStop}%
\bibitem [{\citenamefont {Allahverdyan}\ \emph {et~al.}(2004)\citenamefont
  {Allahverdyan}, \citenamefont {Balian},\ and\ \citenamefont
  {Nieuwenhuizen}}]{allahverdyan2004maximal}%
  \BibitemOpen
  \bibfield  {author} {\bibinfo {author} {\bibfnamefont {A.~E.}\ \bibnamefont
  {Allahverdyan}}, \bibinfo {author} {\bibfnamefont {R.}~\bibnamefont
  {Balian}},\ and\ \bibinfo {author} {\bibfnamefont {T.~M.}\ \bibnamefont
  {Nieuwenhuizen}},\ }\bibfield  {title} {\bibinfo {title} {Maximal work
  extraction from finite quantum systems},\ }\href
  {https://doi.org/10.1209/epl/i2004-10101-2} {\bibfield  {journal} {\bibinfo
  {journal} {Europhysics Letters}\ }\textbf {\bibinfo {volume} {67}},\ \bibinfo
  {pages} {565} (\bibinfo {year} {2004})}\BibitemShut {NoStop}%
\bibitem [{\citenamefont {Touil}\ \emph {et~al.}(2021)\citenamefont {Touil},
  \citenamefont {Çakmak},\ and\ \citenamefont {Deffner}}]{touil2022ergotropy}%
  \BibitemOpen
  \bibfield  {author} {\bibinfo {author} {\bibfnamefont {A.}~\bibnamefont
  {Touil}}, \bibinfo {author} {\bibfnamefont {B.}~\bibnamefont {Çakmak}},\
  and\ \bibinfo {author} {\bibfnamefont {S.}~\bibnamefont {Deffner}},\
  }\bibfield  {title} {\bibinfo {title} {Ergotropy from quantum and classical
  correlations},\ }\href {https://doi.org/10.1088/1751-8121/ac3eba} {\bibfield
  {journal} {\bibinfo  {journal} {Journal of Physics A: Mathematical and
  Theoretical}\ }\textbf {\bibinfo {volume} {55}},\ \bibinfo {pages} {025301}
  (\bibinfo {year} {2021})}\BibitemShut {NoStop}%
\bibitem [{\citenamefont {\ifmmode~\check{S}\else \v{S}\fi{}afr\'anek}\ \emph
  {et~al.}(2023)\citenamefont {\ifmmode~\check{S}\else \v{S}\fi{}afr\'anek},
  \citenamefont {Rosa},\ and\ \citenamefont {Binder}}]{safranek2023work}%
  \BibitemOpen
  \bibfield  {author} {\bibinfo {author} {\bibfnamefont {D.}~\bibnamefont
  {\ifmmode~\check{S}\else \v{S}\fi{}afr\'anek}}, \bibinfo {author}
  {\bibfnamefont {D.}~\bibnamefont {Rosa}},\ and\ \bibinfo {author}
  {\bibfnamefont {F.~C.}\ \bibnamefont {Binder}},\ }\bibfield  {title}
  {\bibinfo {title} {Work extraction from unknown quantum sources},\ }\href
  {https://doi.org/10.1103/PhysRevLett.130.210401} {\bibfield  {journal}
  {\bibinfo  {journal} {Phys. Rev. Lett.}\ }\textbf {\bibinfo {volume} {130}},\
  \bibinfo {pages} {210401} (\bibinfo {year} {2023})}\BibitemShut {NoStop}%
\bibitem [{\citenamefont {Yang}\ \emph
  {et~al.}(2023{\natexlab{a}})\citenamefont {Yang}, \citenamefont {Yang},
  \citenamefont {Alimuddin}, \citenamefont {Salvia}, \citenamefont {Fei},
  \citenamefont {Zhao}, \citenamefont {Nimmrichter},\ and\ \citenamefont
  {Luo}}]{yang2023battery}%
  \BibitemOpen
  \bibfield  {author} {\bibinfo {author} {\bibfnamefont {X.}~\bibnamefont
  {Yang}}, \bibinfo {author} {\bibfnamefont {Y.-H.}\ \bibnamefont {Yang}},
  \bibinfo {author} {\bibfnamefont {M.}~\bibnamefont {Alimuddin}}, \bibinfo
  {author} {\bibfnamefont {R.}~\bibnamefont {Salvia}}, \bibinfo {author}
  {\bibfnamefont {S.-M.}\ \bibnamefont {Fei}}, \bibinfo {author} {\bibfnamefont
  {L.-M.}\ \bibnamefont {Zhao}}, \bibinfo {author} {\bibfnamefont
  {S.}~\bibnamefont {Nimmrichter}},\ and\ \bibinfo {author} {\bibfnamefont
  {M.-X.}\ \bibnamefont {Luo}},\ }\bibfield  {title} {\bibinfo {title} {Battery
  capacity of energy-storing quantum systems},\ }\href
  {https://doi.org/10.1103/PhysRevLett.131.030402} {\bibfield  {journal}
  {\bibinfo  {journal} {Phys. Rev. Lett.}\ }\textbf {\bibinfo {volume} {131}},\
  \bibinfo {pages} {030402} (\bibinfo {year} {2023}{\natexlab{a}})}\BibitemShut
  {NoStop}%
\bibitem [{\citenamefont {Mula}\ \emph {et~al.}(2023)\citenamefont {Mula},
  \citenamefont {Fern\'andez}, \citenamefont {Alvarellos}, \citenamefont
  {Fern\'andez}, \citenamefont {Garc\'{\i}a-Aldea}, \citenamefont {Santalla},\
  and\ \citenamefont {Rodr\'{\i}guez-Laguna}}]{mula2023ergotropy}%
  \BibitemOpen
  \bibfield  {author} {\bibinfo {author} {\bibfnamefont {B.~n.}\ \bibnamefont
  {Mula}}, \bibinfo {author} {\bibfnamefont {E.~M.}\ \bibnamefont
  {Fern\'andez}}, \bibinfo {author} {\bibfnamefont {J.~E.}\ \bibnamefont
  {Alvarellos}}, \bibinfo {author} {\bibfnamefont {J.~J.}\ \bibnamefont
  {Fern\'andez}}, \bibinfo {author} {\bibfnamefont {D.}~\bibnamefont
  {Garc\'{\i}a-Aldea}}, \bibinfo {author} {\bibfnamefont {S.~N.}\ \bibnamefont
  {Santalla}},\ and\ \bibinfo {author} {\bibfnamefont {J.}~\bibnamefont
  {Rodr\'{\i}guez-Laguna}},\ }\bibfield  {title} {\bibinfo {title} {Ergotropy
  and entanglement in critical spin chains},\ }\href
  {https://doi.org/10.1103/PhysRevB.107.075116} {\bibfield  {journal} {\bibinfo
   {journal} {Phys. Rev. B}\ }\textbf {\bibinfo {volume} {107}},\ \bibinfo
  {pages} {075116} (\bibinfo {year} {2023})}\BibitemShut {NoStop}%
\bibitem [{\citenamefont {Erdman}\ \emph {et~al.}(2022)\citenamefont {Erdman},
  \citenamefont {Andolina}, \citenamefont {Giovannetti},\ and\ \citenamefont
  {Noé}}]{erdman2022reinforcement}%
  \BibitemOpen
  \bibfield  {author} {\bibinfo {author} {\bibfnamefont {P.~A.}\ \bibnamefont
  {Erdman}}, \bibinfo {author} {\bibfnamefont {G.~M.}\ \bibnamefont
  {Andolina}}, \bibinfo {author} {\bibfnamefont {V.}~\bibnamefont
  {Giovannetti}},\ and\ \bibinfo {author} {\bibfnamefont {F.}~\bibnamefont
  {Noé}},\ }\href@noop {} {\bibinfo {title} {Reinforcement learning
  optimization of the charging of a dicke quantum battery}} (\bibinfo {year}
  {2022}),\ \Eprint {https://arxiv.org/abs/2212.12397} {arXiv:2212.12397
  [quant-ph]} \BibitemShut {NoStop}%
\bibitem [{\citenamefont {Morrone}\ \emph
  {et~al.}(2023{\natexlab{a}})\citenamefont {Morrone}, \citenamefont {Rossi},
  \citenamefont {Smirne},\ and\ \citenamefont {Genoni}}]{morrone2023charging}%
  \BibitemOpen
  \bibfield  {author} {\bibinfo {author} {\bibfnamefont {D.}~\bibnamefont
  {Morrone}}, \bibinfo {author} {\bibfnamefont {M.~A.~C.}\ \bibnamefont
  {Rossi}}, \bibinfo {author} {\bibfnamefont {A.}~\bibnamefont {Smirne}},\ and\
  \bibinfo {author} {\bibfnamefont {M.~G.}\ \bibnamefont {Genoni}},\ }\bibfield
   {title} {\bibinfo {title} {Charging a quantum battery in a non-markovian
  environment: a collisional model approach},\ }\href
  {https://doi.org/10.1088/2058-9565/accca4} {\bibfield  {journal} {\bibinfo
  {journal} {Quantum Science and Technology}\ }\textbf {\bibinfo {volume}
  {8}},\ \bibinfo {pages} {035007} (\bibinfo {year}
  {2023}{\natexlab{a}})}\BibitemShut {NoStop}%
\bibitem [{\citenamefont {Morrone}\ \emph
  {et~al.}(2023{\natexlab{b}})\citenamefont {Morrone}, \citenamefont {Rossi},\
  and\ \citenamefont {Genoni}}]{morrone2023daemonic}%
  \BibitemOpen
  \bibfield  {author} {\bibinfo {author} {\bibfnamefont {D.}~\bibnamefont
  {Morrone}}, \bibinfo {author} {\bibfnamefont {M.~A.~C.}\ \bibnamefont
  {Rossi}},\ and\ \bibinfo {author} {\bibfnamefont {M.~G.}\ \bibnamefont
  {Genoni}},\ }\href@noop {} {\bibinfo {title} {Daemonic ergotropy in
  continuously-monitored open quantum batteries}} (\bibinfo {year}
  {2023}{\natexlab{b}}),\ \Eprint {https://arxiv.org/abs/2302.12279}
  {arXiv:2302.12279 [quant-ph]} \BibitemShut {NoStop}%
\bibitem [{\citenamefont {Yang}\ \emph
  {et~al.}(2023{\natexlab{b}})\citenamefont {Yang}, \citenamefont {Yang},
  \citenamefont {Alimuddin}, \citenamefont {Salvia}, \citenamefont {Fei},
  \citenamefont {Zhao}, \citenamefont {Nimmrichter},\ and\ \citenamefont
  {Luo}}]{PhysRevLett.131.030402}%
  \BibitemOpen
  \bibfield  {author} {\bibinfo {author} {\bibfnamefont {X.}~\bibnamefont
  {Yang}}, \bibinfo {author} {\bibfnamefont {Y.-H.}\ \bibnamefont {Yang}},
  \bibinfo {author} {\bibfnamefont {M.}~\bibnamefont {Alimuddin}}, \bibinfo
  {author} {\bibfnamefont {R.}~\bibnamefont {Salvia}}, \bibinfo {author}
  {\bibfnamefont {S.-M.}\ \bibnamefont {Fei}}, \bibinfo {author} {\bibfnamefont
  {L.-M.}\ \bibnamefont {Zhao}}, \bibinfo {author} {\bibfnamefont
  {S.}~\bibnamefont {Nimmrichter}},\ and\ \bibinfo {author} {\bibfnamefont
  {M.-X.}\ \bibnamefont {Luo}},\ }\bibfield  {title} {\bibinfo {title} {Battery
  capacity of energy-storing quantum systems},\ }\href
  {https://doi.org/10.1103/PhysRevLett.131.030402} {\bibfield  {journal}
  {\bibinfo  {journal} {Phys. Rev. Lett.}\ }\textbf {\bibinfo {volume} {131}},\
  \bibinfo {pages} {030402} (\bibinfo {year} {2023}{\natexlab{b}})}\BibitemShut
  {NoStop}%
\bibitem [{\citenamefont {Friis}\ and\ \citenamefont
  {Huber}(2018)}]{Friis2018precisionwork}%
  \BibitemOpen
  \bibfield  {author} {\bibinfo {author} {\bibfnamefont {N.}~\bibnamefont
  {Friis}}\ and\ \bibinfo {author} {\bibfnamefont {M.}~\bibnamefont {Huber}},\
  }\bibfield  {title} {\bibinfo {title} {Precision and {W}ork {F}luctuations in
  {G}aussian {B}attery {C}harging},\ }\href
  {https://doi.org/10.22331/q-2018-04-23-61} {\bibfield  {journal} {\bibinfo
  {journal} {{Quantum}}\ }\textbf {\bibinfo {volume} {2}},\ \bibinfo {pages}
  {61} (\bibinfo {year} {2018})}\BibitemShut {NoStop}%
\bibitem [{\citenamefont {Rosa}\ \emph {et~al.}(2020)\citenamefont {Rosa},
  \citenamefont {Rossini}, \citenamefont {Andolina}, \citenamefont {Polini},\
  and\ \citenamefont {Carrega}}]{Rosa2020}%
  \BibitemOpen
  \bibfield  {author} {\bibinfo {author} {\bibfnamefont {D.}~\bibnamefont
  {Rosa}}, \bibinfo {author} {\bibfnamefont {D.}~\bibnamefont {Rossini}},
  \bibinfo {author} {\bibfnamefont {G.~M.}\ \bibnamefont {Andolina}}, \bibinfo
  {author} {\bibfnamefont {M.}~\bibnamefont {Polini}},\ and\ \bibinfo {author}
  {\bibfnamefont {M.}~\bibnamefont {Carrega}},\ }\bibfield  {title} {\bibinfo
  {title} {Ultra-stable charging of fast-scrambling syk quantum batteries},\
  }\href {https://doi.org/10.1007/JHEP11(2020)067} {\bibfield  {journal}
  {\bibinfo  {journal} {Journal of High Energy Physics}\ }\textbf {\bibinfo
  {volume} {2020}},\ \bibinfo {pages} {67} (\bibinfo {year}
  {2020})}\BibitemShut {NoStop}%
\bibitem [{\citenamefont {Rossini}\ \emph {et~al.}(2019)\citenamefont
  {Rossini}, \citenamefont {Andolina},\ and\ \citenamefont
  {Polini}}]{PhysRevB.100.115142}%
  \BibitemOpen
  \bibfield  {author} {\bibinfo {author} {\bibfnamefont {D.}~\bibnamefont
  {Rossini}}, \bibinfo {author} {\bibfnamefont {G.~M.}\ \bibnamefont
  {Andolina}},\ and\ \bibinfo {author} {\bibfnamefont {M.}~\bibnamefont
  {Polini}},\ }\bibfield  {title} {\bibinfo {title} {Many-body localized
  quantum batteries},\ }\href {https://doi.org/10.1103/PhysRevB.100.115142}
  {\bibfield  {journal} {\bibinfo  {journal} {Phys. Rev. B}\ }\textbf {\bibinfo
  {volume} {100}},\ \bibinfo {pages} {115142} (\bibinfo {year}
  {2019})}\BibitemShut {NoStop}%
\bibitem [{\citenamefont {Santos}\ \emph {et~al.}(2019)\citenamefont {Santos},
  \citenamefont {\ifmmode~\mbox{\c{C}}\else \c{C}\fi{}akmak}, \citenamefont
  {Campbell},\ and\ \citenamefont {Zinner}}]{PhysRevE.100.032107}%
  \BibitemOpen
  \bibfield  {author} {\bibinfo {author} {\bibfnamefont {A.~C.}\ \bibnamefont
  {Santos}}, \bibinfo {author} {\bibfnamefont {B.~i. e. i. f. m.~c.}\
  \bibnamefont {\ifmmode~\mbox{\c{C}}\else \c{C}\fi{}akmak}}, \bibinfo {author}
  {\bibfnamefont {S.}~\bibnamefont {Campbell}},\ and\ \bibinfo {author}
  {\bibfnamefont {N.~T.}\ \bibnamefont {Zinner}},\ }\bibfield  {title}
  {\bibinfo {title} {Stable adiabatic quantum batteries},\ }\href
  {https://doi.org/10.1103/PhysRevE.100.032107} {\bibfield  {journal} {\bibinfo
   {journal} {Phys. Rev. E}\ }\textbf {\bibinfo {volume} {100}},\ \bibinfo
  {pages} {032107} (\bibinfo {year} {2019})}\BibitemShut {NoStop}%
\bibitem [{\citenamefont {Shaghaghi}\ \emph {et~al.}(2022)\citenamefont
  {Shaghaghi}, \citenamefont {Singh}, \citenamefont {Benenti},\ and\
  \citenamefont {Rosa}}]{shaghaghi2022micromasers}%
  \BibitemOpen
  \bibfield  {author} {\bibinfo {author} {\bibfnamefont {V.}~\bibnamefont
  {Shaghaghi}}, \bibinfo {author} {\bibfnamefont {V.}~\bibnamefont {Singh}},
  \bibinfo {author} {\bibfnamefont {G.}~\bibnamefont {Benenti}},\ and\ \bibinfo
  {author} {\bibfnamefont {D.}~\bibnamefont {Rosa}},\ }\bibfield  {title}
  {\bibinfo {title} {Micromasers as quantum batteries},\ }\href
  {https://doi.org/10.1088/2058-9565/ac8829} {\bibfield  {journal} {\bibinfo
  {journal} {Quantum Science and Technology}\ }\textbf {\bibinfo {volume}
  {7}},\ \bibinfo {pages} {04LT01} (\bibinfo {year} {2022})}\BibitemShut
  {NoStop}%
\bibitem [{\citenamefont {Shaghaghi}\ \emph {et~al.}(2023)\citenamefont
  {Shaghaghi}, \citenamefont {Singh}, \citenamefont {Carrega}, \citenamefont
  {Rosa},\ and\ \citenamefont {Benenti}}]{shaghaghi2023lossy}%
  \BibitemOpen
  \bibfield  {author} {\bibinfo {author} {\bibfnamefont {V.}~\bibnamefont
  {Shaghaghi}}, \bibinfo {author} {\bibfnamefont {V.}~\bibnamefont {Singh}},
  \bibinfo {author} {\bibfnamefont {M.}~\bibnamefont {Carrega}}, \bibinfo
  {author} {\bibfnamefont {D.}~\bibnamefont {Rosa}},\ and\ \bibinfo {author}
  {\bibfnamefont {G.}~\bibnamefont {Benenti}},\ }\bibfield  {title} {\bibinfo
  {title} {Lossy micromaser battery: Almost pure states in the jaynes--cummings
  regime},\ }\bibfield  {journal} {\bibinfo  {journal} {Entropy}\ }\textbf
  {\bibinfo {volume} {25}},\ \href {https://doi.org/10.3390/e25030430}
  {10.3390/e25030430} (\bibinfo {year} {2023})\BibitemShut {NoStop}%
\bibitem [{\citenamefont {Rodríguez}\ \emph {et~al.}(2023)\citenamefont
  {Rodríguez}, \citenamefont {Rosa},\ and\ \citenamefont
  {Olle}}]{rodríguez2023aidiscovery}%
  \BibitemOpen
  \bibfield  {author} {\bibinfo {author} {\bibfnamefont {C.}~\bibnamefont
  {Rodríguez}}, \bibinfo {author} {\bibfnamefont {D.}~\bibnamefont {Rosa}},\
  and\ \bibinfo {author} {\bibfnamefont {J.}~\bibnamefont {Olle}},\ }\href@noop
  {} {\bibinfo {title} {Ai-discovery of a new charging protocol in a micromaser
  quantum battery}} (\bibinfo {year} {2023}),\ \Eprint
  {https://arxiv.org/abs/2301.09408} {arXiv:2301.09408 [quant-ph]} \BibitemShut
  {NoStop}%
\bibitem [{\citenamefont {Rossini}\ \emph {et~al.}(2020)\citenamefont
  {Rossini}, \citenamefont {Andolina}, \citenamefont {Rosa}, \citenamefont
  {Carrega},\ and\ \citenamefont {Polini}}]{PhysRevLett.125.236402}%
  \BibitemOpen
  \bibfield  {author} {\bibinfo {author} {\bibfnamefont {D.}~\bibnamefont
  {Rossini}}, \bibinfo {author} {\bibfnamefont {G.~M.}\ \bibnamefont
  {Andolina}}, \bibinfo {author} {\bibfnamefont {D.}~\bibnamefont {Rosa}},
  \bibinfo {author} {\bibfnamefont {M.}~\bibnamefont {Carrega}},\ and\ \bibinfo
  {author} {\bibfnamefont {M.}~\bibnamefont {Polini}},\ }\bibfield  {title}
  {\bibinfo {title} {Quantum advantage in the charging process of
  sachdev-ye-kitaev batteries},\ }\href
  {https://doi.org/10.1103/PhysRevLett.125.236402} {\bibfield  {journal}
  {\bibinfo  {journal} {Phys. Rev. Lett.}\ }\textbf {\bibinfo {volume} {125}},\
  \bibinfo {pages} {236402} (\bibinfo {year} {2020})}\BibitemShut {NoStop}%
\bibitem [{\citenamefont {Salvia}\ \emph {et~al.}(2023)\citenamefont {Salvia},
  \citenamefont {Perarnau-Llobet}, \citenamefont {Haack}, \citenamefont
  {Brunner},\ and\ \citenamefont {Nimmrichter}}]{PhysRevResearch.5.013155}%
  \BibitemOpen
  \bibfield  {author} {\bibinfo {author} {\bibfnamefont {R.}~\bibnamefont
  {Salvia}}, \bibinfo {author} {\bibfnamefont {M.}~\bibnamefont
  {Perarnau-Llobet}}, \bibinfo {author} {\bibfnamefont {G.}~\bibnamefont
  {Haack}}, \bibinfo {author} {\bibfnamefont {N.}~\bibnamefont {Brunner}},\
  and\ \bibinfo {author} {\bibfnamefont {S.}~\bibnamefont {Nimmrichter}},\
  }\bibfield  {title} {\bibinfo {title} {Quantum advantage in charging cavity
  and spin batteries by repeated interactions},\ }\href
  {https://doi.org/10.1103/PhysRevResearch.5.013155} {\bibfield  {journal}
  {\bibinfo  {journal} {Phys. Rev. Res.}\ }\textbf {\bibinfo {volume} {5}},\
  \bibinfo {pages} {013155} (\bibinfo {year} {2023})}\BibitemShut {NoStop}%
\bibitem [{\citenamefont {Ferraro}\ \emph {et~al.}(2018)\citenamefont
  {Ferraro}, \citenamefont {Campisi}, \citenamefont {Andolina}, \citenamefont
  {Pellegrini},\ and\ \citenamefont {Polini}}]{PhysRevLett.120.117702}%
  \BibitemOpen
  \bibfield  {author} {\bibinfo {author} {\bibfnamefont {D.}~\bibnamefont
  {Ferraro}}, \bibinfo {author} {\bibfnamefont {M.}~\bibnamefont {Campisi}},
  \bibinfo {author} {\bibfnamefont {G.~M.}\ \bibnamefont {Andolina}}, \bibinfo
  {author} {\bibfnamefont {V.}~\bibnamefont {Pellegrini}},\ and\ \bibinfo
  {author} {\bibfnamefont {M.}~\bibnamefont {Polini}},\ }\bibfield  {title}
  {\bibinfo {title} {High-power collective charging of a solid-state quantum
  battery},\ }\href {https://doi.org/10.1103/PhysRevLett.120.117702} {\bibfield
   {journal} {\bibinfo  {journal} {Phys. Rev. Lett.}\ }\textbf {\bibinfo
  {volume} {120}},\ \bibinfo {pages} {117702} (\bibinfo {year}
  {2018})}\BibitemShut {NoStop}%
\bibitem [{\citenamefont {Zhao}\ \emph {et~al.}(2022)\citenamefont {Zhao},
  \citenamefont {Dou},\ and\ \citenamefont {Zhao}}]{PhysRevResearch.4.013172}%
  \BibitemOpen
  \bibfield  {author} {\bibinfo {author} {\bibfnamefont {F.}~\bibnamefont
  {Zhao}}, \bibinfo {author} {\bibfnamefont {F.-Q.}\ \bibnamefont {Dou}},\ and\
  \bibinfo {author} {\bibfnamefont {Q.}~\bibnamefont {Zhao}},\ }\bibfield
  {title} {\bibinfo {title} {Charging performance of the su-schrieffer-heeger
  quantum battery},\ }\href {https://doi.org/10.1103/PhysRevResearch.4.013172}
  {\bibfield  {journal} {\bibinfo  {journal} {Phys. Rev. Res.}\ }\textbf
  {\bibinfo {volume} {4}},\ \bibinfo {pages} {013172} (\bibinfo {year}
  {2022})}\BibitemShut {NoStop}%
\bibitem [{\citenamefont {Andolina}\ \emph {et~al.}(2018)\citenamefont
  {Andolina}, \citenamefont {Farina}, \citenamefont {Mari}, \citenamefont
  {Pellegrini}, \citenamefont {Giovannetti},\ and\ \citenamefont
  {Polini}}]{PhysRevB.98.205423}%
  \BibitemOpen
  \bibfield  {author} {\bibinfo {author} {\bibfnamefont {G.~M.}\ \bibnamefont
  {Andolina}}, \bibinfo {author} {\bibfnamefont {D.}~\bibnamefont {Farina}},
  \bibinfo {author} {\bibfnamefont {A.}~\bibnamefont {Mari}}, \bibinfo {author}
  {\bibfnamefont {V.}~\bibnamefont {Pellegrini}}, \bibinfo {author}
  {\bibfnamefont {V.}~\bibnamefont {Giovannetti}},\ and\ \bibinfo {author}
  {\bibfnamefont {M.}~\bibnamefont {Polini}},\ }\bibfield  {title} {\bibinfo
  {title} {Charger-mediated energy transfer in exactly solvable models for
  quantum batteries},\ }\href {https://doi.org/10.1103/PhysRevB.98.205423}
  {\bibfield  {journal} {\bibinfo  {journal} {Phys. Rev. B}\ }\textbf {\bibinfo
  {volume} {98}},\ \bibinfo {pages} {205423} (\bibinfo {year}
  {2018})}\BibitemShut {NoStop}%
\bibitem [{\citenamefont {Andolina}\ \emph
  {et~al.}(2019{\natexlab{a}})\citenamefont {Andolina}, \citenamefont {Keck},
  \citenamefont {Mari}, \citenamefont {Campisi}, \citenamefont {Giovannetti},\
  and\ \citenamefont {Polini}}]{PhysRevLett.122.047702}%
  \BibitemOpen
  \bibfield  {author} {\bibinfo {author} {\bibfnamefont {G.~M.}\ \bibnamefont
  {Andolina}}, \bibinfo {author} {\bibfnamefont {M.}~\bibnamefont {Keck}},
  \bibinfo {author} {\bibfnamefont {A.}~\bibnamefont {Mari}}, \bibinfo {author}
  {\bibfnamefont {M.}~\bibnamefont {Campisi}}, \bibinfo {author} {\bibfnamefont
  {V.}~\bibnamefont {Giovannetti}},\ and\ \bibinfo {author} {\bibfnamefont
  {M.}~\bibnamefont {Polini}},\ }\bibfield  {title} {\bibinfo {title}
  {Extractable work, the role of correlations, and asymptotic freedom in
  quantum batteries},\ }\href {https://doi.org/10.1103/PhysRevLett.122.047702}
  {\bibfield  {journal} {\bibinfo  {journal} {Phys. Rev. Lett.}\ }\textbf
  {\bibinfo {volume} {122}},\ \bibinfo {pages} {047702} (\bibinfo {year}
  {2019}{\natexlab{a}})}\BibitemShut {NoStop}%
\bibitem [{\citenamefont {Campaioli}\ \emph {et~al.}(2017)\citenamefont
  {Campaioli}, \citenamefont {Pollock}, \citenamefont {Binder}, \citenamefont
  {C\'eleri}, \citenamefont {Goold}, \citenamefont {Vinjanampathy},\ and\
  \citenamefont {Modi}}]{PhysRevLett.118.150601}%
  \BibitemOpen
  \bibfield  {author} {\bibinfo {author} {\bibfnamefont {F.}~\bibnamefont
  {Campaioli}}, \bibinfo {author} {\bibfnamefont {F.~A.}\ \bibnamefont
  {Pollock}}, \bibinfo {author} {\bibfnamefont {F.~C.}\ \bibnamefont {Binder}},
  \bibinfo {author} {\bibfnamefont {L.}~\bibnamefont {C\'eleri}}, \bibinfo
  {author} {\bibfnamefont {J.}~\bibnamefont {Goold}}, \bibinfo {author}
  {\bibfnamefont {S.}~\bibnamefont {Vinjanampathy}},\ and\ \bibinfo {author}
  {\bibfnamefont {K.}~\bibnamefont {Modi}},\ }\bibfield  {title} {\bibinfo
  {title} {Enhancing the charging power of quantum batteries},\ }\href
  {https://doi.org/10.1103/PhysRevLett.118.150601} {\bibfield  {journal}
  {\bibinfo  {journal} {Phys. Rev. Lett.}\ }\textbf {\bibinfo {volume} {118}},\
  \bibinfo {pages} {150601} (\bibinfo {year} {2017})}\BibitemShut {NoStop}%
\bibitem [{\citenamefont {Seah}\ \emph {et~al.}(2021)\citenamefont {Seah},
  \citenamefont {Perarnau-Llobet}, \citenamefont {Haack}, \citenamefont
  {Brunner},\ and\ \citenamefont {Nimmrichter}}]{seah2021quantum}%
  \BibitemOpen
  \bibfield  {author} {\bibinfo {author} {\bibfnamefont {S.}~\bibnamefont
  {Seah}}, \bibinfo {author} {\bibfnamefont {M.}~\bibnamefont
  {Perarnau-Llobet}}, \bibinfo {author} {\bibfnamefont {G.}~\bibnamefont
  {Haack}}, \bibinfo {author} {\bibfnamefont {N.}~\bibnamefont {Brunner}},\
  and\ \bibinfo {author} {\bibfnamefont {S.}~\bibnamefont {Nimmrichter}},\
  }\bibfield  {title} {\bibinfo {title} {Quantum speed-up in collisional
  battery charging},\ }\href {https://doi.org/10.1103/PhysRevLett.127.100601}
  {\bibfield  {journal} {\bibinfo  {journal} {Phys. Rev. Lett.}\ }\textbf
  {\bibinfo {volume} {127}},\ \bibinfo {pages} {100601} (\bibinfo {year}
  {2021})}\BibitemShut {NoStop}%
\bibitem [{\citenamefont {Gyhm}\ \emph {et~al.}(2022)\citenamefont {Gyhm},
  \citenamefont {\ifmmode~\check{S}\else \v{S}\fi{}afr\'anek},\ and\
  \citenamefont {Rosa}}]{PhysRevLett.128.140501}%
  \BibitemOpen
  \bibfield  {author} {\bibinfo {author} {\bibfnamefont {J.-Y.}\ \bibnamefont
  {Gyhm}}, \bibinfo {author} {\bibfnamefont {D.}~\bibnamefont
  {\ifmmode~\check{S}\else \v{S}\fi{}afr\'anek}},\ and\ \bibinfo {author}
  {\bibfnamefont {D.}~\bibnamefont {Rosa}},\ }\bibfield  {title} {\bibinfo
  {title} {Quantum charging advantage cannot be extensive without global
  operations},\ }\href {https://doi.org/10.1103/PhysRevLett.128.140501}
  {\bibfield  {journal} {\bibinfo  {journal} {Phys. Rev. Lett.}\ }\textbf
  {\bibinfo {volume} {128}},\ \bibinfo {pages} {140501} (\bibinfo {year}
  {2022})}\BibitemShut {NoStop}%
\bibitem [{\citenamefont {Le}\ \emph {et~al.}(2018)\citenamefont {Le},
  \citenamefont {Levinsen}, \citenamefont {Modi}, \citenamefont {Parish},\ and\
  \citenamefont {Pollock}}]{PhysRevA.97.022106}%
  \BibitemOpen
  \bibfield  {author} {\bibinfo {author} {\bibfnamefont {T.~P.}\ \bibnamefont
  {Le}}, \bibinfo {author} {\bibfnamefont {J.}~\bibnamefont {Levinsen}},
  \bibinfo {author} {\bibfnamefont {K.}~\bibnamefont {Modi}}, \bibinfo {author}
  {\bibfnamefont {M.~M.}\ \bibnamefont {Parish}},\ and\ \bibinfo {author}
  {\bibfnamefont {F.~A.}\ \bibnamefont {Pollock}},\ }\bibfield  {title}
  {\bibinfo {title} {Spin-chain model of a many-body quantum battery},\ }\href
  {https://doi.org/10.1103/PhysRevA.97.022106} {\bibfield  {journal} {\bibinfo
  {journal} {Phys. Rev. A}\ }\textbf {\bibinfo {volume} {97}},\ \bibinfo
  {pages} {022106} (\bibinfo {year} {2018})}\BibitemShut {NoStop}%
\bibitem [{\citenamefont {Gyhm}\ and\ \citenamefont
  {Fischer}(2023)}]{gyhm2023beneficial}%
  \BibitemOpen
  \bibfield  {author} {\bibinfo {author} {\bibfnamefont {J.-Y.}\ \bibnamefont
  {Gyhm}}\ and\ \bibinfo {author} {\bibfnamefont {U.~R.}\ \bibnamefont
  {Fischer}},\ }\href@noop {} {\bibinfo {title} {Beneficial and detrimental
  entanglement for quantum battery charging}} (\bibinfo {year} {2023}),\
  \Eprint {https://arxiv.org/abs/2303.07841} {arXiv:2303.07841 [quant-ph]}
  \BibitemShut {NoStop}%
\bibitem [{\citenamefont {Mohan}\ and\ \citenamefont {Pati}(2021)}]{Mohan2021}%
  \BibitemOpen
  \bibfield  {author} {\bibinfo {author} {\bibfnamefont {B.}~\bibnamefont
  {Mohan}}\ and\ \bibinfo {author} {\bibfnamefont {A.~K.}\ \bibnamefont
  {Pati}},\ }\bibfield  {title} {\bibinfo {title} {Reverse quantum speed limit:
  How slowly a quantum battery can discharge},\ }\href
  {https://doi.org/10.1103/PhysRevA.104.042209} {\bibfield  {journal} {\bibinfo
   {journal} {Phys. Rev. A}\ }\textbf {\bibinfo {volume} {104}},\ \bibinfo
  {pages} {042209} (\bibinfo {year} {2021})}\BibitemShut {NoStop}%
\bibitem [{\citenamefont {Garc\'{\i}a-Pintos}\ \emph
  {et~al.}(2020)\citenamefont {Garc\'{\i}a-Pintos}, \citenamefont {Hamma},\
  and\ \citenamefont {del Campo}}]{PhysRevLett.125.040601}%
  \BibitemOpen
  \bibfield  {author} {\bibinfo {author} {\bibfnamefont {L.~P.}\ \bibnamefont
  {Garc\'{\i}a-Pintos}}, \bibinfo {author} {\bibfnamefont {A.}~\bibnamefont
  {Hamma}},\ and\ \bibinfo {author} {\bibfnamefont {A.}~\bibnamefont {del
  Campo}},\ }\bibfield  {title} {\bibinfo {title} {Fluctuations in extractable
  work bound the charging power of quantum batteries},\ }\href
  {https://doi.org/10.1103/PhysRevLett.125.040601} {\bibfield  {journal}
  {\bibinfo  {journal} {Phys. Rev. Lett.}\ }\textbf {\bibinfo {volume} {125}},\
  \bibinfo {pages} {040601} (\bibinfo {year} {2020})}\BibitemShut {NoStop}%
\bibitem [{\citenamefont {Joshi}\ and\ \citenamefont
  {Mahesh}(2022)}]{PhysRevA.106.042601}%
  \BibitemOpen
  \bibfield  {author} {\bibinfo {author} {\bibfnamefont {J.}~\bibnamefont
  {Joshi}}\ and\ \bibinfo {author} {\bibfnamefont {T.~S.}\ \bibnamefont
  {Mahesh}},\ }\bibfield  {title} {\bibinfo {title} {Experimental investigation
  of a quantum battery using star-topology nmr spin systems},\ }\href
  {https://doi.org/10.1103/PhysRevA.106.042601} {\bibfield  {journal} {\bibinfo
   {journal} {Phys. Rev. A}\ }\textbf {\bibinfo {volume} {106}},\ \bibinfo
  {pages} {042601} (\bibinfo {year} {2022})}\BibitemShut {NoStop}%
\bibitem [{\citenamefont {Quach}\ \emph {et~al.}(2022)\citenamefont {Quach},
  \citenamefont {McGhee}, \citenamefont {Ganzer}, \citenamefont {Rouse},
  \citenamefont {Lovett}, \citenamefont {Gauger}, \citenamefont {Keeling},
  \citenamefont {Cerullo}, \citenamefont {Lidzey},\ and\ \citenamefont
  {Virgili}}]{doi:10.1126/sciadv.abk3160}%
  \BibitemOpen
  \bibfield  {author} {\bibinfo {author} {\bibfnamefont {J.~Q.}\ \bibnamefont
  {Quach}}, \bibinfo {author} {\bibfnamefont {K.~E.}\ \bibnamefont {McGhee}},
  \bibinfo {author} {\bibfnamefont {L.}~\bibnamefont {Ganzer}}, \bibinfo
  {author} {\bibfnamefont {D.~M.}\ \bibnamefont {Rouse}}, \bibinfo {author}
  {\bibfnamefont {B.~W.}\ \bibnamefont {Lovett}}, \bibinfo {author}
  {\bibfnamefont {E.~M.}\ \bibnamefont {Gauger}}, \bibinfo {author}
  {\bibfnamefont {J.}~\bibnamefont {Keeling}}, \bibinfo {author} {\bibfnamefont
  {G.}~\bibnamefont {Cerullo}}, \bibinfo {author} {\bibfnamefont {D.~G.}\
  \bibnamefont {Lidzey}},\ and\ \bibinfo {author} {\bibfnamefont
  {T.}~\bibnamefont {Virgili}},\ }\bibfield  {title} {\bibinfo {title}
  {Superabsorption in an organic microcavity: Toward a quantum battery},\
  }\href {https://doi.org/10.1126/sciadv.abk3160} {\bibfield  {journal}
  {\bibinfo  {journal} {Science Advances}\ }\textbf {\bibinfo {volume} {8}},\
  \bibinfo {pages} {eabk3160} (\bibinfo {year} {2022})},\ \Eprint
  {https://arxiv.org/abs/https://www.science.org/doi/pdf/10.1126/sciadv.abk3160}
  {https://www.science.org/doi/pdf/10.1126/sciadv.abk3160} \BibitemShut
  {NoStop}%
\bibitem [{\citenamefont {Campaioli}\ \emph
  {et~al.}(2018{\natexlab{a}})\citenamefont {Campaioli}, \citenamefont
  {Pollock},\ and\ \citenamefont {Vinjanampathy}}]{Campaioli2018}%
  \BibitemOpen
  \bibfield  {author} {\bibinfo {author} {\bibfnamefont {F.}~\bibnamefont
  {Campaioli}}, \bibinfo {author} {\bibfnamefont {F.~A.}\ \bibnamefont
  {Pollock}},\ and\ \bibinfo {author} {\bibfnamefont {S.}~\bibnamefont
  {Vinjanampathy}},\ }\bibinfo {title} {Quantum batteries},\ in\ \href
  {https://doi.org/10.1007/978-3-319-99046-0_8} {\emph {\bibinfo {booktitle}
  {Thermodynamics in the Quantum Regime: Fundamental Aspects and New
  Directions}}},\ \bibinfo {editor} {edited by\ \bibinfo {editor}
  {\bibfnamefont {F.}~\bibnamefont {Binder}}, \bibinfo {editor} {\bibfnamefont
  {L.~A.}\ \bibnamefont {Correa}}, \bibinfo {editor} {\bibfnamefont
  {C.}~\bibnamefont {Gogolin}}, \bibinfo {editor} {\bibfnamefont
  {J.}~\bibnamefont {Anders}},\ and\ \bibinfo {editor} {\bibfnamefont
  {G.}~\bibnamefont {Adesso}}}\ (\bibinfo  {publisher} {Springer International
  Publishing},\ \bibinfo {address} {Cham},\ \bibinfo {year} {2018})\ pp.\
  \bibinfo {pages} {207--225}\BibitemShut {NoStop}%
\bibitem [{\citenamefont {Crescente}\ \emph {et~al.}(2020)\citenamefont
  {Crescente}, \citenamefont {Carrega}, \citenamefont {Sassetti},\ and\
  \citenamefont {Ferraro}}]{Crescente_2020}%
  \BibitemOpen
  \bibfield  {author} {\bibinfo {author} {\bibfnamefont {A.}~\bibnamefont
  {Crescente}}, \bibinfo {author} {\bibfnamefont {M.}~\bibnamefont {Carrega}},
  \bibinfo {author} {\bibfnamefont {M.}~\bibnamefont {Sassetti}},\ and\
  \bibinfo {author} {\bibfnamefont {D.}~\bibnamefont {Ferraro}},\ }\bibfield
  {title} {\bibinfo {title} {Charging and energy fluctuations of a driven
  quantum battery},\ }\href {https://doi.org/10.1088/1367-2630/ab91fc}
  {\bibfield  {journal} {\bibinfo  {journal} {New Journal of Physics}\ }\textbf
  {\bibinfo {volume} {22}},\ \bibinfo {pages} {063057} (\bibinfo {year}
  {2020})}\BibitemShut {NoStop}%
\bibitem [{\citenamefont {Bengtsson}\ and\ \citenamefont
  {{\.Z}yczkowski}(2017)}]{bengtsson2017geometry}%
  \BibitemOpen
  \bibfield  {author} {\bibinfo {author} {\bibfnamefont {I.}~\bibnamefont
  {Bengtsson}}\ and\ \bibinfo {author} {\bibfnamefont {K.}~\bibnamefont
  {{\.Z}yczkowski}},\ }\href@noop {} {\emph {\bibinfo {title} {Geometry of
  quantum states: an introduction to quantum entanglement}}}\ (\bibinfo
  {publisher} {Cambridge university press},\ \bibinfo {year}
  {2017})\BibitemShut {NoStop}%
\bibitem [{\citenamefont {Levitin}\ and\ \citenamefont
  {Toffoli}(2009)}]{PhysRevLett.103.160502}%
  \BibitemOpen
  \bibfield  {author} {\bibinfo {author} {\bibfnamefont {L.~B.}\ \bibnamefont
  {Levitin}}\ and\ \bibinfo {author} {\bibfnamefont {T.}~\bibnamefont
  {Toffoli}},\ }\bibfield  {title} {\bibinfo {title} {Fundamental limit on the
  rate of quantum dynamics: The unified bound is tight},\ }\href
  {https://doi.org/10.1103/PhysRevLett.103.160502} {\bibfield  {journal}
  {\bibinfo  {journal} {Phys. Rev. Lett.}\ }\textbf {\bibinfo {volume} {103}},\
  \bibinfo {pages} {160502} (\bibinfo {year} {2009})}\BibitemShut {NoStop}%
\bibitem [{\citenamefont {Uffink}(1993)}]{uffink1993}%
  \BibitemOpen
  \bibfield  {author} {\bibinfo {author} {\bibfnamefont {J.}~\bibnamefont
  {Uffink}},\ }\bibfield  {title} {\bibinfo {title} {{The rate of evolution of
  a quantum state}},\ }\href {https://doi.org/10.1119/1.17368} {\bibfield
  {journal} {\bibinfo  {journal} {American Journal of Physics}\ }\textbf
  {\bibinfo {volume} {61}},\ \bibinfo {pages} {935} (\bibinfo {year} {1993})},\
  \Eprint
  {https://arxiv.org/abs/https://pubs.aip.org/aapt/ajp/article-pdf/61/10/935/11694314/935\_1\_online.pdf}
  {https://pubs.aip.org/aapt/ajp/article-pdf/61/10/935/11694314/935\_1\_online.pdf}
  \BibitemShut {NoStop}%
\bibitem [{Note1()}]{Note1}%
  \BibitemOpen
  \bibinfo {note} {This behavior could be already observed in previous
  literature, see Fig.~3 in Supplemental Material of~\cite
  {PhysRevLett.120.060409}}\BibitemShut {NoStop}%
\bibitem [{\citenamefont {Campaioli}\ \emph
  {et~al.}(2018{\natexlab{b}})\citenamefont {Campaioli}, \citenamefont
  {Pollock}, \citenamefont {Binder},\ and\ \citenamefont
  {Modi}}]{PhysRevLett.120.060409}%
  \BibitemOpen
  \bibfield  {author} {\bibinfo {author} {\bibfnamefont {F.}~\bibnamefont
  {Campaioli}}, \bibinfo {author} {\bibfnamefont {F.~A.}\ \bibnamefont
  {Pollock}}, \bibinfo {author} {\bibfnamefont {F.~C.}\ \bibnamefont
  {Binder}},\ and\ \bibinfo {author} {\bibfnamefont {K.}~\bibnamefont {Modi}},\
  }\bibfield  {title} {\bibinfo {title} {Tightening quantum speed limits for
  almost all states},\ }\href {https://doi.org/10.1103/PhysRevLett.120.060409}
  {\bibfield  {journal} {\bibinfo  {journal} {Phys. Rev. Lett.}\ }\textbf
  {\bibinfo {volume} {120}},\ \bibinfo {pages} {060409} (\bibinfo {year}
  {2018}{\natexlab{b}})}\BibitemShut {NoStop}%
\bibitem [{\citenamefont {Giovannetti}\ \emph {et~al.}(2004)\citenamefont
  {Giovannetti}, \citenamefont {Lloyd},\ and\ \citenamefont
  {Maccone}}]{VittorioGiovannetti_2004}%
  \BibitemOpen
  \bibfield  {author} {\bibinfo {author} {\bibfnamefont {V.}~\bibnamefont
  {Giovannetti}}, \bibinfo {author} {\bibfnamefont {S.}~\bibnamefont {Lloyd}},\
  and\ \bibinfo {author} {\bibfnamefont {L.}~\bibnamefont {Maccone}},\
  }\bibfield  {title} {\bibinfo {title} {The speed limit of quantum unitary
  evolution},\ }\href {https://doi.org/10.1088/1464-4266/6/8/028} {\bibfield
  {journal} {\bibinfo  {journal} {Journal of Optics B: Quantum and
  Semiclassical Optics}\ }\textbf {\bibinfo {volume} {6}},\ \bibinfo {pages}
  {S807} (\bibinfo {year} {2004})}\BibitemShut {NoStop}%
\bibitem [{\citenamefont {Deffner}(2017)}]{Deffner.2017}%
  \BibitemOpen
  \bibfield  {author} {\bibinfo {author} {\bibfnamefont {S.}~\bibnamefont
  {Deffner}},\ }\bibfield  {title} {\bibinfo {title} {Geometric quantum speed
  limits: a case for wigner phase space},\ }\href
  {https://doi.org/10.1088/1367-2630/aa83dc} {\bibfield  {journal} {\bibinfo
  {journal} {New Journal of Physics}\ }\textbf {\bibinfo {volume} {19}},\
  \bibinfo {pages} {103018} (\bibinfo {year} {2017})}\BibitemShut {NoStop}%
\bibitem [{\citenamefont {Gyhm}\ \emph {et~al.}(2023)\citenamefont {Gyhm},
  \citenamefont {Rosa},\ and\ \citenamefont
  {\v{S}afr\'{a}nek}}]{gyhm2023to_appear}%
  \BibitemOpen
  \bibfield  {author} {\bibinfo {author} {\bibfnamefont {J.-Y.}\ \bibnamefont
  {Gyhm}}, \bibinfo {author} {\bibfnamefont {D.}~\bibnamefont {Rosa}},\ and\
  \bibinfo {author} {\bibfnamefont {D.}~\bibnamefont {\v{S}afr\'{a}nek}},\
  }\href@noop {} {\bibinfo {title} {To appear}} (\bibinfo {year}
  {2023})\BibitemShut {NoStop}%
\bibitem [{\citenamefont {Jozsa}(1994)}]{jozsa1994fidelity}%
  \BibitemOpen
  \bibfield  {author} {\bibinfo {author} {\bibfnamefont {R.}~\bibnamefont
  {Jozsa}},\ }\bibfield  {title} {\bibinfo {title} {Fidelity for mixed quantum
  states},\ }\href {https://doi.org/10.1080/09500349414552171} {\bibfield
  {journal} {\bibinfo  {journal} {Journal of Modern Optics}\ }\textbf {\bibinfo
  {volume} {41}},\ \bibinfo {pages} {2315} (\bibinfo {year} {1994})},\ \Eprint
  {https://arxiv.org/abs/https://doi.org/10.1080/09500349414552171}
  {https://doi.org/10.1080/09500349414552171} \BibitemShut {NoStop}%
\bibitem [{\citenamefont {{Nielsen}}\ and\ \citenamefont
  {{Chuang}}(2010)}]{nielsen2010quantum}%
  \BibitemOpen
  \bibfield  {author} {\bibinfo {author} {\bibfnamefont {M.~A.}\ \bibnamefont
  {{Nielsen}}}\ and\ \bibinfo {author} {\bibfnamefont {I.~L.}\ \bibnamefont
  {{Chuang}}},\ }\href@noop {} {\emph {\bibinfo {title} {{Quantum Computation
  and Quantum Information}}}}\ (\bibinfo  {publisher} {Cambridge university
  press},\ \bibinfo {year} {2010})\BibitemShut {NoStop}%
\bibitem [{\citenamefont {Hübner}(1992)}]{HUBNER1992explicit}%
  \BibitemOpen
  \bibfield  {author} {\bibinfo {author} {\bibfnamefont {M.}~\bibnamefont
  {Hübner}},\ }\bibfield  {title} {\bibinfo {title} {Explicit computation of
  the bures distance for density matrices},\ }\href
  {https://doi.org/https://doi.org/10.1016/0375-9601(92)91004-B} {\bibfield
  {journal} {\bibinfo  {journal} {Physics Letters A}\ }\textbf {\bibinfo
  {volume} {163}},\ \bibinfo {pages} {239} (\bibinfo {year}
  {1992})}\BibitemShut {NoStop}%
\bibitem [{\citenamefont {Rastegin}(2007)}]{Rastegin_2007}%
  \BibitemOpen
  \bibfield  {author} {\bibinfo {author} {\bibfnamefont {A.~E.}\ \bibnamefont
  {Rastegin}},\ }\bibfield  {title} {\bibinfo {title} {Trace distance from the
  viewpoint of quantum operation techniques},\ }\href
  {https://doi.org/10.1088/1751-8113/40/31/026} {\bibfield  {journal} {\bibinfo
   {journal} {Journal of Physics A: Mathematical and Theoretical}\ }\textbf
  {\bibinfo {volume} {40}},\ \bibinfo {pages} {9533} (\bibinfo {year}
  {2007})}\BibitemShut {NoStop}%
\bibitem [{\citenamefont {Pucha\l{}a}\ and\ \citenamefont
  {Miszczak}(2009)}]{PhysRevA.79.024302}%
  \BibitemOpen
  \bibfield  {author} {\bibinfo {author} {\bibfnamefont {Z.}~\bibnamefont
  {Pucha\l{}a}}\ and\ \bibinfo {author} {\bibfnamefont {J.~A.}\ \bibnamefont
  {Miszczak}},\ }\bibfield  {title} {\bibinfo {title} {Bound on trace distance
  based on superfidelity},\ }\href {https://doi.org/10.1103/PhysRevA.79.024302}
  {\bibfield  {journal} {\bibinfo  {journal} {Phys. Rev. A}\ }\textbf {\bibinfo
  {volume} {79}},\ \bibinfo {pages} {024302} (\bibinfo {year}
  {2009})}\BibitemShut {NoStop}%
\bibitem [{\citenamefont {Juli\`a-Farr\'e}\ \emph {et~al.}(2020)\citenamefont
  {Juli\`a-Farr\'e}, \citenamefont {Salamon}, \citenamefont {Riera},
  \citenamefont {Bera},\ and\ \citenamefont
  {Lewenstein}}]{PhysRevResearch.2.023113}%
  \BibitemOpen
  \bibfield  {author} {\bibinfo {author} {\bibfnamefont {S.}~\bibnamefont
  {Juli\`a-Farr\'e}}, \bibinfo {author} {\bibfnamefont {T.}~\bibnamefont
  {Salamon}}, \bibinfo {author} {\bibfnamefont {A.}~\bibnamefont {Riera}},
  \bibinfo {author} {\bibfnamefont {M.~N.}\ \bibnamefont {Bera}},\ and\
  \bibinfo {author} {\bibfnamefont {M.}~\bibnamefont {Lewenstein}},\ }\bibfield
   {title} {\bibinfo {title} {Bounds on the capacity and power of quantum
  batteries},\ }\href {https://doi.org/10.1103/PhysRevResearch.2.023113}
  {\bibfield  {journal} {\bibinfo  {journal} {Phys. Rev. Res.}\ }\textbf
  {\bibinfo {volume} {2}},\ \bibinfo {pages} {023113} (\bibinfo {year}
  {2020})}\BibitemShut {NoStop}%
\bibitem [{Note2()}]{Note2}%
  \BibitemOpen
  \bibinfo {note} {Note that in the charging power bound, Eq.~\protect \eqref
  {bound_of_power}, we also assume that the battery Hamiltonian ${\protect \hat
  {H}}$ is centered around zero, meaning that the absolute values of the
  maximal and minimal eigenvalues are the same. If this does not hold, we can
  make the bound tighter by centering it by an additive constant.}\BibitemShut
  {Stop}%
\bibitem [{\citenamefont {Andolina}\ \emph
  {et~al.}(2019{\natexlab{b}})\citenamefont {Andolina}, \citenamefont {Keck},
  \citenamefont {Mari}, \citenamefont {Giovannetti},\ and\ \citenamefont
  {Polini}}]{andolina2019quantum}%
  \BibitemOpen
  \bibfield  {author} {\bibinfo {author} {\bibfnamefont {G.~M.}\ \bibnamefont
  {Andolina}}, \bibinfo {author} {\bibfnamefont {M.}~\bibnamefont {Keck}},
  \bibinfo {author} {\bibfnamefont {A.}~\bibnamefont {Mari}}, \bibinfo {author}
  {\bibfnamefont {V.}~\bibnamefont {Giovannetti}},\ and\ \bibinfo {author}
  {\bibfnamefont {M.}~\bibnamefont {Polini}},\ }\bibfield  {title} {\bibinfo
  {title} {Quantum versus classical many-body batteries},\ }\href
  {https://doi.org/10.1103/PhysRevB.99.205437} {\bibfield  {journal} {\bibinfo
  {journal} {Phys. Rev. B}\ }\textbf {\bibinfo {volume} {99}},\ \bibinfo
  {pages} {205437} (\bibinfo {year} {2019}{\natexlab{b}})}\BibitemShut
  {NoStop}%
\bibitem [{\citenamefont {Anandan}\ and\ \citenamefont
  {Aharonov}(1990)}]{PhysRevLett.65.1697}%
  \BibitemOpen
  \bibfield  {author} {\bibinfo {author} {\bibfnamefont {J.}~\bibnamefont
  {Anandan}}\ and\ \bibinfo {author} {\bibfnamefont {Y.}~\bibnamefont
  {Aharonov}},\ }\bibfield  {title} {\bibinfo {title} {Geometry of quantum
  evolution},\ }\href {https://doi.org/10.1103/PhysRevLett.65.1697} {\bibfield
  {journal} {\bibinfo  {journal} {Phys. Rev. Lett.}\ }\textbf {\bibinfo
  {volume} {65}},\ \bibinfo {pages} {1697} (\bibinfo {year}
  {1990})}\BibitemShut {NoStop}%
\bibitem [{\citenamefont {Deffner}\ and\ \citenamefont
  {Campbell}(2017)}]{Deffner_2017}%
  \BibitemOpen
  \bibfield  {author} {\bibinfo {author} {\bibfnamefont {S.}~\bibnamefont
  {Deffner}}\ and\ \bibinfo {author} {\bibfnamefont {S.}~\bibnamefont
  {Campbell}},\ }\bibfield  {title} {\bibinfo {title} {Quantum speed limits:
  from heisenberg’s uncertainty principle to optimal quantum control},\
  }\href {https://doi.org/10.1088/1751-8121/aa86c6} {\bibfield  {journal}
  {\bibinfo  {journal} {Journal of Physics A: Mathematical and Theoretical}\
  }\textbf {\bibinfo {volume} {50}},\ \bibinfo {pages} {453001} (\bibinfo
  {year} {2017})}\BibitemShut {NoStop}%
\bibitem [{\citenamefont {Mandelstam}\ and\ \citenamefont
  {Tamm}(1991)}]{Mandelstam1991}%
  \BibitemOpen
  \bibfield  {author} {\bibinfo {author} {\bibfnamefont {L.}~\bibnamefont
  {Mandelstam}}\ and\ \bibinfo {author} {\bibfnamefont {I.}~\bibnamefont
  {Tamm}},\ }\bibinfo {title} {The uncertainty relation between energy and time
  in non-relativistic quantum mechanics},\ in\ \href
  {https://doi.org/10.1007/978-3-642-74626-0_8} {\emph {\bibinfo {booktitle}
  {Selected Papers}}},\ \bibinfo {editor} {edited by\ \bibinfo {editor}
  {\bibfnamefont {B.~M.}\ \bibnamefont {Bolotovskii}}, \bibinfo {editor}
  {\bibfnamefont {V.~Y.}\ \bibnamefont {Frenkel}},\ and\ \bibinfo {editor}
  {\bibfnamefont {R.}~\bibnamefont {Peierls}}}\ (\bibinfo  {publisher}
  {Springer Berlin Heidelberg},\ \bibinfo {address} {Berlin, Heidelberg},\
  \bibinfo {year} {1991})\ pp.\ \bibinfo {pages} {115--123}\BibitemShut
  {NoStop}%
\bibitem [{\citenamefont {Margolus}\ and\ \citenamefont
  {Levitin}(1998)}]{MARGOLUS1998188}%
  \BibitemOpen
  \bibfield  {author} {\bibinfo {author} {\bibfnamefont {N.}~\bibnamefont
  {Margolus}}\ and\ \bibinfo {author} {\bibfnamefont {L.~B.}\ \bibnamefont
  {Levitin}},\ }\bibfield  {title} {\bibinfo {title} {The maximum speed of
  dynamical evolution},\ }\href
  {https://doi.org/https://doi.org/10.1016/S0167-2789(98)00054-2} {\bibfield
  {journal} {\bibinfo  {journal} {Physica D: Nonlinear Phenomena}\ }\textbf
  {\bibinfo {volume} {120}},\ \bibinfo {pages} {188} (\bibinfo {year}
  {1998})},\ \bibinfo {note} {proceedings of the Fourth Workshop on Physics and
  Consumption}\BibitemShut {NoStop}%
\bibitem [{\citenamefont {Pfeifer}(1993)}]{PhysRevLett.70.3365}%
  \BibitemOpen
  \bibfield  {author} {\bibinfo {author} {\bibfnamefont {P.}~\bibnamefont
  {Pfeifer}},\ }\bibfield  {title} {\bibinfo {title} {How fast can a quantum
  state change with time?},\ }\href
  {https://doi.org/10.1103/PhysRevLett.70.3365} {\bibfield  {journal} {\bibinfo
   {journal} {Phys. Rev. Lett.}\ }\textbf {\bibinfo {volume} {70}},\ \bibinfo
  {pages} {3365} (\bibinfo {year} {1993})}\BibitemShut {NoStop}%
\bibitem [{\citenamefont {Pati}\ \emph {et~al.}(2023)\citenamefont {Pati},
  \citenamefont {Mohan}, \citenamefont {Sahil},\ and\ \citenamefont
  {Braunstein}}]{pati2023exact}%
  \BibitemOpen
  \bibfield  {author} {\bibinfo {author} {\bibfnamefont {A.~K.}\ \bibnamefont
  {Pati}}, \bibinfo {author} {\bibfnamefont {B.}~\bibnamefont {Mohan}},
  \bibinfo {author} {\bibnamefont {Sahil}},\ and\ \bibinfo {author}
  {\bibfnamefont {S.~L.}\ \bibnamefont {Braunstein}},\ }\href
  {https://arxiv.org/abs/2305.03839} {\bibinfo {title} {Exact quantum speed
  limits}} (\bibinfo {year} {2023}),\ \Eprint
  {https://arxiv.org/abs/2305.03839} {arXiv:2305.03839 [quant-ph]} \BibitemShut
  {NoStop}%
\bibitem [{\citenamefont {Funo}\ \emph {et~al.}(2019)\citenamefont {Funo},
  \citenamefont {Shiraishi},\ and\ \citenamefont {Saito}}]{Funo_2019}%
  \BibitemOpen
  \bibfield  {author} {\bibinfo {author} {\bibfnamefont {K.}~\bibnamefont
  {Funo}}, \bibinfo {author} {\bibfnamefont {N.}~\bibnamefont {Shiraishi}},\
  and\ \bibinfo {author} {\bibfnamefont {K.}~\bibnamefont {Saito}},\ }\bibfield
   {title} {\bibinfo {title} {Speed limit for open quantum systems},\ }\href
  {https://doi.org/10.1088/1367-2630/aaf9f5} {\bibfield  {journal} {\bibinfo
  {journal} {New Journal of Physics}\ }\textbf {\bibinfo {volume} {21}},\
  \bibinfo {pages} {013006} (\bibinfo {year} {2019})}\BibitemShut {NoStop}%
\bibitem [{\citenamefont {del Campo}\ \emph {et~al.}(2013)\citenamefont {del
  Campo}, \citenamefont {Egusquiza}, \citenamefont {Plenio},\ and\
  \citenamefont {Huelga}}]{PhysRevLett.110.050403}%
  \BibitemOpen
  \bibfield  {author} {\bibinfo {author} {\bibfnamefont {A.}~\bibnamefont {del
  Campo}}, \bibinfo {author} {\bibfnamefont {I.~L.}\ \bibnamefont {Egusquiza}},
  \bibinfo {author} {\bibfnamefont {M.~B.}\ \bibnamefont {Plenio}},\ and\
  \bibinfo {author} {\bibfnamefont {S.~F.}\ \bibnamefont {Huelga}},\ }\bibfield
   {title} {\bibinfo {title} {Quantum speed limits in open system dynamics},\
  }\href {https://doi.org/10.1103/PhysRevLett.110.050403} {\bibfield  {journal}
  {\bibinfo  {journal} {Phys. Rev. Lett.}\ }\textbf {\bibinfo {volume} {110}},\
  \bibinfo {pages} {050403} (\bibinfo {year} {2013})}\BibitemShut {NoStop}%
\bibitem [{\citenamefont {Sun}\ \emph {et~al.}(2021)\citenamefont {Sun},
  \citenamefont {Peng}, \citenamefont {Hu},\ and\ \citenamefont
  {Zheng}}]{PhysRevLett.127.100404}%
  \BibitemOpen
  \bibfield  {author} {\bibinfo {author} {\bibfnamefont {S.}~\bibnamefont
  {Sun}}, \bibinfo {author} {\bibfnamefont {Y.}~\bibnamefont {Peng}}, \bibinfo
  {author} {\bibfnamefont {X.}~\bibnamefont {Hu}},\ and\ \bibinfo {author}
  {\bibfnamefont {Y.}~\bibnamefont {Zheng}},\ }\bibfield  {title} {\bibinfo
  {title} {Quantum speed limit quantified by the changing rate of phase},\
  }\href {https://doi.org/10.1103/PhysRevLett.127.100404} {\bibfield  {journal}
  {\bibinfo  {journal} {Phys. Rev. Lett.}\ }\textbf {\bibinfo {volume} {127}},\
  \bibinfo {pages} {100404} (\bibinfo {year} {2021})}\BibitemShut {NoStop}%
\bibitem [{\citenamefont {Impens}\ \emph {et~al.}(2021)\citenamefont {Impens},
  \citenamefont {D'Angelis}, \citenamefont {Pinheiro},\ and\ \citenamefont
  {Gu\'ery-Odelin}}]{PhysRevA.104.052620}%
  \BibitemOpen
  \bibfield  {author} {\bibinfo {author} {\bibfnamefont {F.}~\bibnamefont
  {Impens}}, \bibinfo {author} {\bibfnamefont {F.~M.}\ \bibnamefont
  {D'Angelis}}, \bibinfo {author} {\bibfnamefont {F.~A.}\ \bibnamefont
  {Pinheiro}},\ and\ \bibinfo {author} {\bibfnamefont {D.}~\bibnamefont
  {Gu\'ery-Odelin}},\ }\bibfield  {title} {\bibinfo {title} {Time scaling and
  quantum speed limit in non-hermitian hamiltonians},\ }\href
  {https://doi.org/10.1103/PhysRevA.104.052620} {\bibfield  {journal} {\bibinfo
   {journal} {Phys. Rev. A}\ }\textbf {\bibinfo {volume} {104}},\ \bibinfo
  {pages} {052620} (\bibinfo {year} {2021})}\BibitemShut {NoStop}%
\bibitem [{\citenamefont {Meng}\ \emph {et~al.}(2015)\citenamefont {Meng},
  \citenamefont {Wu},\ and\ \citenamefont {Guo}}]{Meng2015}%
  \BibitemOpen
  \bibfield  {author} {\bibinfo {author} {\bibfnamefont {X.}~\bibnamefont
  {Meng}}, \bibinfo {author} {\bibfnamefont {C.}~\bibnamefont {Wu}},\ and\
  \bibinfo {author} {\bibfnamefont {H.}~\bibnamefont {Guo}},\ }\bibfield
  {title} {\bibinfo {title} {Minimal evolution time and quantum speed limit of
  non-markovian open systems},\ }\href {https://doi.org/10.1038/srep16357}
  {\bibfield  {journal} {\bibinfo  {journal} {Scientific Reports}\ }\textbf
  {\bibinfo {volume} {5}},\ \bibinfo {pages} {16357} (\bibinfo {year}
  {2015})}\BibitemShut {NoStop}%
\bibitem [{\citenamefont {Wu}\ and\ \citenamefont {Yu}(2020)}]{Wu2020}%
  \BibitemOpen
  \bibfield  {author} {\bibinfo {author} {\bibfnamefont {S.-x.}\ \bibnamefont
  {Wu}}\ and\ \bibinfo {author} {\bibfnamefont {C.-s.}\ \bibnamefont {Yu}},\
  }\bibfield  {title} {\bibinfo {title} {Quantum speed limit based on the bound
  of bures angle},\ }\href {https://doi.org/10.1038/s41598-020-62409-w}
  {\bibfield  {journal} {\bibinfo  {journal} {Scientific Reports}\ }\textbf
  {\bibinfo {volume} {10}},\ \bibinfo {pages} {5500} (\bibinfo {year}
  {2020})}\BibitemShut {NoStop}%
\bibitem [{\citenamefont {Mohan}\ and\ \citenamefont {Pati}(2022)}]{Mohan2022}%
  \BibitemOpen
  \bibfield  {author} {\bibinfo {author} {\bibfnamefont {B.}~\bibnamefont
  {Mohan}}\ and\ \bibinfo {author} {\bibfnamefont {A.~K.}\ \bibnamefont
  {Pati}},\ }\bibfield  {title} {\bibinfo {title} {Quantum speed limits for
  observables},\ }\href {https://doi.org/10.1103/PhysRevA.106.042436}
  {\bibfield  {journal} {\bibinfo  {journal} {Phys. Rev. A}\ }\textbf {\bibinfo
  {volume} {106}},\ \bibinfo {pages} {042436} (\bibinfo {year}
  {2022})}\BibitemShut {NoStop}%
\bibitem [{\citenamefont {Uhlmann}(1992)}]{UHLMANN1992329}%
  \BibitemOpen
  \bibfield  {author} {\bibinfo {author} {\bibfnamefont {A.}~\bibnamefont
  {Uhlmann}},\ }\bibfield  {title} {\bibinfo {title} {An energy dispersion
  estimate},\ }\href
  {https://doi.org/https://doi.org/10.1016/0375-9601(92)90555-Z} {\bibfield
  {journal} {\bibinfo  {journal} {Physics Letters A}\ }\textbf {\bibinfo
  {volume} {161}},\ \bibinfo {pages} {329} (\bibinfo {year}
  {1992})}\BibitemShut {NoStop}%
\bibitem [{\citenamefont {Deffner}\ and\ \citenamefont
  {Lutz}(2013)}]{Deffner_2013}%
  \BibitemOpen
  \bibfield  {author} {\bibinfo {author} {\bibfnamefont {S.}~\bibnamefont
  {Deffner}}\ and\ \bibinfo {author} {\bibfnamefont {E.}~\bibnamefont {Lutz}},\
  }\bibfield  {title} {\bibinfo {title} {Energy–time uncertainty relation for
  driven quantum systems},\ }\href
  {https://doi.org/10.1088/1751-8113/46/33/335302} {\bibfield  {journal}
  {\bibinfo  {journal} {Journal of Physics A: Mathematical and Theoretical}\
  }\textbf {\bibinfo {volume} {46}},\ \bibinfo {pages} {335302} (\bibinfo
  {year} {2013})}\BibitemShut {NoStop}%
\bibitem [{\citenamefont {Braunstein}\ and\ \citenamefont
  {Caves}(1994)}]{PhysRevLett.72.3439}%
  \BibitemOpen
  \bibfield  {author} {\bibinfo {author} {\bibfnamefont {S.~L.}\ \bibnamefont
  {Braunstein}}\ and\ \bibinfo {author} {\bibfnamefont {C.~M.}\ \bibnamefont
  {Caves}},\ }\bibfield  {title} {\bibinfo {title} {Statistical distance and
  the geometry of quantum states},\ }\href
  {https://doi.org/10.1103/PhysRevLett.72.3439} {\bibfield  {journal} {\bibinfo
   {journal} {Phys. Rev. Lett.}\ }\textbf {\bibinfo {volume} {72}},\ \bibinfo
  {pages} {3439} (\bibinfo {year} {1994})}\BibitemShut {NoStop}%
\bibitem [{\citenamefont {Mondal}\ \emph {et~al.}(2016)\citenamefont {Mondal},
  \citenamefont {Datta},\ and\ \citenamefont {Sazim}}]{MONDAL2016689}%
  \BibitemOpen
  \bibfield  {author} {\bibinfo {author} {\bibfnamefont {D.}~\bibnamefont
  {Mondal}}, \bibinfo {author} {\bibfnamefont {C.}~\bibnamefont {Datta}},\ and\
  \bibinfo {author} {\bibfnamefont {S.}~\bibnamefont {Sazim}},\ }\bibfield
  {title} {\bibinfo {title} {Quantum coherence sets the quantum speed limit for
  mixed states},\ }\href
  {https://doi.org/https://doi.org/10.1016/j.physleta.2015.12.015} {\bibfield
  {journal} {\bibinfo  {journal} {Physics Letters A}\ }\textbf {\bibinfo
  {volume} {380}},\ \bibinfo {pages} {689} (\bibinfo {year}
  {2016})}\BibitemShut {NoStop}%
\bibitem [{\citenamefont {Braunstein}\ and\ \citenamefont
  {Milburn}(1995)}]{PhysRevA.51.1820}%
  \BibitemOpen
  \bibfield  {author} {\bibinfo {author} {\bibfnamefont {S.~L.}\ \bibnamefont
  {Braunstein}}\ and\ \bibinfo {author} {\bibfnamefont {G.~J.}\ \bibnamefont
  {Milburn}},\ }\bibfield  {title} {\bibinfo {title} {Dynamics of statistical
  distance: Quantum limits for two-level clocks},\ }\href
  {https://doi.org/10.1103/PhysRevA.51.1820} {\bibfield  {journal} {\bibinfo
  {journal} {Phys. Rev. A}\ }\textbf {\bibinfo {volume} {51}},\ \bibinfo
  {pages} {1820} (\bibinfo {year} {1995})}\BibitemShut {NoStop}%
\bibitem [{\citenamefont {Campaioli}\ \emph
  {et~al.}(2019{\natexlab{a}})\citenamefont {Campaioli}, \citenamefont {Sloan},
  \citenamefont {Modi},\ and\ \citenamefont {Pollock}}]{physRevA.100.062328}%
  \BibitemOpen
  \bibfield  {author} {\bibinfo {author} {\bibfnamefont {F.}~\bibnamefont
  {Campaioli}}, \bibinfo {author} {\bibfnamefont {W.}~\bibnamefont {Sloan}},
  \bibinfo {author} {\bibfnamefont {K.}~\bibnamefont {Modi}},\ and\ \bibinfo
  {author} {\bibfnamefont {F.~A.}\ \bibnamefont {Pollock}},\ }\bibfield
  {title} {\bibinfo {title} {Algorithm for solving unconstrained unitary
  quantum brachistochrone problems},\ }\href
  {https://doi.org/10.1103/PhysRevA.100.062328} {\bibfield  {journal} {\bibinfo
   {journal} {Phys. Rev. A}\ }\textbf {\bibinfo {volume} {100}},\ \bibinfo
  {pages} {062328} (\bibinfo {year} {2019}{\natexlab{a}})}\BibitemShut
  {NoStop}%
\bibitem [{\citenamefont {Campaioli}\ \emph
  {et~al.}(2019{\natexlab{b}})\citenamefont {Campaioli}, \citenamefont {Sloan},
  \citenamefont {Modi},\ and\ \citenamefont
  {Pollock}}]{campaioli2019algorithm}%
  \BibitemOpen
  \bibfield  {author} {\bibinfo {author} {\bibfnamefont {F.}~\bibnamefont
  {Campaioli}}, \bibinfo {author} {\bibfnamefont {W.}~\bibnamefont {Sloan}},
  \bibinfo {author} {\bibfnamefont {K.}~\bibnamefont {Modi}},\ and\ \bibinfo
  {author} {\bibfnamefont {F.~A.}\ \bibnamefont {Pollock}},\ }\bibfield
  {title} {\bibinfo {title} {Algorithm for solving unconstrained unitary
  quantum brachistochrone problems},\ }\href
  {https://doi.org/10.1103/PhysRevA.100.062328} {\bibfield  {journal} {\bibinfo
   {journal} {Phys. Rev. A}\ }\textbf {\bibinfo {volume} {100}},\ \bibinfo
  {pages} {062328} (\bibinfo {year} {2019}{\natexlab{b}})}\BibitemShut
  {NoStop}%
\end{thebibliography}%

\end{document}